\documentclass[11pt]{article}

\usepackage{amsmath,amssymb}
\usepackage{graphicx,color}
\usepackage{mathrsfs}

\def\tr{{\,\mathrm{tr}\,}}

\numberwithin{equation}{section}
\begin{document}

\quad 
\vspace{-2.0cm}

\begin{flushright}
\parbox{3cm}
{
{\bf Sept 2009}\hfill \\
YITP-09-59 \hfill \\
 }
\end{flushright}

\vspace*{2.5cm}

\begin{center}
\Large\bf 
{\fontfamily{pbk}{\bfseries{\scshape{\selectfont 
M2-branes Theories without 3+1 Dimensional Parents\\ via Un-Higgsing}}}}\\
\end{center}
\vspace*{1.0cm}
\centerline{\large 
\textbf{Masato Taki}
}
\begin{center}
\textit{Yukawa Institute for Theoretical Physics, Kyoto University, Kyoto 606-8502, Japan}
\vspace*{0.2cm}

\begin{center}
{taki@yukawa.kyoto-u.ac.jp}\\
\end{center}

\end{center}

\vspace*{0.7cm}

\centerline{\bf \textbf{Abstract}} 

\vspace*{0.5cm}
$\mathcal{N}=2$ quiver Chern-Simons theory has lately attracted attention as the 
world volume theory of multiple M2 branes on a Calabi-Yau 4-fold.
We study the connection between the stringy derivation of M2 brane theories and the forward algorithm
which gives the toric Calabi-Yau 4-fold as the moduli space of the quiver theory.
Then the existence of the 3+1 dimensional parent, which is the consistent 3+1 dimensional superconformal
theory with the same quiver diagram, is crucial for stringy derivation of M2 brane theories.
We also investigate the construction of M2 brane theories that do not have 3+1 dimensional parents.
The un-Higgsing procedure plays a key role to construct these M2 brane theories. 
We find some $\mathcal{N}=2$ quiver Chern-Simons theories which correspond to interesting Calabi-Yau singularities.
\vspace*{1.0cm}

\vfill

\thispagestyle{empty}
\setcounter{page}{0}

\newpage

\section{Introduction}

Various studies have been done on the physics of D-branes.
An effective theory on the world volume of coincided D-branes has been studied well:
open strings attached to these D-branes give the degree of freedom of the world volume theory
and we find a supersymmetric gauge theory which describe it.
Meanwhile, there has been only a little understanding of the low energy physics of M-branes
since little is known about degree of freedom on M-theory branes.

Recently, there have been important progress toward the understanding of the world volume physics of multiple M2 branes.
Even the world volume theory of the M2 branes probing the simplest background $\mathbb{C}^4$ was not known until quite recently.
An obstruction to construct the theory was the requirement of the maximal supersymmetry $\mathcal{N}=8$ in $2+1$ dimensions:
$\mathcal{N}> 3$ supersymmetry was difficult to realize on $2+1$ dimensional field theory Lagrangians.
A clue to the solution of the problem is the work of Schwarz\cite{Schwarz:2004yj}
where it was pointed out that the introduction of the supersymmetric Chern-Simons term 
enables us to construct theories withe extended supersymmetry.
Inspired by this observation, Bagger-Lambert \cite{Bagger:2007jr}\cite{Bagger:2007vi} and Gustavsson \cite{Gustavsson:2007vu}
found a superconformal Chern-Simons theory 
, which we call the BLG theory,
with manifest $\mathcal{N}=8$ supersymmetry and $SO(8)$ R-symmetry.
$3$-algebra plays an important role in their Lagrangian description,
which is unusual structure from the viewpoint of field theory.
However, the action was rewritten in \cite{VanRaamsdonk:2008ft}
as a $SU(2)\times SU(2)$ Chern-Simons theory instead of the $SO(8)$ gauge group of BLG.
This formulation does not require $3$-algebra, and therefore quiver Chern-Simons theories attracted attentions.
$\mathcal{N}=4$ quiver Chern-Simons theories of the type were constructed by Gaiotto-Witten \cite{Gaiotto:2008sd}\cite{Hosomichi:2008jd}
extending $\mathcal{N}=2$ theories of \cite{Gaiotto:2007qi}. 
Theories with $\mathcal{N}=5,6$  supersymmetry were given by \cite{Hosomichi:2008jb}.

The moduli space of the BLG theory for a specific choice of its Chern-Simons level 
is $\textrm{Sym}^2(\mathbb{R}^8/\mathbb{Z}_2)$ \cite{Lambert:2008et}\cite{Distler:2008mk},
and thus it is believed that the BLG theory with the Chern-Simons level describes two M2 branes probing the $\mathbb{R}^8/\mathbb{Z}_2$ singularity.
However, its moduli space for a generic Chern-Simons level lacks interpretation as a singularity probed by M2 branes.
It is therefore very difficult to construct M2 brane theories for more complicated singularities, such as the toric Calabi-Yau 4-folds.
Meanwhile, Aharony, Bergman, Jafferis and Maldacena \cite{Aharony:2008ug} introduced 
a superconformal $SU(N)\times SU(N)$ (or $U(N)\times U(N)$) Chern-Simons theory 
with manifest $\mathcal{N}=6$ supersymmetry \cite{Benna:2008zy}.
It is thought that supersymmetry of this theory would be enhanced to  $\mathcal{N}=8$ for $k=1,2$. 
Since the moduli space of the theory for the quantized Chern-Simons levels $(k,-k)$ is $\textrm{Sym}^N(\mathbb{C}^4/\mathbb{Z}_k)$,
this theory  is a candidate for the M2 brane theory of $\mathbb{C}^4/\mathbb{Z}_k$ for  an arbitrary choice of $(k,-k)$.
Moreover, the ABJM formulation is very compatible with the extension to M2 brane theories for more complicated backgrounds.

It is also very interesting to generalize the ABJM theory to theories with less supersymmetry.
The world volume theory of M2 branes on a Calabi-Yau 4-fold $X=C(Y)$, which is a cone over  a 7 dimensional Sasaki-Einstein manifolds $Y$,
is believed to be dual to M-theory on the background $AdS_4\times Y$, which preserves $\mathcal{N}=2$ supersymmetry.
For this reason, $\mathcal{N}=2$ quiver Chern-Simons theories 
have been investigated as candidates for the duals of these backgrounds \cite{Martelli:2008si}\cite{Ueda:2008hx}\cite{Hanany:2008cd}.
It is not so easy to construct a Chern-Simons theory whose moduli space is a Calabi-Yau 4-fold.
Then the so-called ``brane tiling" (or ``dimer model") method 
\cite{Hanany:2005ve}\cite{Franco:2005rj}\cite{Franco:2005sm} provides a powerful way to find
a large class of $\mathcal{N}=2$ quiver Chern-Simons theories which are expected to be dual to  $AdS_4\times Y$
\cite{Hanany:2008cd}\cite{Hanany:2008fj}\cite{Franco:2008um}.
M-theory crystals \cite{Lee:2006hw}\cite{Lee:2007kv}\cite{Kim:2007ic} 
also play a key role to understand these quiver Chern-Simons theories \cite{Imamura:2008qs}.
In general, many quiver Chern-Simons theories are associated with a single Calabi-Yau 4-fold
\cite{Franco:2009sp}\cite{Davey:2009sr}\cite{Davey:2009qx},
which is an analogue of the toric duality 
\cite{Feng:2000mi}\cite{Feng:2001xr}\cite{Beasley:2001zp}\cite{Feng:2001bn}\cite{Feng:2002zw} of D3 brane theories.
This phenomena would be very important to understand the low energy physics of M2 branes.
Therefore we study and derive these toric phases of M2 brane theories by using the dimer model and the stringy derivation of M2 brane theories.

Many quiver Chern-Simons theories have been constructed by using parent superconformal theories in $3+1$ dimensions:
we construct a quiver Chern-Simons theory by adopting the quiver of the parent. 
However it was found that every quiver Chern-Simons theory cannot have a parent theory.
This is because the constraints of quiver theory in $3+1$ dimensions coming from the vanishing $\beta$-functions
are absent in superconformal quiver Chern-Simons theories.
Thus we study quiver Chern-Simons theories without parents in order to survey the "landscape" of M2 brane theories. 

In the first half of this article we study the relation between the forward algorithm and stringy derivation of M2 brane theories.
The forward algorithm, which is a method to determine the geometry of the moduli space of a quiver gauge theory,
 for M2 brane theories has been developed in \cite{Martelli:2008si}\cite{Ueda:2008hx}\cite{Hanany:2008cd}.
The dimer model description of a quiver Chern-Simons theory plays an important role to formulate the effective forward algorithm.
The dimer model is a dual graph of a quiver diagram and the Kasteleyn matrix of the dimer gives the toric data of the moduli space \cite{Hanany:2008fj}.
Meanwhile Aganagic \cite{Aganagic:2009zk} found a string theoretical derivation of M2 brane theories, which gives an inverse algorithm.
We study therefore the relation between these two algorithms.
Then we find that the existence of the $3+1$ dimensional parent theory is crucial to relate these two approach.

In the latter half of the paper we introduce the notion of the grandparent theory 
and we construct many theories without a consistent parent theory.
Then the un-Higgsing procedure plays a key role to construct these theories.
In this paper we utilize mainly the specific un-Higgsing of "doubling" type for 
studying quiver Chern-Simons  theories without a consistent parent theory.
This un-Higgsing method is applied to  quiver theories
in recent works \cite{Davey:2009bp}\cite{Benishti:2009ky}.

This paper is organized as follows.
In section 2, we give a brief reviews on the quiver Chern-Simons theory, its moduli space, the forward algorithm and a stringy origin of M2 brane theories.
A relation between the forward algorithm and the stringy derivation of M2 brane theories is discussed in section 3.
In section 4, we introduce the useful idea of the grandparent theory in order to derive phases of a M2 brane theory from the corresponding Calabi-Yau 4-fold.
The notion of "un-Higgsing" also plays a key role in this section. 
Using the idea developed in the previous section, we derive many M2 brane theories by un-Higgsing orbifold grandparents in section 5.
In section 6 we derive three phases of $C(Q^{111})$ theory.
Conclusions are found in section 7. In appendix A, we discuss the cofactor expansion formula
of the permanent.

\section{M2-branes Theories on Calabi-Yau Four-fold singuralities}

In this section, we give a brief review on the world volume theories of M2-branes probing Calabi-Yau four-fold singuralities. 
It is believed that $\mathcal{N}=2$ superconformal quiver Chern-Simons theories in three dimension realize these theories. 
Many techniques have been developed in order to obtain a quiver Chern-Simons theory from a corresponding toric Calabi-Yau geometry and vice versa.

\subsection{$\mathcal{N}=2$ quiver Chern-Simons theories and Higgs branch}

To begin with, we review on construction of a $\mathcal{N}=2$ quiver Chern-Simons theory action 
by using $\mathcal{N}=2$ superfield in $2+1$ dimensions.
Details would be found in \cite{Benna:2008zy}\cite{Martelli:2008si} for example.
The Lagrangian of $\mathcal{N}=2$ quiver Chern-Simons theory is
\begin{align}
S_{\textrm{CS}}&=\sum_{a=1}^{G} \frac{k_a}{4\pi} \int d^3x \int d^4\theta \int_{0}^{1}dt \tr 
[V_a \bar{\mathcal{D}}^{\alpha} (e^{tV_a}\mathcal{D}_{\alpha}e^{-tV_a} )]\nonumber\\
&=\sum_{a=1}^{G}
\int \frac{k_a}{4\pi} \tr [A_a\wedge dA_a+\frac{2}{3}A_a\wedge A_a\wedge A_a -\bar{\chi}_a\chi_a+2D_a\sigma_a].
\end{align}
Here $G$ is the number of the gauge group factors $\prod_{a=1}^{G} U(N_a)$,
and $V_a$ is a vector superfield for the a-th gauge group $U(N_a)$.
$D$ and $\sigma$ are auxiliary fields of the multiplets.
In this article we study theories with gauge factors of the same rank $N_1=N_2=\cdots =N$.
See \cite{Benna:2008zy}\cite{Martelli:2008si} for more general cases.
\begin{align}
S_{\textrm{matter}}&=-\sum_{X_{ab}} \int d^3x \int d^4\theta 
\tr {X_{ab}}^{\dag}e^{-V_a}X_{ab}e^{V_a}+\left[i\int d^2\theta W(X_{ab}) +\textrm{c.c.}\right]\nonumber\\
&=\sum_{X_{ab}} \int d^3x 
\tr \left[ \mathcal{D}_{\mu}{X_{ab}}^{\dag}\mathcal{D}^{\mu}X_{ab}-
\left| \sigma_a X_{ab}-X_{ab}\sigma_b \right|^2 
+D_aX_{ab}{X_{ab}}^{\dag}-D_b{X_{ba}}^{\dag}X_{ba} \right] \nonumber\\
&\qquad\qquad\qquad\qquad\quad
-\sum_{i=1}^{E} \int d^3x \tr \left[ F_iF_i^{\dag}- \frac{\partial W}{\partial \phi_i}F_i 
-\frac{\partial W}{\partial \phi_i}^{\dag}F_i^{\dag}  \right]+\textrm{ fermions}.
\end{align}
$X_{ab}$ is a chiral matter superfield which transforms as the bifundamental representation under the gauge factors $U(N_a)\times U(N_b)$.
The matter fields are also denoted by $\Phi_i$ and $\mathcal{E}=\{\Phi \}=\{ X \}$ is the set of the matter fields. 
The index $i$ runs from $1$ to $E$,
where $E=|\mathcal{E}|$ is the number of the matter fields.

Extensive work has been done to study the moduli space of the supersymmetric gauge theories in 4 and 3 dimensions
\cite{Acharya:1998db}\cite{Morrison:1998cs}\cite{Beasley:1999uz}\cite{Feng:2000mi}\cite{Feng:2001xr}\cite{Feng:2002zw}\cite{Hanany:2005ve}\cite{Franco:2005rj}\cite{Franco:2006gc}\cite{Martelli:2008si}\cite{Ueda:2008hx}\cite{Hanany:2008cd}\cite{Hanany:2008fj}\cite{Franco:2008um}\cite{Hanany:2008gx}\cite{Franco:2009sp}\cite{Davey:2009sr}\cite{Davey:2009qx}\cite{Hanany:2009vx}\cite{Davey:2009bp}\cite{Hewlett:2009bx}.
An important point is that the Higgs branch of a $\prod U(N)$  gauge theory for branes probing a toric singularity $\mathcal{M}$
is the symmetric product $\textrm{Sym}^N \mathcal{M}$ of the abelian moduli space.
We focus on  $U(1)^G$ Chern-Simons theories, since we are now interested in the Calabi-Yau geometry itself $\mathcal{M}$
which a brane probes.
Moreover we study moduli spaces at classical level.
The reason classical analysis is sufficient to study the geometry is 
because it is believed that the moduli space does not be modified by quantum corrections under the toric condition. 
Thus the moduli spaces we study in this article are classical ones of abelian theories.

Let us continue to study the action of a $\mathcal{N}=2$ quiver Chern-Simons theory for the abelian gauge group $U(1)^G$.
The scalar potential of this theory is given by
\begin{align}
V=\sum_{i=1}^{E}  \left|\frac{\partial W}{\partial \phi_i}\right|^2
-\sum_{a=1}^{G}\frac{k_a}{2\pi} D_a\sigma_a
+\sum_{X_{ab}}\left| \sigma_a X_{ab}-X_{ab}\sigma_b \right|^2
-\sum_{X_{ab}} {X_{ab}}^{\dag}X_{ab}(D_a-D_b).
\end{align}
Here we have integrated out the auxiliary fields $F_i$.
The first term is the F-term potential, and the others come from the D-terms.

The structure of the F-term equations of quiver Chern-Simons theory
\begin{align}
F_i^{\dag}=\frac{\partial W}{\partial \phi_i}=0
\end{align}
is completely the same as that of $\mathcal{N}=1$ quiver gauge theories.
The set of the solutions is referred to as the master space \cite{Forcella:2008bb}
\begin{align}
\mathcal{F}=\{ \partial_iW(\phi_i)=0 \}\subset \mathbb{C}^E.
\end{align}
This algebraic variety for our theory gives a toric Calabi-Yau manifold.
The perfect matching variables, as we will see, is very useful to solve the F-term equation,
and we can construct the master space as a symplectic quotient:
\begin{align}
\mathcal{F}=\mathbb{C}^c//U(1)^{c-G-2}.
\end{align}

Turning now to the D-term equations, diference with  $3+1$ dimensional quiver gauge theories will be clear.
We rewrite the third term of the scalar potential as follows:
\begin{align}
\sum_{X_{ab}} {X_{ab}}^{\dag}X_{ab}(D_a-D_b)
=\sum_{a=1}^{G}D_a \left[  \sum_{b=1}^{G}{X_{ab}}^{\dag}X_{ab}
-\sum_{b=1}^{G}X_{ba}{X_{ba}}^{\dag} \right]
=\sum_{a=1}^{G}D_a \mu_a(X).
\end{align}
Here we introduce the moment map $\mu_a$ for the a-th gauge group:
\begin{align}
\mu_a(X)\equiv 
\sum_{b=1}^{G}X_{ab}^{\dag}X_{ab}
-\sum_{b=1}^{G}X_{ba}X_{ba}^{\dag}.
\end{align}
The equation of motion of the auxiliary field $D$ is given by
\begin{align}
\label{momentmap}
\mu_a(X)=\frac{k_a \sigma_a}{2\pi}.
\end{align}
We can regard $\zeta_a\equiv k_a\sigma /2\pi$ as an analogue of the FI parameter of $3+1$ dimensional theories.
One essential differene is that $\zeta$ is not a parameter but a vacuum expectation value (VEV) of the auxiliary field $\sigma_a$.
The fields $D_a$ are linear in the action, which is different from the case of  $3+1$ dimensional gauge theories, 
and they play therefore a role of  Lagrange multipliers.
We can therefore integrate out the auxiliary field $D$ with this equation (\ref{momentmap}).
After integrating out $D$, the D-term potential becomes
\begin{align}
V_{\textrm{D-terms}}=\sum_{X_{ab}}\left| \sigma_a X_{ab}-X_{ab}\sigma_b \right|^2.
\end{align}
Then we obtain the D-term equations
\begin{align}
\label{Dterm1}
X_{ab}(\sigma_a-\sigma_b)=0.
\end{align}

In this article we study, following \cite{Martelli:2008si},  the special branch on which all the VEV's of matter fields satisfy $X_{ab}\neq 0$.
We can rewrite this condition using (\ref{Dterm1})
\begin{align}
\sigma_1=\cdots =\sigma_G=\sigma\in \mathbb{R}^{*}.
\end{align}
This is the so-called Higgs branch which admits the interpretation as the moduli space of an M2 brane on a Calabi-Yau.

Meanwhile let us sum up (\ref{momentmap}) over $a=1, \cdots ,G$.
Since $\sum \mu_a=0$ follows from the definition of $\mu_a$, we find
\begin{align}
\sum_{a=1}^{G} k_a \sigma_a=0.
\end{align}
On the Higgsed branch, this relation implies
\begin{align}
\sum_{a=1}^{G} k_a \sigma =0.
\end{align}
The necessary condition for the existence of the branch $\sigma\neq 0$ is therefore given by
\begin{align}
 \sum_{a=1}^{G} k_a =0.
\end{align}
We will study such an assignment of the Chern-Simons levels troughtout the paper.
We also impose the following condition for simplicity:
\begin{align}
\textrm{gcd}(\{ k_a \})=1.
\end{align}
From these conditions, the over all $U(1)$ is decoupled from the theory.
Moreover the "FI-term" $\zeta_a\propto\sigma k_a$ pick out the special $U(1)$ along the direction of the 
Chern-Simons level vector $k$.
There exist therefore the remaining $U(1)^{G-2}$ in this theory.
Thus the moduli space can be computed as the symplectic quotient of the master space:
\begin{align}
\mathcal{M}=\mathcal{F}//U(1)^{G-2}.
\end{align}
Here we impose $G-2$ D-term conditions and gauge symmetry as the symplectic quotient on the solution space of the F-term equations.

\subsection{M2 brane theories, dimer models and moduli spaces}
\begin{figure}[htbp]
\begin{center}
\includegraphics[width=10cm,bb=0 0 501 263]{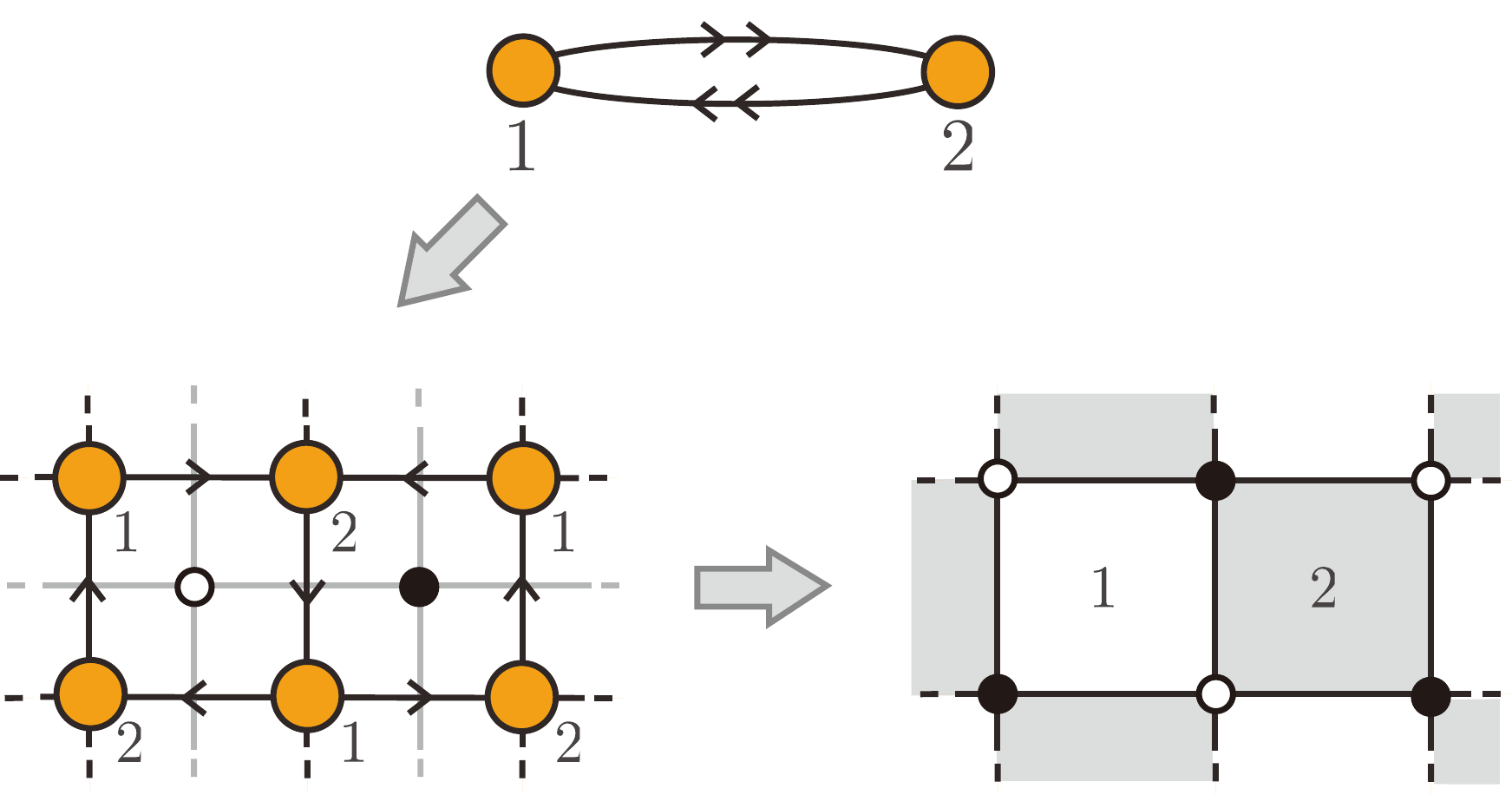}\\
\end{center}
\caption{The quiver diagram, the periodic quiver diagram and the dimer model of the ABJM theory.}
\label{def1}
\end{figure}
The brane tilings, or the dimer models, have been an important tool to study quiver gauge theories.
A dimer model of our interest is a bipartite graph on a 2-torus 
and we shall describe it as a graph consists of black and white nodes and edges on the fundamental domain of the torus.
We will see many dimer models of concrete examples in the following sections.
This diagram encodes the information on a quiver theory effectively.
It is easy to associate a dimer with a quiver diagram:
the dimer model corresponding to a quiver is defined as the dual diagram of the periodic quiver diagram as Fig.\ref{def1}.
Each edge labelled by $i$ corresponds to a bifundamental matter chiral superfield $\Phi_i\in \mathcal{E}$ of the quiver theory 
and each face labelled by $a$ is associated with a gauge group $U(1)_a$.
This assignment is very natural since the dimer model is the dual graph of the quiver diagram.
We can assign the gauge charges to matters as follows.
At first we assign an orientation on the diagram: 
we define clockwise orientation around white nodes and we define anti-clockwise orientation around black nodes.
\begin{figure}[htbp]
\begin{center}
\includegraphics[width=4cm,bb=0 0 171 172]{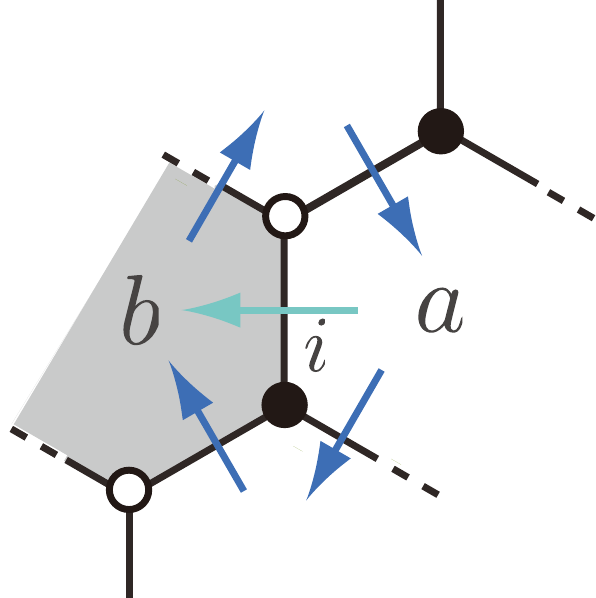}\\
\end{center}
\caption{The assignment of the elements $d_{ai}=-d_{bi}=1$ of the incidence matrix for an edge $i$.}
\label{def2}
\end{figure}
Let us consider the part of the dimer model Fig.\ref{def2}.
The edge $i$ crosses with the orientation arrow from the face $a$ to $b$,
and then we define the $U(1)_a$ charge $d_{ai}$ of the matter field $\Phi_i$ as follows:
\begin{align}
d_{ai}=-d_{bi}=1, \textrm{ otherwise } d_{ci}=0.
\end{align}
In this case $\Phi_i$ transforms as the bifundamental representation under $U(1)_a\times U(1)_b$.
We also denote  $\Phi_i$  as $X_{ab}$.
These $U(1)$ charges form the $G\times E$ incidence matrix $d$.
We can rewrite the Chern-Simons levels by using this matrix and integers $n_i$:
\begin{align}
k=d\cdot n.
\end{align}

In this paper we assume that the quiver Chern-Simons theories satisfy the toric condition: 
each matter field appears in the superpotential precisely twice with opposite sign.
Under this assumption, we can construct the superpotential of the theory from the dimer model:
\begin{align}
\label{defsuperpot}
W=\sum_{\circ \in \mathcal{W}}\prod_{i\in \mathcal{E}_{\circ}}\Phi_i
-\sum_{\bullet \in \mathcal{B}}\prod_{i\in \mathcal{E}_{\bullet}}\Phi_i.
\end{align}
Here $\mathcal{W}$ and $\mathcal{B}$ are the sets of the white and the black nodes.
$\mathcal{E}_{\circ}$ indicates the set of the edges which are attached to the node $\circ$.
The same is true for the black node $\bullet$.

The dimer model associated with a quiver theory not only is very useful  for the description of the quiver theory,  
it also gives the effective way to investigate the moduli space of the quiver gauge theory.
The forward algorithm is the method to derive the moduli space from the dimer model (or quiver theory).
We introduce the Kasteleyn matrix of the dimer model for the purpose.
The row of the matrix indices the black nodes of the dimer model and the column indices the white nodes.
The $(\bullet,\circ)$-component of the Kasteleyn matrix is defined by
\begin{align}
K_{\bullet\circ}(x,y,z)=\sum_{i_{\bullet\circ}}\Phi_{i_{\bullet\circ}}w_{n_{i_{\bullet\circ}}}(x,y)z^{n_{i_{\bullet\circ}}}.
\end{align}
Here $i_{\bullet\circ}$ is an edge which connects the nodes $\bullet$ and $\circ$.
If the edge $i$ crosses the boundary of the fundamental domain, the weight $w_i$ gets the facotr $x$ or $y$ (or $x^{-1}$ or $y^{-1}$ 
according to the orientation).
The weight is one if the edge does not cross the boundary.

Let us formulate the fast forward algorithm with the Kasteleyn matrix.
The permanent of the Kasteleyn matrix, in particular, is very useful to compute the moduli space:
\begin{align}
\textrm{perm}K_{\bullet\circ}(x,y,z)=\sum_{\alpha=1}^{c}p_{\alpha}x^{u_{\alpha}}y^{v_{\alpha}}z^{q_{\alpha}}
\end{align}
See Appendix.\ref{sec:perm} for the definition of $\textrm{perm}$.
The point is that $\textrm{perm}K$
gives the points $(u_{\alpha},v_{\alpha},q_{\alpha})$ of the toric diagram of the moduli space.
We refer to the monomial $p_{\alpha}$ of the fundamental fields $\Phi_i$ as the perfect matchings.

The perfect matching matrix $P$ is also an important object in the forward algorithm.
This $E\times c$ matrix is defined by
\begin{align}
&P_{i\alpha}=1 \textrm{ if } \Phi_{i}\in p_{\alpha},\nonumber\\
&P_{i\alpha}=0 \textrm{ otherwise}.\nonumber
\end{align}
The kernel of the matrix $Q_F\equiv \textrm{ker} P$ gives the charge matrix of the perfect matchings,
and therefore we obtain the GLSM description of the Master space:
\begin{align}
\mathcal{F}=\mathbb{C}^c//Q_F.
\end{align}
Next we impose the D-term conditions on the space.
Let us recall that the incidence matrix $d$ encodes the $U(1)$ charges 
and the perfect matching matrix transform the matter fields into the perfect matchings. 
Thus we can compute the $U(1)$ charges $\tilde{Q}$ of the perfect matchings:
\begin{align}
d=\tilde{Q}\cdot P^{t}.
\end{align}
Only the $U(1)^{G-2}$ symmetry which is orthogonal to $1$ and $k$ direction appears in the D-term charge matrix
since the two linear combinations of $U(1)$'s for weights $1$ and $k$ are decoupled from the theory.
The D-term charge is therefore given  by
\begin{align}
Q_D=\textrm{ker} C\cdot \tilde{Q},
\end{align}
where
\begin{align}
C&=\left(\begin{array}{cccc}1 & 1 & \cdots & 1 \\k_1 & k_2 & \cdots & 0\end{array}\right)\nonumber\\
&=\left(\begin{array}{c}1^t \\k^t \end{array}\right).
\end{align}
Using these data, we can construct the moduli space as a symplectic quotient:
\begin{align}
\mathcal{M}=\mathcal{F}//Q_D=(\mathbb{C}^c//Q_F)//Q_D.
\end{align}
This means that the integral kernel of the total charge ${}^tQ_t=({}^tQ_D,{}^tQ_F)$ gives the matrix into which
the points of the toric diagram are collected:
\begin{align}
G=\textrm{ker}Q_t.
\end{align}

The equivalence between the Kasteleyn  matrix method and the symplectic quotient approach was shown in \cite{Davey:2009sr}.

\subsection{Stringy derivation of M2 brane theories}
As we saw the algorithm to derive the moduli space from a quiver Chern-Simons theory,
we review method to determine the a quiver Chern-Simons theory associated with M2 branes on a Calabi-Yau by utilizing string theory.
This inverse algorithm was developed in \cite{Aganagic:2009zk}.

Let us consider M2 branes on the following Calabi-Yau 4-fold $\mathcal{M}$:
\begin{align}
\sum_{\alpha =1}^{c} Q_{l\alpha} |X_{\alpha}|^2=r_l, \quad l=1,2,\cdots,c-4.
\end{align}
We collect the charge vectors $Q_{l\alpha}$ into the $(c-4)\times c$ matrix $Q_t$ which we defined before.
The chiral fields $X_{\alpha}$  are divided by the $U(1)$ gauge groups 
\begin{align}
X_{\alpha}\to e^{i\lambda_lQ_{l\alpha}} X_{\alpha}.
\end{align}

We can reconstruct the 4-fold as a fibration over a Calabi-Yau 3-fold 
by adding new variables $r_0\in \mathbb{R}$ and  $\theta_0\in \mathbb{R}$.
First we introduce an additional charge $Q_0$ which satisfies
\begin{align}
\sum_{\alpha} Q_{0\alpha} |X_{\alpha}|^2 =r_0
\end{align}
and the toric condition $\sum_{\alpha}Q_{0\alpha} =0$.
We can avoid changing the geometry by dividing the additional gauge symmetry:
\begin{align}
&\theta_0 \to \theta_0 + \lambda_0,\nonumber\\
&X_{\alpha}\to e^{i\lambda_0Q_{0\alpha}}X_\alpha.
\end{align}
We define the vector $q$ 
\begin{align}
\sum_\alpha Q_{0\alpha} q_\alpha=1,\quad \sum_\alpha Q_{l\alpha} q_\alpha=0.
\end{align}
We represent them in matrix notation as $Q_0\cdot q=0$ and $Q_t\cdot q=1_{(c-4)}$.

By fixing $r_0$ and  $\theta_0$, we can view the 4-fold as a $\mathbb{R}$ and $S^1$ fibration over the CAlabi-Yau 3-fold.
The base is defined by
\begin{align}
&\sum_{\alpha =1}^{c} Q_{l\alpha} |X_{\alpha}|^2=r_l,\\
&\sum_{\alpha =1}^{c} Q_{0\alpha} |X_{\alpha}|^2=r_0.
\end{align}
This prescription is the starting point of Aganagic's argument in \cite{Aganagic:2009zk}.

Let us compactify M-theory on the circle fibered over the base 3-fold $\mathcal{M}_3$.
The resulting Type IIA superstring theory on  $\mathcal{M}_3\times \mathbb{R}$ contains D2 branes 
and RR 2-form fluxes $G_{(2)}=dA_{(2)}=\sum q_{\alpha} \omega_{\alpha}$, 
which are induced from the non-trivial curvature of the fibration \cite{Aganagic:2009zk}.

D2 branes on the singularity  $\mathcal{M}_3$ decay into fractional branes and
these fractional branes imply the non-trivian quiver gauge theory on $2+1$ dimensional world volume.
However we expect that the resulting theory is a Chern-Simons theory since our set-up originates from M2 branes on singularity.
The point is that the flux through vanishing cycles induces the Chern-Simons terms.
Let us consider the fractional brane which is a wrapped D4 brane on a vanishing cycle $\Delta_a$:
\begin{align}
\sum_\alpha Q_{a\alpha} |X_\alpha|^2 =t_a.
\end{align}
Then the  Wess-Zumino term on the world volume implies the Chern-Simons levels corresponding to gauge factor $\Delta_a$ as
\begin{align}
k_a &=\int_{\Delta_a} G_{(2)}\nonumber\\
&=\sum_{\alpha} Q_{a\alpha} q_\alpha =Q\cdot q.
\end{align}
Here we use
\begin{align}
\int_{\Delta_a} \omega_\alpha^{(2)}=Q_{a\alpha},
\end{align}
for the 2-cycle $\Delta_a$.
By assuming that the kinetic terms of the gauge fields vanish at IR, 
we obtain the quivr Chern-Simons theory on the world volume of D2 (or M2) branes.
This is the outline of the stringy derivation which was found in \cite{Aganagic:2009zk}.

\section{Forward Algorithm and Stringy Construction}

\subsection{Fractional brane and perfect matchings}

\subsubsection*{\bf Aganagic's construction of Calabi-Yau 3-fold and M2 brane theory}

Let us recall the inverse algorithm of Aganagic which we reviewed in the previous section.
In the inverse algorithm, we start with a Calabi-Yau 4-fold which a M2 brane probes.
First we choose a charge vector $Q_0$ which satisfies ${}^tQ_0\cdot 1_G=0$.
By adding these GLSM charges ${Q_0}_{\alpha}$ to the original GLSM charges $Q_{\mu \alpha}$ 
which define the original Calabi-Yau 4-fold, 
we obtain a Calabi-Yau 3-fold which serves as a $3+1$ dimensional parent.
The inverse algorithm in $3+1$ dimensions is well understood.
In this way we obtain a quiver diagram of the M2 brane theory using this parents.
Next we define a $G\times 1$ vector $q$ as a solution of the following constraints:
\begin{align}
\label{defq}
{}^tQ_0\cdot q=1,\quad {}^tQ_{\mu}\cdot q=0.
\end{align}
In addition, we can get the fractional charge matrix $Q_{a\alpha}$ of Calabi-Yau 3-fold.
We can therefore find the Chern-Simons level vector $k$ of the M2 brane theory using these data:
\begin{align}
\label{cslofag}
k=Q\cdot q.
\end{align}

\subsubsection*{\bf Forward algorithm for Calabi-Yau 4-fold}
Next let us summarize the forward algorithm. 
The forward algorithm provides a way to determine the Calabi-Yau geometry of the moduli space 
from the quiver gauge theory which describes the world volume theory of branes on a toric Calabi-Yau singularity.
In the forward algorithm, we start with the dimer model which describes a quiver Chern-Simons theory.
Using the prescription which we have reviewed in the previous section, we derive the incidence matrix $d$ and
the perfect matching matrix $P$.
The relations 
\begin{align}
\label{csloffa}
d=\tilde{Q}\cdot{}^tP,\quad k=d\cdot n
\end{align}
imply the charge matrix $Q$ and the integral vector $n$.
These data give the toric diagram of the moduli space.

\subsubsection*{\bf Forward algorithm and Calabi-Yau 3-fold - Proposal}

Having seen the inverse and forward algorithms, we now able to study relation between them.
Comparing (\ref{cslofag}) and  (\ref{csloffa}), we propose the relation between these two approach:
\begin{align}
Q=\tilde{Q},\quad q={}^tP\cdot n.
\end{align}
This means that the matrix $\tilde{Q}$ gives fractional brane charges with respect to the perfect matchings.
In addition, the integers $q_{\alpha}$ give the third coordinates of the 3 dimensional toric diagram which 
would be projected out when we derive the 2 dimensional toric diagram of the parent.
In other words, the vector $q$ satisfies $Q_t\cdot q=0$ as follows:
\begin{align}
&Q_F\cdot q=(Q_F\cdot {}^tP)\cdot n=0,\nonumber\\
&Q_D\cdot q=Q_D\cdot {}^tP\cdot n={\textrm{Ker}}(C)\cdot \tilde{Q}\cdot {}^tP\cdot n={\textrm{Ker}}(C)\cdot k=0.\nonumber
\end{align}
To show the last equality, we use ${}^tv\cdot k=0$ for $v\in {}^t{\textrm{Ker}}(C)$.
Thus $q_{\alpha}$ gives the third coordinate of the point which corresponds to a perfect matching $p_{\alpha}$. 

In order to relate the forward algorithm with the above-mentioned inverse algorithm,
we also have to represent the charge vector $Q_0$ in the language of the forward algorithm.
The point is that $Q_0$ corresponds to the $U(1)$ charges of GLSM fields for Calabi-Yau 3-fold 
which is eliminated from the set of $U(1)$ charges in the original Calabi-Yau 4-fold.
Similarly, we pick $G-2$ baryonic $U(1)$'s from $U(1)^G$  when we compute the mesonic moduli space.
It is therefore natural that $U(1)$ defined by $Q_0$ is precisely one of the two remaining $U(1)$'s.
In the forward algorithm, the baryonic $U(1)$'s are projected onto the hyperplane $\mathbb{Z}^{G-2}$ orthogonal to $1_c$ and $k$ 
via  $Q_D={\textrm{Ker}}(C)\cdot \tilde{Q}$. Hence, 
\begin{align}
{}^t\tilde{Q}\cdot 1_c,\quad {}^t\tilde{Q}\cdot k,
\end{align}
are the charges of the remainder which are translated into the language of the perfect matchings.
${}^t\tilde{Q}\cdot 1_c$ cannot be $Q_0$, 
since this $U(1)$ is already encoded in the charge matrix of F-terms ${}^t\tilde{Q}\cdot 1_c\in {\textrm{Ker}}(P)=Q_F$.
We can show it by using ${}^td=P\cdot {}^t\tilde{Q}$ and $\sum_a d_{ai}=0$:
\begin{align}
P\cdot ({}^t\tilde{Q}\cdot 1_G)={}^td\cdot 1_c=0_E.
\end{align}
Then, ${}^t\tilde{Q}\cdot k$ is the only candidate for the charge $Q_0$ which specifies the Calabi-Yau 3-fold and the parent.
We define the following charge vector for our purpose:
\begin{align}
\hat{Q}_{0}={}^t\tilde{Q}\cdot k.
\end{align}
It is not so hard to prove that the sum of the charges is zero ${}^t\hat{Q}_{0}\cdot 1_c=0$ by using ${}^tP\cdot 1_E\propto 1_c$ \cite{Davey:2009sr}. 
We have to investigate the inner product of $\hat{Q}_0$ and $q$, since $Q_0$ must satisfy the first equation of (\ref{defq}).
The answer is given by
\begin{align}
{}^t\hat{Q}_0\cdot q={}^tk \cdot \tilde{Q}\cdot {}^tP\cdot n={}^tk\cdot k.
\end{align}
Thus we find a key property of $\hat{Q}_{0}$:
\begin{align}
{}^t\hat{Q}_{0}\cdot q = k^2.
\end{align}
Meanwhile $Q_0$ is defined by (\ref{defq}).
It is therefore natural to define the charge vector $Q_0$ as follows:
\begin{align}
Q_0=\frac{1}{k^2}\hat{Q}_0.
\end{align}
A problem now arises:
the definition of $Q_0$ would not give a integral vector for generic case.
We study this problem by taking simple theories for example.
Then we propose that the integral answer $\hat{Q}_0/{k^2}\in \mathbb{Z}^c$ involves a M2 brane theory with a consistent parent.

\subsection{$\mathbb{C}^4$ theory}

\subsubsection*{\bf Phase I: ABJM theory $\mathscr{C}$}

\begin{figure}[htbp]
\begin{center}
\includegraphics[width=5cm,bb=0 0 204 204]{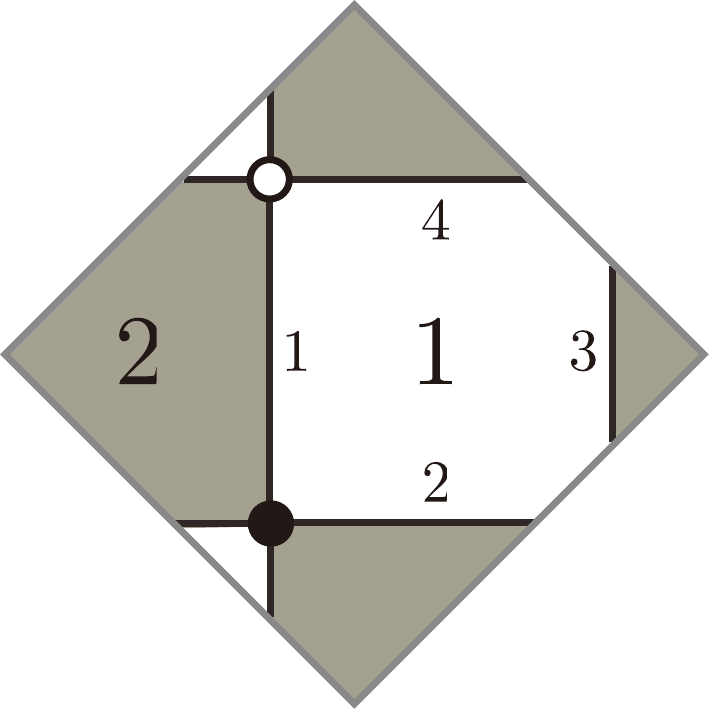}\\
\end{center}
\caption{The dimer model of the ABJM theory.}
\label{ABJMdimer}
\end{figure}
\begin{figure}[htbp]
\begin{center}
\includegraphics[width=4cm,bb=0 0 168 56]{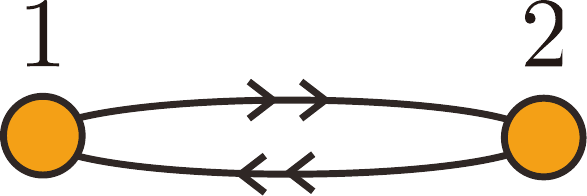}\\
\end{center}
\caption{The quiver of the ABJM theory. The Chern-Simons levels are $k_1=-k_2=1$.}
\label{ABJMquiver}
\end{figure}

Let us analyze the well-studied M2 brane theory for $\mathbb{C}^4$: the ABJM theory.
The dimer model and the quiver diagram of the ABJM theory are shown in Fig.\ref{ABJMdimer} and Fig.\ref{ABJMquiver} respectively.
Since this dimer model consists of two square tiles, we shall refer to it as the chessboard model $\mathscr{C}$ following \cite{Davey:2009sr}.
An important point is that Fig.\ref{ABJMquiver} is the same quiver diagram as the well-known Klebanov-Witten theory \cite{Klebanov:1998hh}.
Hence the Klevanov-Witten theory is the $3+1$ dimensional parent for the ABJM theory.
We expect that the ABJM theory is related with the conifold, 
since the Klebanov-Witten theory is a worldvolume theory of D3 branes on the Calabi-Yau 3-fold
\footnote[1]{See \cite{Maruyoshi:2009uk} for a new Seiberg dual description of the Klebanov-Witten theory with $N_c=2$.}.
The relation has realized in \cite{Martelli:2008si}\cite{Aganagic:2009zk}: a fibration over the conifold gives the Calabi-Yau 4-fold $\mathbb{C}^4$.
We review the fact here on the way to the study of the existence of the integer charges $Q_0$.

The Chern-Simons levels we study here are given by ${}^t {k}=(1,-1)$. 
Following the formalism we reviewed in the previous section, 
we can associate the incident matrix with the dimer model:
\begin{equation}
\label{dconifold}
d=\begin{array}{c|cccc}
& \phi_1 & \phi_2 & \phi_3 & \phi_4
\\ \hline 1 & 1 & -1 & 1 & -1 
\\ 2 &-1 & 1 & -1 &1\end{array}.
\end{equation}
Using the relation $ {k}=d\cdot{}^t n$, we find the integer vector $n$: 
\begin{align}
{}^t {n}=(0,0,1,0).
\end{align}
The dimer model Fig.\ref{ABJMdimer} also gives the Kasteleyn matrix.
Since there are one white node and black node in the dimer, 
the Kasteleyn matrix is a $1\times 1$ matrix.
\begin{align}
K=\phi_1+\phi_2x+\phi_3xyz+\phi_4y.
\end{align}
Thus the perfect matchings are given by
\begin{align}
p_1=X_{12}^1=\Phi_1,\quad p_2=X_{12}^2=\Phi_3,\quad p_3=X_{21}^2=\Phi_4,\quad p_4=X_{21}^1=\Phi_2.
\end{align}
There exists therefore a one to one correspondence between matter fields and perfect matchings.
The perfect matching matrix is given by
\begin{align}
\label{Pconifold}
P=\left(\begin{array}{cccc}1 & 0 & 0 & 0 \\0 & 0 & 0 & 1 \\0 & 1 & 0 & 0 \\0 & 0 & 1 & 0\end{array}\right).
\end{align}
The relation $d=\tilde{Q}\cdot {}^tP$ implies the charge matrix of the model:
\begin{align}
\tilde{Q}=\left(\begin{array}{cccc}1 & 1 & -1 & -1 \\-1 & -1 & 1 & 1\end{array}\right).
\end{align}
Then $\hat{Q}_0={}^t \tilde{Q}\cdot k$ and ${}^t q={}^t n\cdot P$ gives
\begin{align}
\label{q0ABJM}
&{}^t \hat{Q}_0=(2,2,-2,-2),\nonumber\\
&{}^t q=(0,1,0,0).\nonumber
\end{align}
Hence we find that the relation ${}^t \hat{Q}_0 \cdot q=2=k^2$ holds as expeted. 
These result means that we can find an integer charge vector
\begin{align}
{}^t Q_0&=\frac{1}{k^2}{}^t \hat{Q}_0\nonumber\\
&=(1,1,-1,-1) \in \mathbb{Z}^4,
\end{align}
which satisfies ${}^t Q_0 \cdot q=1$.
Notice that the $U(1)$ charge vector for GLSM fields is precisely the charges for conifold.
In other word, the $U(1)$ quotient of Calabi-Yau 4-fold $\mathbb{C}^4$ implies the conifold $\mathbb{C}^4//U(1)_{Q_0}=\mathcal{C}$ 
which gives the quiver diagram of ABJM theory as a $3+1$ dimensional parent.
We propose that the existence of the integer charges such as (\ref{q0ABJM}) for the ABJM theory is a peculiar feature of the M2 brane theories
which have consistent $3+1$ dimensional parent theories.
As we will see in the following examples, M2 brane theories without consistent parents do not implies the integer charges.

\subsubsection*{\bf Phase II: dual ABJM theory $\mathscr{D}_1 \mathscr{H}_1$}

\begin{figure}[htbp]
\begin{center}
\includegraphics[width=6cm,bb=0 0 213 125]{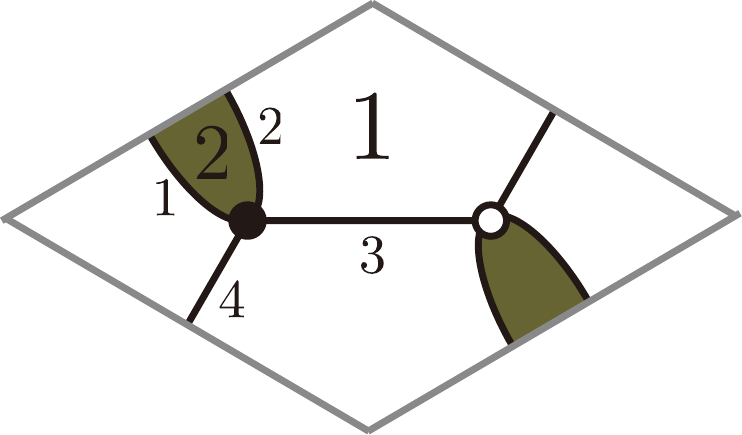}\\
\end{center}
\caption{The dimer model of the dual ABJM theory.}
\label{dualABJMdimer}
\end{figure}
\begin{figure}[htbp]
\begin{center}
\includegraphics[width=5cm,bb=0 0 199 51]{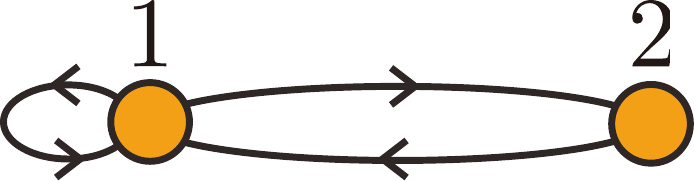}\\
\end{center}
\caption{The quiver of the dual ABJM theory. The Chern-Simons levels are $k_1=-k_2=1$.}
\label{dualABJMquiver}
\end{figure}

It was found in \cite{Hanany:2008fj} that a quiver Chern-Simons theory 
which has no cocsistent parents also describes the world volume theory of M2 branes on $\mathbb{C}^4$.
The dimer model of this theory is drawn in Fig.\ref{dualABJMdimer}.
This model is called the one double-bonded one-hexagon model  $\mathscr{D}_1 \mathscr{H}_1$ \cite{Davey:2009sr} 
since the dimer model consists of tiles in the shape of a hexagon with a double bond.
The quiver diagram is shown in Fig.\ref{dualABJMquiver}.
The Chern-Simons levels are given by ${}^tk=(1,-1)$. 
The incidence matrix of the dimer graph is
\begin{equation}
d=\begin{array}{c|cccc}
& \phi_1 & \phi_2 & \phi_3 & \phi_4
\\ \hline 1 & -1 & 1 & d_{13} & d_{14}
\\ 2 &1 & -1 & 0 &0\end{array}.
\end{equation}
Here there exists an ambiguity of a choice of components $d_{13} $ and $d_{14}$.
Though we can choose them zero by following \cite{Davey:2009sr}, we leave them ambiguous.
Then we find the integer vector $n$ which satisfy $k={}^td\cdot n$:
\begin{align}
{}^tn=(0,0,1,0).
\end{align}
The perfect matching matrix is 
\begin{align}
P=\left(\begin{array}{cccc}0 & 0 & 1 & 0 \\ 1 & 0 & 0 & 0 \\0 & 0 & 0 & 1 \\0 & 1 & 0 & 0\end{array}\right),
\end{align}
where we use the $1\times 1$ Kasteleyn matrix $K=\phi_1x^{-1}+\phi_2x^{-1}z+\phi_3+\phi_4y$.

We can find an integral solution of the equation $d={}^t\tilde{Q}\cdot P$:
\begin{align}
\tilde{Q}=\left(\begin{array}{cccc}1 & d_{14} & -1 & d_{13} \\ -1 & 0 & 1 & 0 \end{array}\right).
\end{align}
Then we obtain
\begin{align}
&{}^t \hat{Q}_0={}^tk\cdot \tilde{Q}=(2,d_{14},-2,d_{13}),\nonumber\\
&{}^t q={}^tn\cdot n=(1,0,0,0).\nonumber
\end{align}
They satisfy ${}^tq\cdot \hat{Q}_0=2=k^2$ as expected.

An important point is that for generic $d_{13}$ and $d_{14}$ the relation
\begin{align}
Q_0=\frac{1}{k^2}\hat{Q}_0
\end{align}
involves a fractional charge vector $Q_0$.
With the above-mentioned choice $d_{13}=d_{14}=0$, $Q_0$ would be a integer charge vector.
Notice that not every M2 brane theory without a parent  implies the integral $Q_0$.
Therefore the dual ABJM model $\mathscr{D}_1 \mathscr{H}_1$ is a peculiar example 
in that it does not have a consistent parent but gives a integral $Q_0$ under specific conditions.
Below, we investigate some phases without consistent parents which give non-integral charges $Q_0$.

\subsection{$\mathbb{C}\times \mathcal{C}$ theory}

Next let us discuss the phases of the $\mathbb{C}\times \mathcal{C}$ theory which was obtained in \cite{Davey:2009sr}.

\subsubsection*{\bf Phase I: The $\mathscr{D}_1\mathscr{C}$ model}
\begin{figure}[htbp]
\begin{center}
\includegraphics[width=5cm,bb=0 0 204 204]{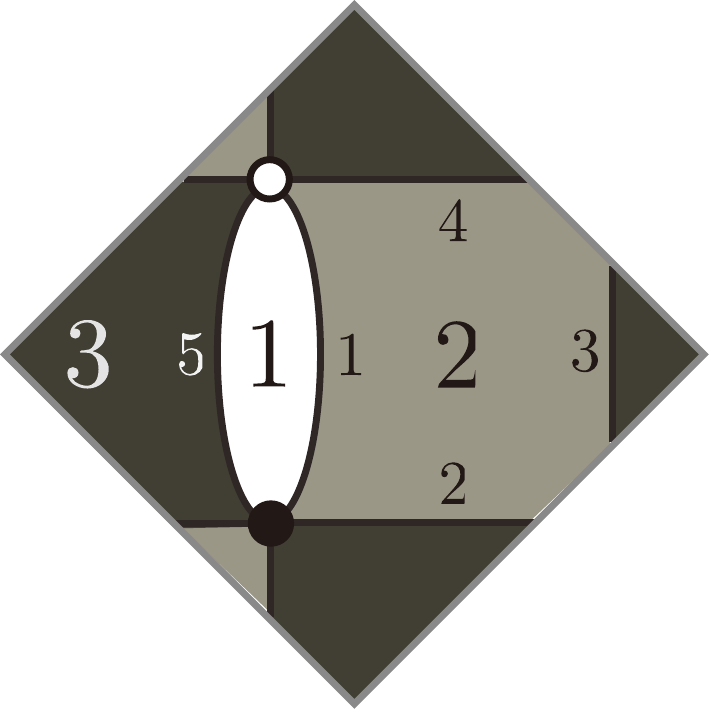}\\
\end{center}
\caption{The dimer model of the Phase I of the $\mathbb{C}\times \mathcal{C}$ theory $\mathscr{D}_1\mathscr{C}$.}
\label{D1Cdimer}
\end{figure}
\begin{figure}[htbp]
\begin{center}
\includegraphics[width=4cm,bb=0 0 168 146]{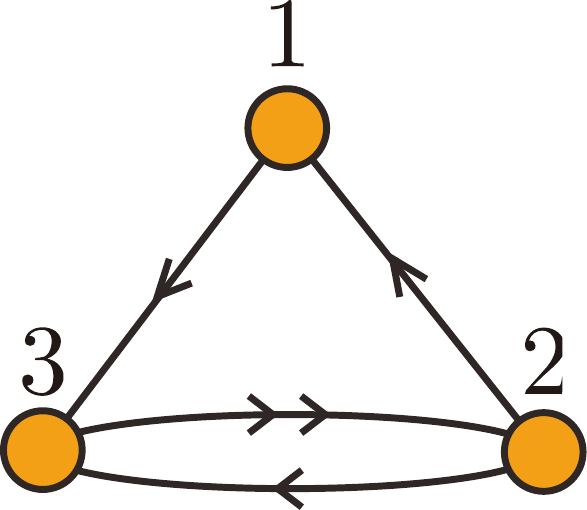}\\
\end{center}
\caption{The quiver of  the Phase II of the $\mathbb{C}\times \mathcal{C}$ theory.
The Chern-Simons levels are ${}^tk=(-1,1,0)$.}
\label{D1Cquiver}
\end{figure}

The dimer model of the Phase I of $\mathbb{C}\times \mathcal{C}$ theory 
is shown in Fig.\ref{D1Cdimer}.
It is called the one double-bonded chessboard model  $\mathscr{D}_1\mathscr{C}$ \cite{Davey:2009sr}.
The quiver diagram corresponding to the dimer is given in Fig.\ref{D1Cquiver}.
The Chern-Simons levels we study here are ${}^tk=(-1,1,0)$.
This theory also does not have a $3+1$ dimensional parent, since the node $1$ of the quiver has $N_f=N_c$ flavors. 
The incidence matrix of the model is given by
\begin{equation}
d=\begin{array}{c|ccccc}
& \phi_1 & \phi_2 & \phi_3 & \phi_4 &\phi_5
\\ \hline 1 & -1& 0 & 0 & 0  &1
\\ 2            &1 & -1 & 1 & -1&0
\\ 3            &0 & 1  & -1 &1&-1
\end{array}.
\end{equation}
Thus we can choose the vector $n$ as
\begin{align}
{}^tn=(1,0,0,0,0).
\end{align}
The Kasteleyn matrix of the model is also $1\times 1$:
\begin{align}
K=\phi_1+\phi_2x+\phi_3xyz+\phi_4y +\phi_5.
\end{align}
The perfect matching matrix is therefore given by
\begin{align}
P=\left(\begin{array}{ccccc}
0 & 0 & 1 & 0 &0
\\ 1 & 0 & 0 & 0 &0
\\0 & 0 & 0 & 1&0
 \\0 & 1 & 0 & 0&0
  \\0 & 0 & 0 & 0&1
 \end{array}\right).
\end{align}
It is easy to solve the equation $d=\tilde{Q}\cdot {}^tP$ for this matrix $P$.
We find the following integral solution:
\begin{align}
\tilde{Q}=
\left(\begin{array}{ccccc}
   0 & 0 & -1 & 0 & 1
\\ -1 & -1 & 1 & 1 & 0
\\ 1 & 1 & 0 & -1& -1
 \end{array}\right).
\end{align}
We are now able to compute $\hat{Q}_0$ and $ q$ using these data.
\begin{align}
&{}^t \hat{Q}_0={}^tk\cdot \tilde{Q}=(-1,-1,2,1,-1),\nonumber\\
&{}^t q={}^tn\cdot n=(0,0,1,0,0).\nonumber
\end{align}
They satisfy ${}^tq\cdot \hat{Q}_0=2=k^2$ as expected.
We find that the charge vector $Q_0$, which is defined by $Q_0=\frac{1}{k^2}\hat{Q}_0$, is not integral,
which might be a sign of inconsistency of the dimer model in the viewpoint of $3+1$ dimensional gauge theory.
Notice that the inconsistency does not give us trouble since we study $2+1$ dimensional Chern-Simons theory.

\subsubsection*{\bf Phase II: The $\mathscr{H}_2$ model}
\begin{figure}[htbp]
\begin{center}
\includegraphics[width=3cm,bb=0 0 93 159]{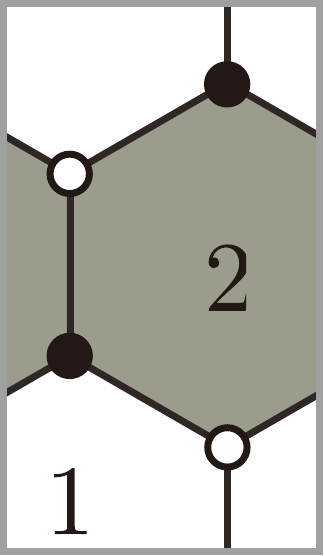}\\
\end{center}
\caption{The dimer model of the Phase I of the $\mathbb{C}\times \mathcal{C}$ theory $\mathscr{H}_2$.}
\label{H2dimer}
\end{figure}
\begin{figure}[htbp]
\begin{center}
\includegraphics[width=6cm,bb=0 0 240 55]{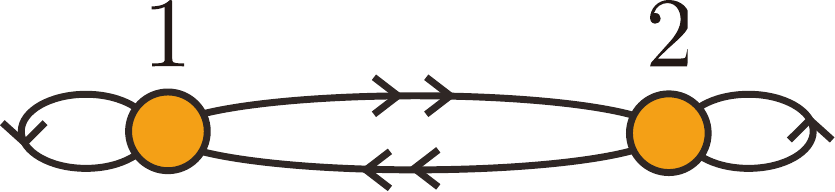}\\
\end{center}
\caption{The quiver of  the Phase II of the $\mathbb{C}\times \mathcal{C}$ theory.
The Chern-Simons levels are ${}^tk=(-1,1)$.}
\label{H2quiver}
\end{figure}

The dimer model of the Phase II of $\mathbb{C}\times \mathcal{C}$ theory,
which  is called the two hexagon model  $\mathscr{H}_2$ \cite{Davey:2009sr},
is shown in Fig.\ref{H2dimer}.
The quiver diagram corresponding to the dimer is given in Fig.\ref{H2quiver}.
The Chern-Simons levels for the phase are ${}^tk=(1,-1)$.
The incidence matrix of the dimer model is given by
\begin{equation}
d=\begin{array}{c|cccccc}
& \phi_1 & \phi_2 & \phi_3 & \phi_4 &\phi_5&\phi_6
\\ \hline 1 & d_{11} & -1 & 1 & 1  &-1 &0
\\ 2            &0 & 1 & -1 & -1&1&d_{25}
\end{array}.
\end{equation}
Thus we can choose the vector $n$ as
\begin{align}
{}^tn=(0,0,1,0,0,0).
\end{align}

The Kasteleyn matrix of the model is $2\times 2$, and the permanent of the perfect matching matrix is  \cite{Davey:2009sr}
\begin{equation}
P=\left(\begin{array}{ccccc}
   0 & 0 & 0 & 0 &1
\\ 0 & 1 & 0 & 1 &0
\\ 1 & 0 & 1 & 0&0
\\ 1& 0 & 0 & 1&0
\\ 0& 1 & 1& 0&1
\\ 0 & 0 & 0 & 0 &1
\end{array}\right).
\end{equation}
Then we can solve the equation $d=\tilde{Q}\cdot {}^tP$ as follows:
\begin{equation}
\tilde{Q}=
\left(\begin{array}{ccccc}
   \tilde{Q}_1 & \tilde{Q}_1-2& 1-\tilde{Q}_1 & 1-\tilde{Q}_1 &0
\\ \tilde{Q}_2 & \tilde{Q}_2+2 & -1-\tilde{Q}_2 & -1-\tilde{Q}_2 &0
\end{array}\right).
\end{equation}
Here we choose $d_{11}=d_{26}=0$.
Let us compute ${Q}_0$ and $ q$ using these data.
\begin{align}
&{}^t \hat{Q}_0={}^tk\cdot \tilde{Q}=(\tilde{Q}_1-\tilde{Q}_2,\tilde{Q}_1-\tilde{Q}_2-4,2-\tilde{Q}_1+\tilde{Q}_2,2-\tilde{Q}_1+\tilde{Q}_2,0),\nonumber\\
&{}^t q={}^tn\cdot n=(1,0,0,1,0,0).\nonumber
\end{align}
They satisfy
\begin{align}
{}^tq\cdot \hat{Q}_0&=(\tilde{Q}_1-\tilde{Q}_2)+(2-\tilde{Q}_1+\tilde{Q}_2)\nonumber\\
&=2=k^2.
\end{align}
It is now easy to see that the charge vector
\begin{align}
Q_0=\frac{1}{k^2}\hat{Q}_0
\end{align}
is not integral for generic $\tilde{Q}_1$ and $\tilde{Q}_2$.
However we find a integral $Q_0$ for a specific choice of $\tilde{Q}_{1,2}$.
When substituting $\tilde{Q}_1=-\tilde{Q}_2=1$, for instance, we find
\begin{align}
{}^tQ_0=(1,-1,0,0,0).
\end{align}
In this way this theory, which has a consistent $3+1$ dimensional parent, implies a integral charge $Q_0$.

\subsubsection*{\bf Phase III: The $\mathscr{D}_2\mathscr{H}_1$ model}
\begin{figure}[htbp]
\begin{center}
\includegraphics[width=6cm,bb=0 0 213 125]{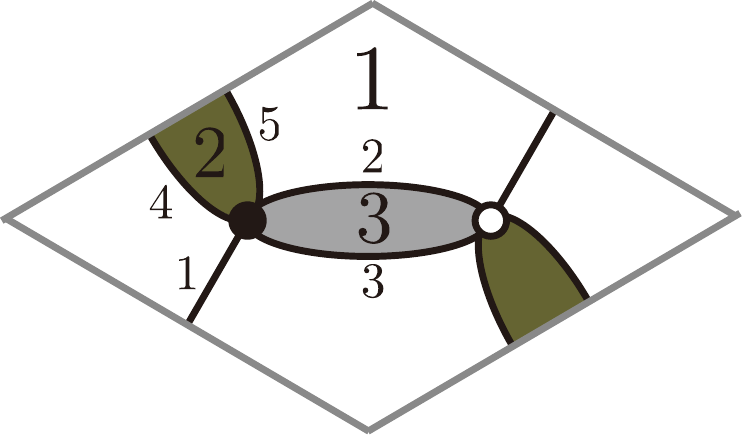}\\
\end{center}
\caption{The dimer model of the Phase III of the $\mathbb{C}\times \mathcal{C}$ theory $\mathscr{D}_2\mathscr{H}_1$.}
\label{D2H1dimer}
\end{figure}
\begin{figure}[htbp]
\begin{center}
\includegraphics[width=5cm,bb=0 0 191 144]{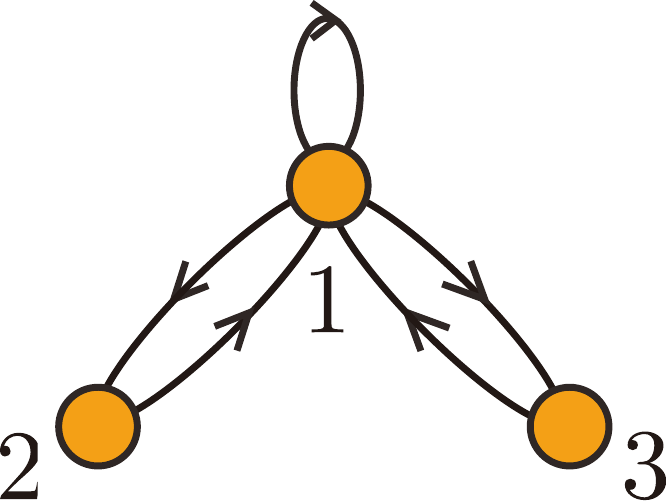}\\
\end{center}
\caption{The quiver of  the Phase III of the $\mathbb{C}\times \mathcal{C}$ theory.
The Chern-Simons levels are ${}^tk=(0,1,-1)$.}
\label{D2H1quiver}
\end{figure}

The dimer model of the Phase III of $\mathbb{C}\times \mathcal{C}$ theory 
is shown in Fig.\ref{D2H1dimer}.
We refer to it as the two double-bonded one-hexagon model  $\mathscr{D}_2\mathscr{H}_1$ \cite{Davey:2009sr}.
The quiver diagram corresponding to the dimer is given in Fig.\ref{D2H1quiver}.
The Chern-Simons levels in this phase are given by ${}^tk=(0,1,-1)$.
The incidence matrix of the dimer model is given by
\begin{equation}
d=\begin{array}{c|ccccc}
& \phi_1 & \phi_2 & \phi_3 & \phi_4 &\phi_5
\\ \hline 1 & d_{11} & 1 & -1 & 1  &-1
\\ 2            &0 & 0 & 0 & -1&1
\\ 3            &0 & -1 & 1 & 0&0
\end{array}.
\end{equation}
We can solve the equation $k=d\cdot n$, and we find the following integral solution:
\begin{align}
{}^tn=(0,1,0,0,1).
\end{align}
The Kasteleyn matrix of the model is the following $1\times 1$ matrix:
\begin{align}
K=\phi_1y+\phi_2z+\phi_3+\phi_4x +\phi_5x.
\end{align}
Therefore perfect matchings are just the matter fields.
The perfect matchi matrix is given by
\begin{align}
P=\left(\begin{array}{ccccc}
0 & 0 & 0 & 0 &1
\\ 0& 1 & 0 & 0 &0
\\0 & 0 & 0 & 0&0
 \\ 0 & 0 & 0 & 1&0
  \\0 & 0 & 1 & 0&0
 \end{array}\right).
\end{align}
We find the following integral solutions of  the equation $d=\tilde{Q}\cdot {}^tP$:
\begin{equation}
\tilde{Q}=
\left(\begin{array}{ccccc}
   1 & 1& -1 &-1 & \tilde{Q}_{15}
\\ -1&0&1&0&0
\\ 0&-1&0&1&0
\end{array}\right).
\end{equation}
Then we find ${Q}_0$ and $ q$ by using these data:
\begin{align}
&{}^t \hat{Q}_0={}^tk\cdot \tilde{Q}=(-1,1,1,-1,0),\nonumber\\
&{}^t q={}^tn\cdot n=(0,1,1,0,0).\nonumber
\end{align}
They satisfy ${}^tq\cdot \hat{Q}_0=2=k^2$.
In this case we obtain the fractional charge vector:
\begin{align}
&{}^t Q_0=\left(0,\frac{1}{2},\frac{1}{2},0,0\right).
\end{align}
This might be a reflection of the fact that a $3+1$ dimensional parent theory with the quiver Fig.\ref{D2H1quiver}
 gives rise to inconsistency.

\subsection{Consistent parents and the inverse algorithm}
As we have observed using simple examples,
$Q_0$ would be fractional for a theory without a consistent parent.
We expect that a dimer model with a consistent parent theory implies the integral charge vector $Q_0$:
\begin{align}
Q_0=\frac{1}{k^2}\hat{Q}_0\in \mathbb{Z}^c.
\end{align}
It would be interesting to prove the integer property
using consistency conditions on dimer models \cite{Hanany:2005ss}\cite{Gulotta:2008ef}.
We leave it as an interesting open problem.

For such a integral charge, it is straightforward to define the parent Calabi-Yau 4-fold
as a symplectic quotient of the original 4-fold: 
\begin{align}
\mathcal{M}^{\textrm{qCS}}//U(1)_{Q_0}=\mathcal{M}^{\textrm{parent}}.
\end{align}

In this section we have studied relation between the forward and inverse algorithm using dimer model,
and we found characteristic feature of a quiver Chern-Simons theory without a consistent parent.
It would be desirable to understand the relation from the string theory viewpoint, 
such as mirror symmetry\cite{Feng:2005gw}.

\section{$3+1$ Dimensional Grandparents, Un-Higgsings, and Theories Without Consistent Parents}

In this section, we point out that  
world volume theories of M2 branes on a toric Calabi-Yau 4-fold
are derived from a special class of quiver gauge theories in $3+1$ dimensions which are associated with the Calabi-Yau 4-fold.
We call those $3+1$ dimensional grandparent theories.
Each way to project a toric diagram of a Calabi-Yau 4-fold on a plane gives each grandparent theory.
In general, the projected toric diagram has multiplicities 
which cannot  be derived from  any $3+1$ dimensional theories.
Therefore we eliminate few points (or multiplicities) from the diagram and define a grandparent whose toric diagram can be realized as a moduli space
of a certain $3+1$ dimensional theory.
In the following sections, we work out many examples of the procedure.

In some cases, an M2 brane theory is provided with a consistent $3+1$ dimensional parent quiver gauge theory,
where the word "consistent" here means that the theory describes a SCFT when flows to IR.
This condition constraints the number of flavors  for each node of the quiver gauge theory.
For an M2 brane theory which has such a parent theory, the 3+1 dimensional parent is just its grandparents by definition.

A grandparent theory leads to an M2 brane theory even if the theory does not have any parent theory.
In this case, we add points to the toric diagram of the grandparent theory in order to recover the projected one of the 4-fold.
We employ the so-called un-Higgsing procedure \cite{Feng:2002fv} to increase toric points, 
especially inconsistent un-Higgsing \cite{Hanany:2005ss} 
which does not change the area of a toric diagram but increases the number of faces of a dimer model.
We refer to the procedure as "un-Higgsing" 
because with this operation the number of the $U(1)$ gauge symmetry and GLSM fields is increase 
in the gauged linear sigma model description of the toric 3-fold.
As the result, points and multiplicities are added to the toric diagram.
In general, a way of projection and un-Higgsing of points is not unique.
This ambiguity implies rich landscape of the M2 brane theories.
As we will see, a special types of un-Higgsing which is known as "doubling" \cite{Davey:2009bp} plays a role in this article.

By turning on Chern-Simons levels of an un-Higgsed theory, we can uplift a 2 dimensional toric diagram 
and construct a 3 dimensional one.
Thus an appropriate choice of Chern-Simons levels leads to the toric diagram of the original Calabi-Yau 4-fold which M2 branes probe.
Using these methods, we can construct many quiver Chern-Simons theories which would describe M2 brane theories.
The point is that our scheme is applicable to M2 brane theories
whether the parents are consistent quiver gauge theories in 4 dimension or not.

We demonstrate the construction of M2 brane theories in some concrete examples using their grandparents.
Before discussing new theories, we study two phases proposed in \cite{Davey:2009sr} from our viewpoint.

\subsection{$\mathbb{C}^4$ theory}

In the previous section, we study two quiver theories and their dimers whose abelian moduli spaces are  $\mathbb{C}^4$.
Using the forward algorithm, we can show that the abelian mesonic moduli spaces of these theories are really $\mathbb{C}^4$ \cite{Davey:2009sr}.
At this stage, origin of these theories in the perspective of the Calabi-Yau  4-fold $\mathbb{C}^4$ is unclear.
Therefore, we propose $3+1$ dimensional grandparents and their un-Higgsings 
in order to give a detailed explanation of origin of  these theories from the Calabi-Yau  4-fold.
\begin{figure}[htbp]
\begin{center}
\includegraphics[width=9cm,bb=0 0 306 99]{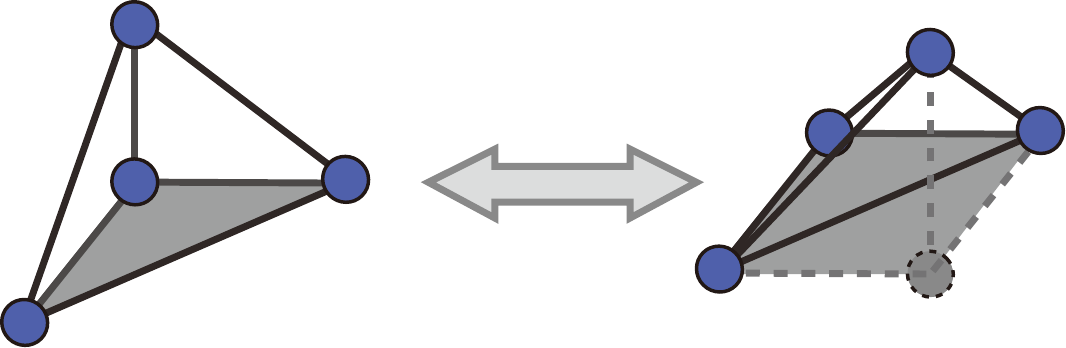}\\
\end{center}
\caption{A $SL(3,\mathbb{Z})$ transformation of the toric diagram of $\mathbb{C}^4$.}
\label{C4SL3Z}
\end{figure}
Our starting point is that 
we can change the shape of the base of the 3 dimensional toric diagram of $\mathbb{C}^4$ by
using $SL(3,\mathbb{Z})$ transformation as  Fig.\ref{C4SL3Z}.
One choice of the shape is a right-angled triangle, and the other is a regular square.
Then we project the 3 dimensional toric diagram onto the base plane.
We regard a projected diagram as a 2 dimensional toric diagram of a Calabi-Yau 3-fold associated with this Calabi-Yau 4-fold.
The 2 dimensional toric diagram with the shape of a triangle without internal points 
is precisely the one of the Calabi-Yau 3-fold  $\mathbb{C}^3$ up to multiplicities of its GLSM fields.
The regular square is precisely the toric diagram of the conifold.

\begin{figure}[htbp]
\begin{center}
\includegraphics[width=9cm,bb=0 0 447 191]{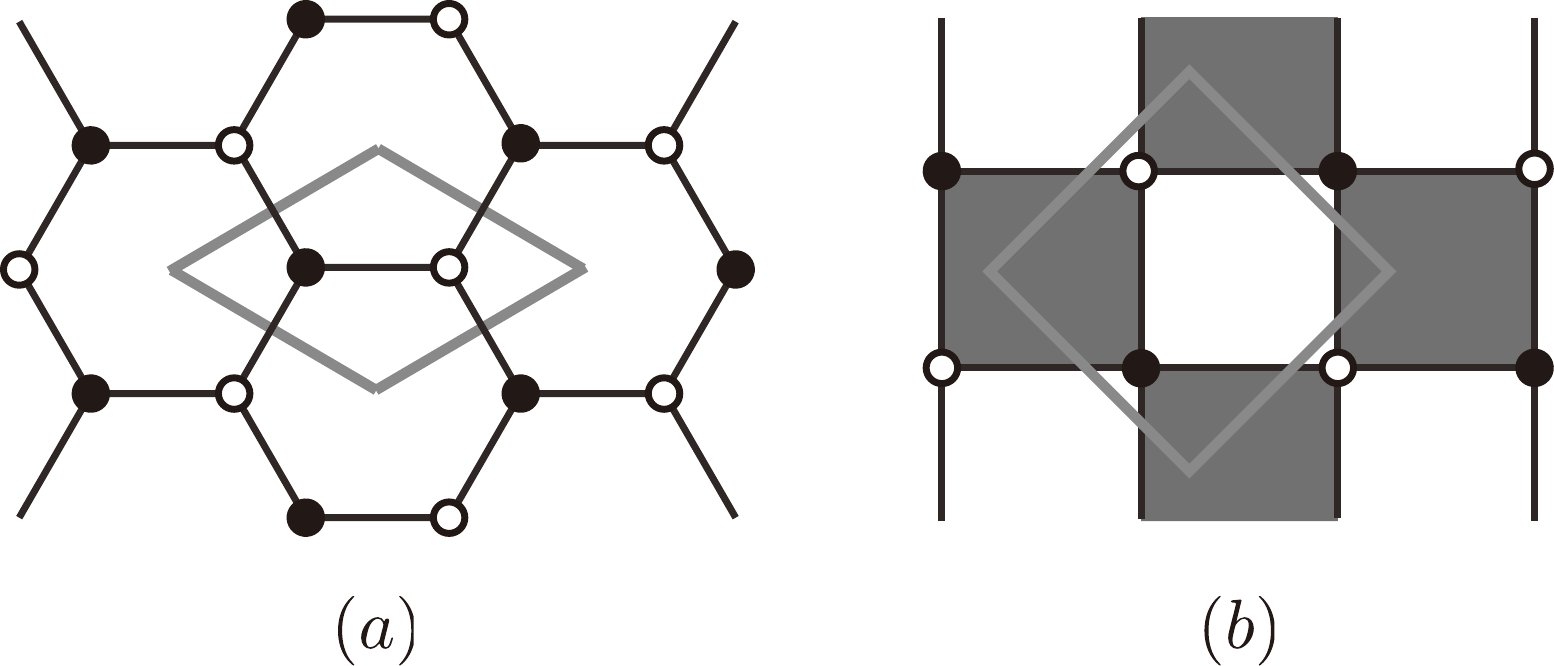}\\
\end{center}
\caption{Dimers of the two projected toric diagrams.}
\label{C4granpdimer}
\end{figure}

It is well kwown that $\mathcal{N}=4$ super Yang-Mills theory in $3+1$ dimensions 
is the world-volume theory of D3 branes probing $\mathbb{C}^3$.
On the one hand, the world-volume theory for the conifold is the well-known Klebanov-Witten theory \cite{Klebanov:1998hh}.
We call these two theories the $3+1$ dimensional grandparents theories of the M2 brane theories for $\mathbb{C}^4$.
The dimer models of the grandparents are tilings of hexagons and squares as Fig.\ref{C4granpdimer}.
As we shall see in this section, the two phases of the M2 brane theory for $\mathbb{C}^4$ originate from these two grandparents.

\subsubsection*{\bf Phase I: ABJM theory $\mathscr{C}$}

\begin{figure}[htbp]
\begin{center}
\includegraphics[width=9cm,bb=0 0 315 163]{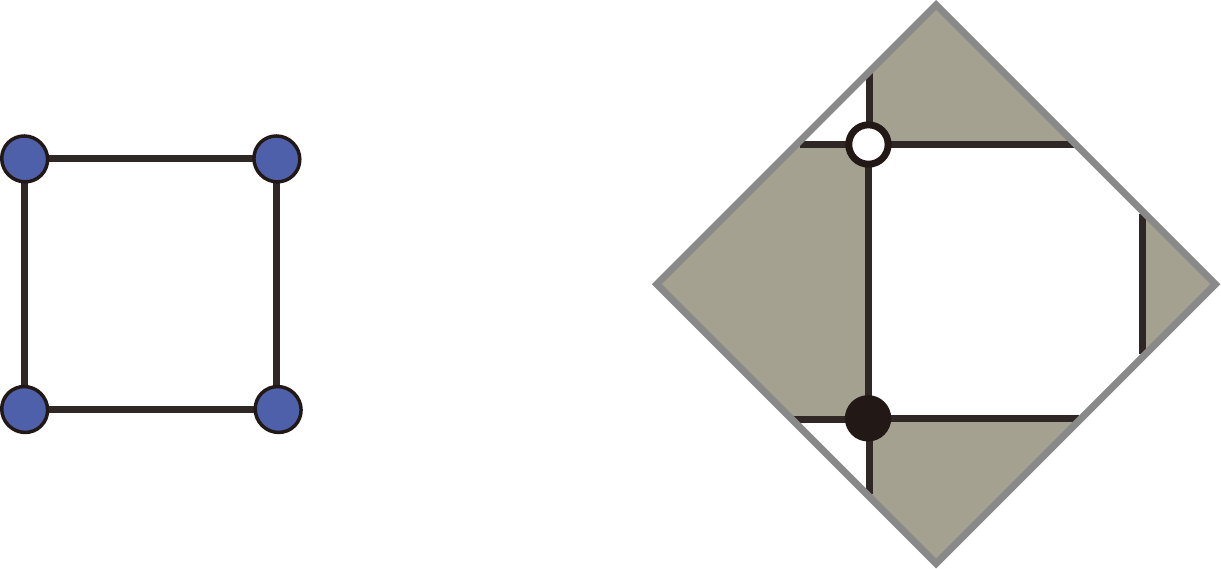}\\
\end{center}
\caption{The toric diagram and the dimer model for the conifold (grand)parent.}
\label{conifoldABJMdimer}
\end{figure}

Let us start with the canonical example known as the ABJM theory. 
We choose a specific projection of the toric diagram of $\mathscr{C}$ which is shown in the right of Fig.\ref{C4SL3Z}.
This projected toric diagram in 2 dimensions involves the ABJM phase of the $\mathbb{C}^4$ theory.
The grandparent theory emerges from the projection is the conifold theory, 
since the 2 dimensional diagram is in the shape of a square.
Moreover the conifold grandparent is precisely a parent theory of  $\mathbb{C}^4$ theory,
since every point in the projected toric diagram has multiplicity 1, which is the same as the conifold theory.
The toric diagram and its dimer model are shown in Fig.\ref{conifoldABJMdimer}.
Let us turn to derive the Chern-Simons levels of the $\mathbb{C}^4$ theory.
We shall assign these levels in order that one point of the toric diagram of the parent theory is uplifted and lifted the toric diagram forms a tetrahedron.
Recall that the Chern-Simons level vector $k$ is determined by the matrix $d$ and $n$ by using the relation $k=d\cdot n$.
Since the matrix $d$ is derived from the dimer of the parent theory, 
all we have to determine is the choice of the vector $n$.

We follow the convention of the previous section.
Let us uplift the point associated with the perfect matching $p_1$,
in other words we choose $n$ so as to satisfy $q=(1,0,0,0)$.
Notice that the coordinates for the third axis are given by $q={}^tP\cdot n$,
where the perfect matching matrix is the same as (\ref{Pconifold}).
We can solve the equation as follows:
\begin{align}
{}^tn=(1,0,0,0).
\end{align}
This vector and the incidence matrix (\ref{dconifold}) give $k$, 
and therefore we obtain the Chern-Simons levels which make the toric diagram of the moduli space a tetrahedron:
\begin{align}
{}^tk=(1,-1).
\end{align}

\subsubsection*{\bf Phase II: dual ABJM theory $\mathscr{D}_1 \mathscr{H}_1$}
\begin{figure}[htbp]
\begin{center}
\includegraphics[width=9cm,bb=0 0 358 138]{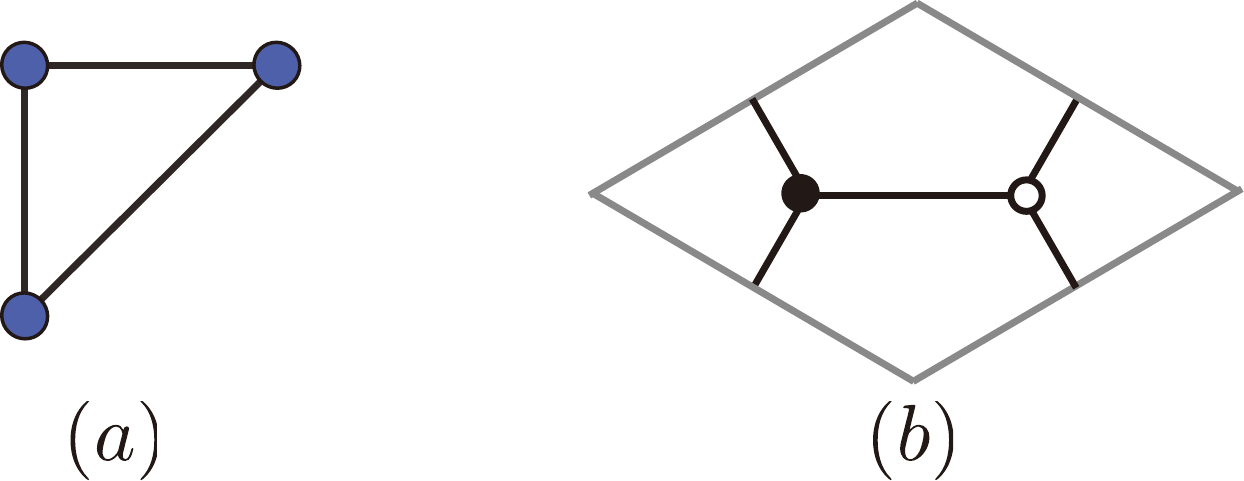}\\
\end{center}
\caption{The toric diagram and the dimer model for the $\mathbb{C}^3$ grandparent.}
\label{C3ABJMdimer}
\end{figure}
\begin{figure}[htbp]
\begin{center}
\includegraphics[width=9cm,bb=0 0 367 144]{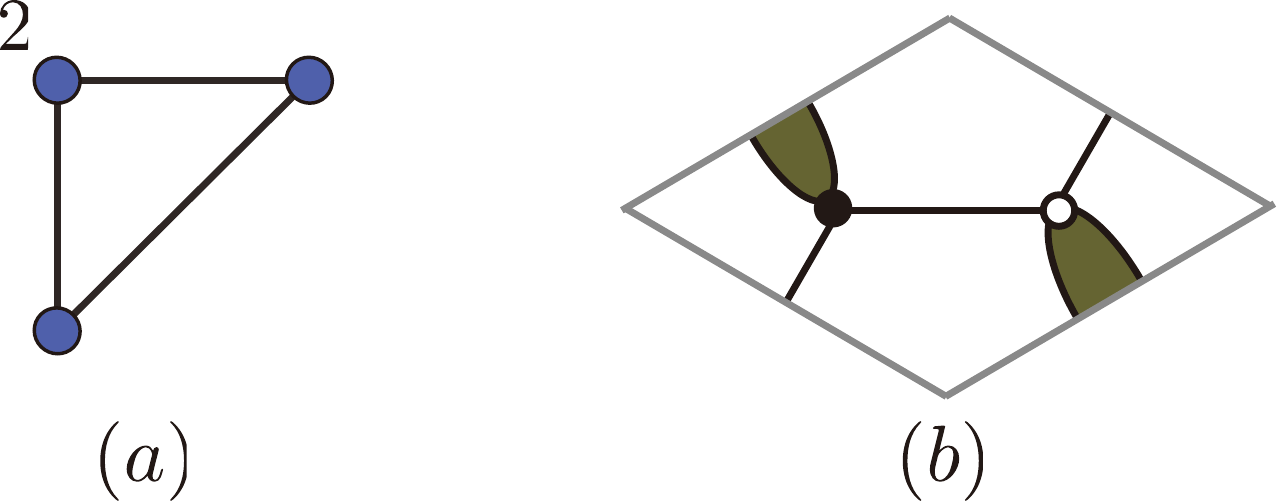}\\
\end{center}
\caption{The toric diagram and the dimer model for the $\mathbb{C}^3$ grandparent.}
\label{unHigC3ABJMdimer}
\end{figure}

Let us focus on  the grandparent theory associated with $\mathbb{C}^3$.
The toric diagram and the dimer model for the $\mathbb{C}^3$ grandparent are shown in Fig.\ref{C3ABJMdimer}.
Since there exists a point on the top of the tetrahedron, 
the projection involves a point with multiplicity 2 in the 2 dimensional toric diagram of the grandparents, which is shown in the left of Fig.\ref{C4SL3Z}.
We have therefore to un-Higgs  the grandparent in order to introduce an additional toric point (perfect matchings). 
One of the simplest choice of un-Higgsing is the so-called "doubling" \cite{Davey:2009bp} 
which introduce a double-bounded edge into the dimer model.
The resulting dimer model is denoted in Fig.\ref{unHigC3ABJMdimer}.
The un-Higgsed model implies an additional perfect matching,
and the perfect matching matrix becomes the $4\times 4$ identity matrix.
Thus one of the points of the 2 dimensional diagram is doubled by un-Higgsing the dimer model.
Following the convention of the previous section, the perfect matching matrix of Fig.\ref{unHigC3ABJMdimer} is given by
\begin{align}
P=\left(\begin{array}{cccc} 0 &  0 & 1 & 0  \\1 &0   &0   & 0 \\  0&   0&  0 & 1 \\  0&   1&   0& 0 \end{array}\right).
\end{align}

Next we lift one of the doubled perfect matching, $p_1$ for example, in order to transform the triangle diagram into the tetrahedron.
Recall that the third coordinate of a point $p_{\alpha}$ is given by $q_{\alpha}=\sum P_{i\alpha}n_i$.
We can therefore uplift the perfect matching by turning on the Chern-Simons levels by ${}^tn=(0,0,1,0)$:
\begin{align}
{}^tq={}^tP\cdot n=(1,0,0,0).
\end{align}
This choice of $n$ corresponds to the Chern-Simons levels $(k_1,k_2)={}^tn\cdot d=(1,-1)$.
In this way, by turning on an appropriate Chern-Simons levels,  the projected toric diagram on a plane 
is lifted to the 3 dimensional diagram of the Calabi-Yau 4-fold.

\section{Un-Higgsings of (3+1) Dimensional Grandparents}

In this section, we study the un-Higgsing procedure in concrete examples.
We especially focus on grandparent theories whose dimer models are the so-called $n$ hexagons $\mathscr{H}_n$.
The moduli spaces of the grandparent theories contain abelian orbifolds of $\mathbb{C}^3$ like $\mathbb{C}^2/\mathbb{Z}_N \times \mathbb{C}$.
Then we un-Higgs the grandparents and obtain quiver Chern-Simons theories
whose moduli spaces are abelian orbifolds of $\mathbb{C}^4$, $C(dP_3)\times\mathbb{C}$ and so forth.
See \cite{Imamura:2008nn}\cite{Terashima:2008ba}\cite{Imamura:2008dt} for orbifold projection of the ABJM theory.
In this section we study the orbifold theory by using the dimer model approach. 
Some of the resulting Chern-Simons theories describes new phases of M2 brane theories.

\subsection{ $\mathbb{C}^2/ \mathbb{Z}_N\times \mathbb{C}$ grandparent}

The orbifold  $\mathbb{C}^2/ \mathbb{Z}_N\times \mathbb{C}$ is one of the well-studied examples of toric Calabi-Yau 3-fold.
We un-Higgs these toric Calabi-Yau 3-folds, which play a role of grandparent theories for M2-brane theories. 
Applying the forward algorithm, we find out that these grandparents implies abelian orbifolds of  $\mathbb{C}^4$.

\subsubsection*{ Un-Higgsing of $\mathbb{C}^2/ \mathbb{Z}_3\times \mathbb{C}$ grandparent}
\begin{figure}[htbp]
\begin{center}
\includegraphics[width=10cm,bb=0 0 490 240]{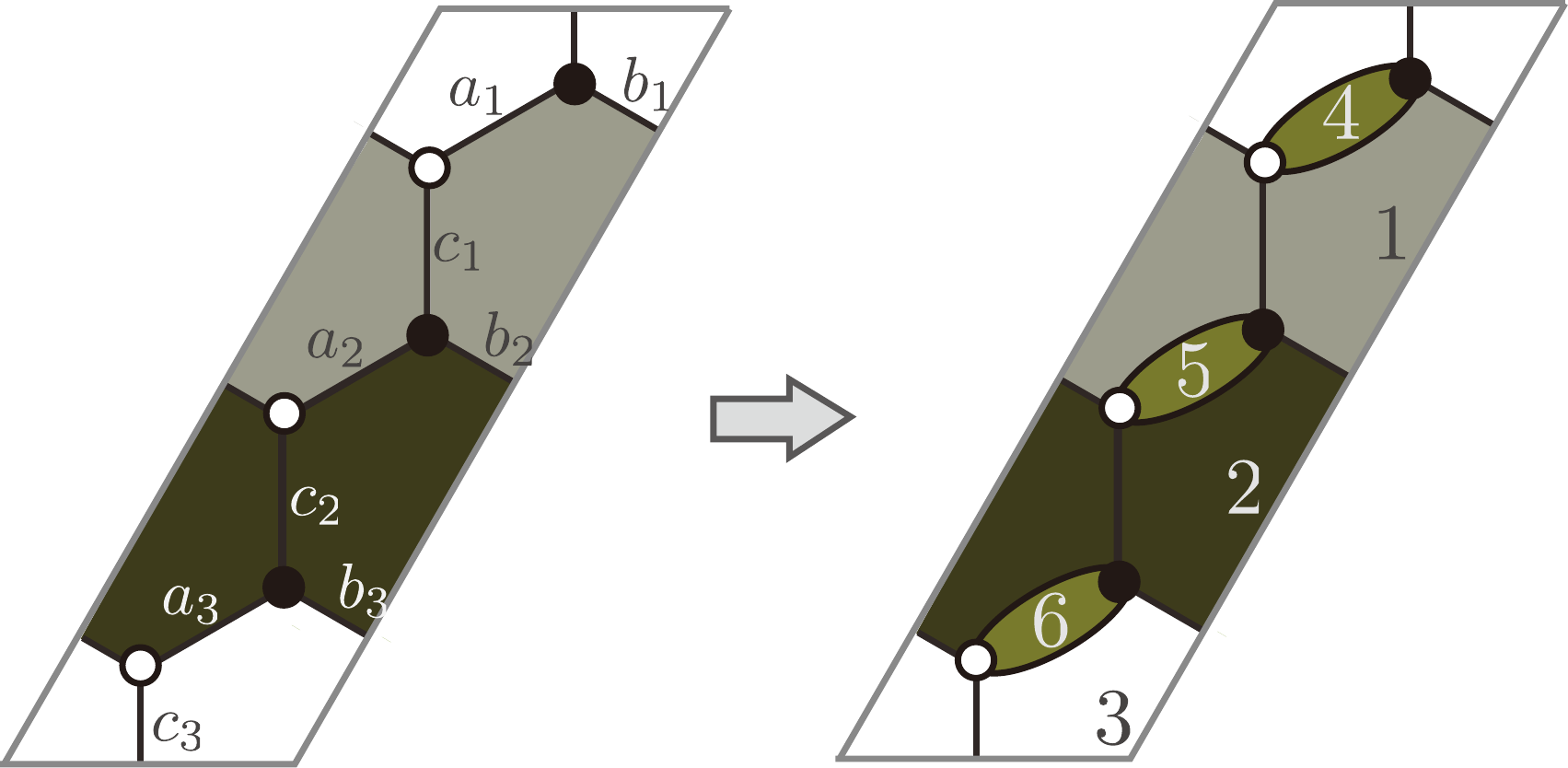}\\
\end{center}
\caption{The dimer of $\mathbb{C}^2/ \mathbb{Z}_3\times \mathbb{C}$ grandparent (left) and its un-Higgsing (right).}
\label{c3zn}
\end{figure}
\begin{figure}[htbp]
\begin{center}
\includegraphics[width=8cm,bb=0 0 271 108]{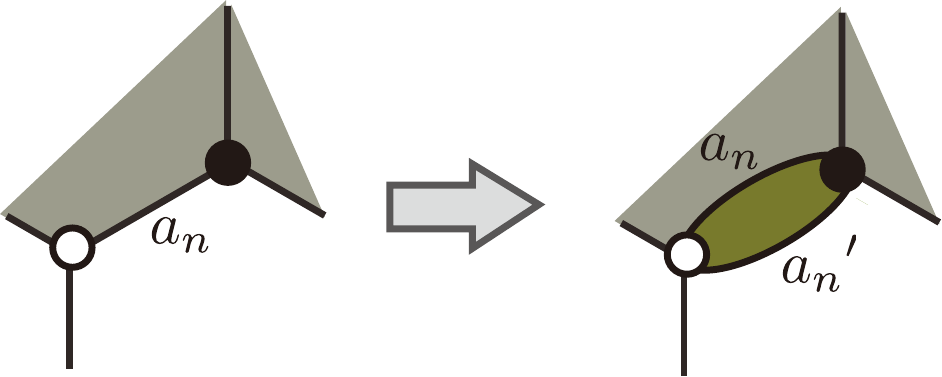}\\
\end{center}
\caption{The toric diagram of  $\mathbb{C}^2/ \mathbb{Z}_3 \times \mathbb{C}$ grandparent and their perfect matchings}
\label{c3znunhiggs}
\end{figure}
\begin{figure}[htbp]
\begin{center}
\includegraphics[width=5cm,bb=0 0 219 121]{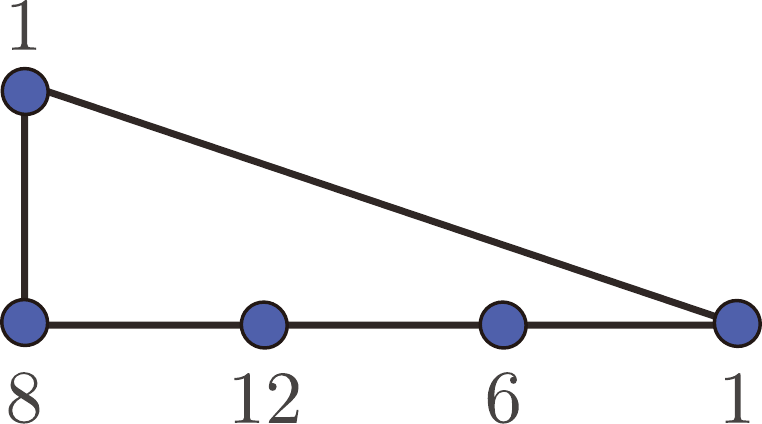}\\
\end{center}
\caption{The un-Higgsed toric diagram and multiplicities}
\label{c3znunhigtoric}
\end{figure}
\begin{figure}[htbp]
\begin{center}
\includegraphics[width=12cm,bb=0 0 571 214]{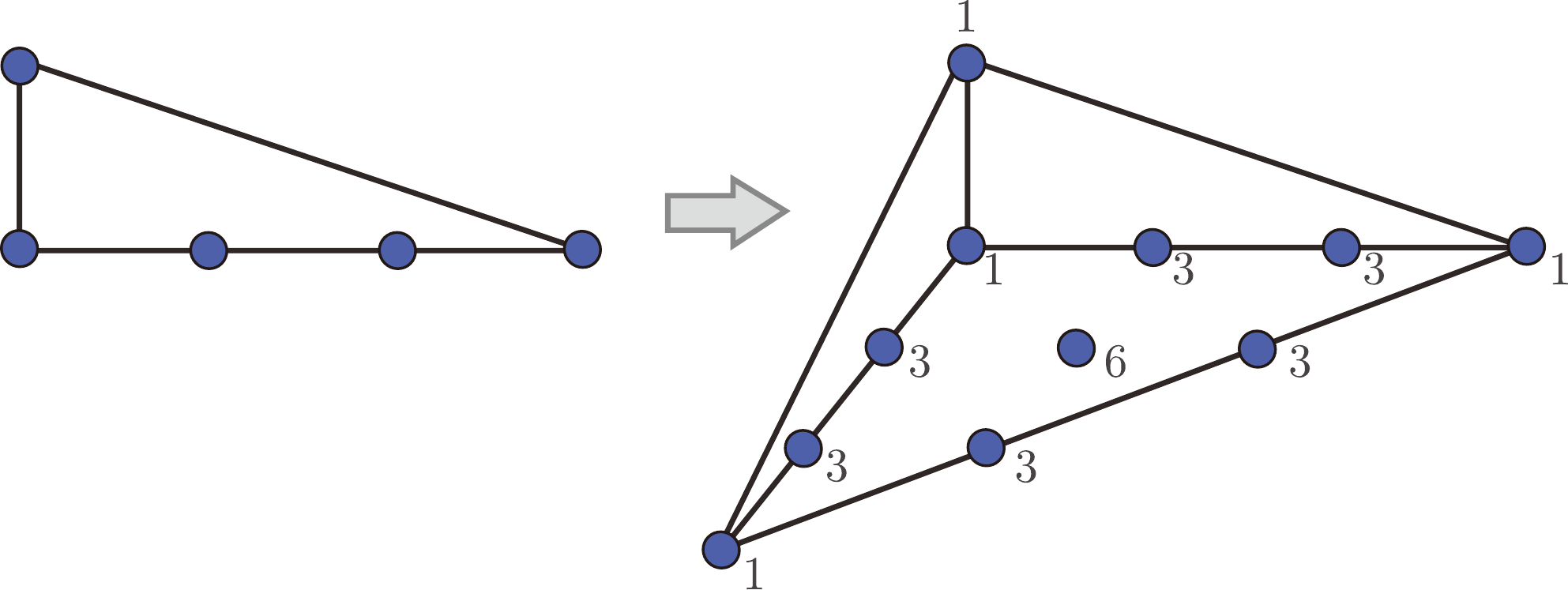}\\
\end{center}
\caption{The lift to the toric diagram of the Calabi-Yau 4-fold}
\label{c3z3z3c}
\end{figure}

We first study an un-Higgsed theory for $\mathbb{C}^2/ \mathbb{Z}_3\times \mathbb{C}$ grandparent.
As we shall see, we get  the M2-brane theory for  $\mathbb{C}^3/( \mathbb{Z}_3 \times \mathbb{Z}_3) \times \mathbb{C}$ 
which was constructed in \cite{Franco:2008um}.
We study the un-Higgsing of the dimer of $\mathbb{C}^2/ \mathbb{Z}_3\times \mathbb{C}$  which is illustrated in Fig.\ref{c3zn}.
The Kasteleyn matrix of the grandparent $\mathbb{C}^2/ \mathbb{Z}_3\times \mathbb{C}$ theory is
\begin{align}
K=\left(
\begin{array}{ccc}a_1+b_1x & 0 & c_3y \\c_1 & a_2+b_2x & 0 \\0 & c_2 & a_3+b_3x
\end{array}\right).
\end{align}
It is easy to compute the permanent of the Kasteleyn matrix:
\begin{align}
\textrm{perm}K&=(a_1+b_1x)( a_2+b_2x)(a_3+b_3x)+c_1c_2c_3y\nonumber\\
&=a_1a_2a_3+x(b_1a_2a_3+a_1b_2a_3+a_1a_2b_3)\nonumber\\
&\quad +x^2(a_1b_2b_3+b_1a_2b_3+b_1b_2a_3)+x^3b_1b_2b_3+yc_1c_2c_3.
\end{align}
This polynomial consists of 9 monomials.
\begin{figure}[hbp]
\begin{center}
\includegraphics[width=6cm,bb=0 0 247 146]{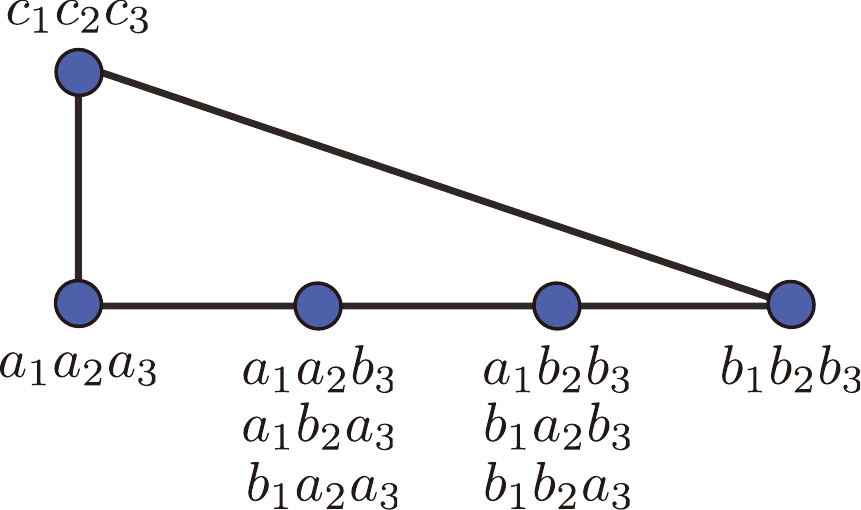}\\
\end{center}
\caption{The toric diagram of  $\mathbb{C}^2/ \mathbb{Z}_3\times \mathbb{C}$ grandparent and its perfect matchings}
\label{c3zntoric}
\end{figure}
Recall that each monomial coefficient of the term $x^iy^j$ corresponds to a point $(i,j)$, 
to which is referred as a perfect matching, of the toric diagram.
Thus we obtain the toric diagram Fig.\ref{c3zntoric} with total multiplicity 9.
Then, we un-Higgs the fields $a_n$ as Fig.\ref{c3znunhiggs}.
The effects of the un-Higgsing are captured by the replacement $a_m \to a_m+a^{\prime}_m z^{n_{a^{\prime}_m}}$ in the Kasteleyn matrix.
Here we turn on $n_i$ for the un-Higgsed fields $a^{\prime}_m$. 
Then the total number of perfect matchings becomes $28=3^3+1$ as Fig.\ref{c3znunhigtoric}.
Turning on $n_i$'s, the 2-dimensional toric diagram is lifted to the corresponding 3-dimensional diagram.
We can compute the 3-dimensional toric diagram by introducing a third coordinate $z$ into the Kasteleyn matrix. 
Let us study the following simple case:
\begin{align}
&n_i=1\quad \textrm{for}\quad i={a^{\prime}}_m,\nonumber\\
&n_i=1\quad \textrm{otherwise}.\nonumber
\end{align}
Then we can calculate the un-Higgsed version of $\textrm{perm}K$ by replacing $a_m$ to $a_m+a^{\prime}_m z$.
The multilcity of the point $(i,j,k)$ is encoded into the coefficient of the term $x^iy^jz^k$ in $\textrm{perm}K$.
Thus we get the moduli space of the un-Higgsed quiver Chern-Simons theory.
The right side of Fig.\ref{c3z3z3c} denotes the toric diagram of the moduli space.
An integer in the figure denotes the multiplicity of each point of the toric diagram.
Notice that we can recover the multiplicities of Fig.\ref{c3znunhigtoric} by projecting the 3-dimensional toric diagram onto the original plane.
The toric Calabi-Yau 4-fold corresponding to Fig.\ref{c3z3z3c} is the abelian orbifold 
$\mathbb{C}^3/( \mathbb{Z}_3 \times \mathbb{Z}_3) \times \mathbb{C}$.

The dimer model encodes the information of the quiver Chern-Simons theory: the quiver diagram and the superpotential.
The superpotential of the un-Higgsed model is
\begin{align}
W&=\tr\left( \phi_1(X_{14}X_{43}X_{31}-X_{12}X_{25}X_{51}\right)\nonumber\\
&\qquad\qquad +\phi_2\left(X_{25}X_{51}X_{12}-X_{32}X_{63}X_{62}\right)
+\phi_3\left(X_{36}X_{62}X_{23}-X_{31}X_{14}X_{43})\right).
\end{align}
Here $\phi_a$ is the adjoint field for the a-th gauge group.
It is easy to compute the incidence matrix $d_{ai}$ and the quiver diagram associated with the dimer model Fig.\ref{c3zn}.
\begin{figure}[htbp]
\begin{center}
\includegraphics[width=5cm,bb=0 0 256 236]{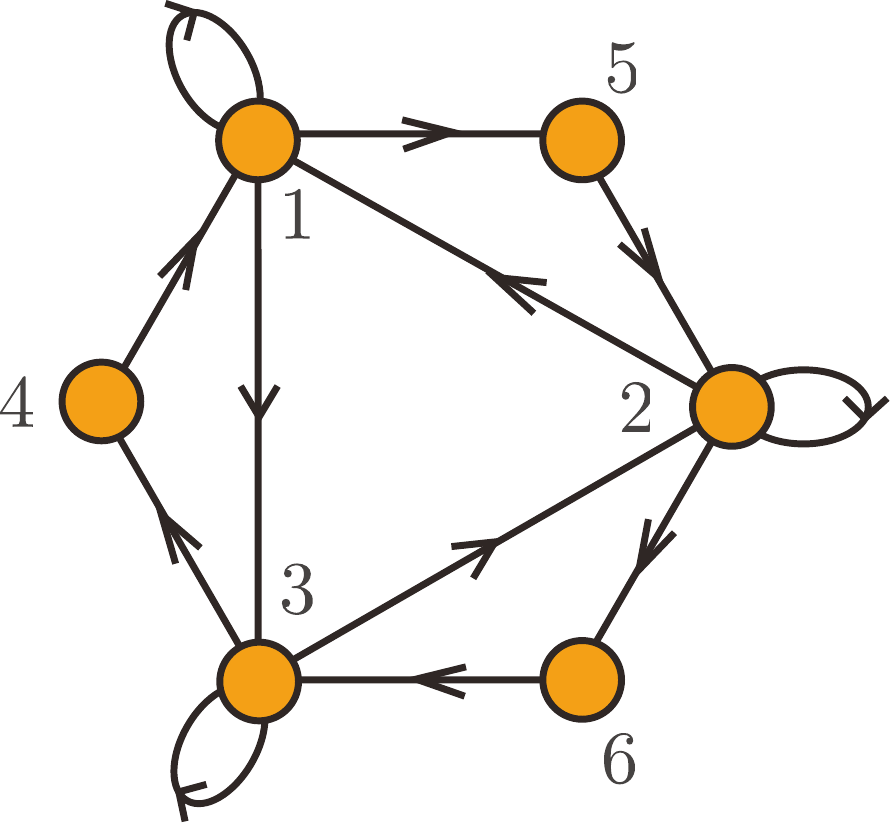}\\
\end{center}
\caption{The quiver Chern-Simons theory of M2-branes probing $\mathbb{C}^3/( \mathbb{Z}_3 \times \mathbb{Z}_3) \times \mathbb{C}$.
The Chern-Simons level vector is ${}^t k=(1,1,1,-1,-1,-1)$.}
\label{c3z3z3cquiver}
\end{figure}
Fig.\ref{c3z3z3cquiver} is the quiver diagram of the dimer model.
Using the matrix $d$, the choice of the vector $n$, and the relation $k=d \cdot n$,
we can determine the choice of  the Chern-Simons levels 
under which the moduli space of the quiver theory becomes $\mathbb{C}^3/( \mathbb{Z}_3 \times \mathbb{Z}_3) \times \mathbb{C}$.
For instance, the Chern-Simons level $k_1$ of the first gauge group is given by
\begin{align}
k_1=\sum_{i\in\{\textrm{matters}\}}d_{1i}n_i=d_{1{a^{\prime}}_1}n_{{a^{\prime}}_1}=1.
\end{align}
Thus we obtains the choice of the Chern-Simons level vector of our interest:
\begin{align}
{}^t k=(1,1,1,-1,-1,-1).
\end{align}
Then we get the quiver Chern-Simons theory whose abelian moduli space is 
$\mathbb{C}^3/( \mathbb{Z}_3 \times \mathbb{Z}_3) \times \mathbb{C}$.
Therefore this theory is the world volume theory of the M2-branes probing the toric singularity 
$\mathbb{C}^3/( \mathbb{Z}_3 \times \mathbb{Z}_3) \times \mathbb{C}$.
Fig.\ref{c3z3z3cquiver} is the quiver diagram of the $\mathbb{C}^3/( \mathbb{Z}_3 \times \mathbb{Z}_3) \times \mathbb{C}$ theory.
This is precisely the theory which was proposed in \cite{Franco:2008um}.
We have re-derived it by using the un-Higgsing approach.

In the rest of this section we apply the un-Higgsing method for several grandparent theories in order to realize that 
this approach provides an effective method to produce M2-brane theories with and without (3+1)-dimensional parents.
Then we find out some new quiver Chern-Simons theories which describe the world volume theories of M2-branes.

\subsubsection*{ Un-Higgsing of $\mathbb{C}^2/ \mathbb{Z}_N \times \mathbb{C}$ grandparent}

\begin{figure}[htbp]
\begin{center}
\includegraphics[width=5cm,bb=0 0 329 328]{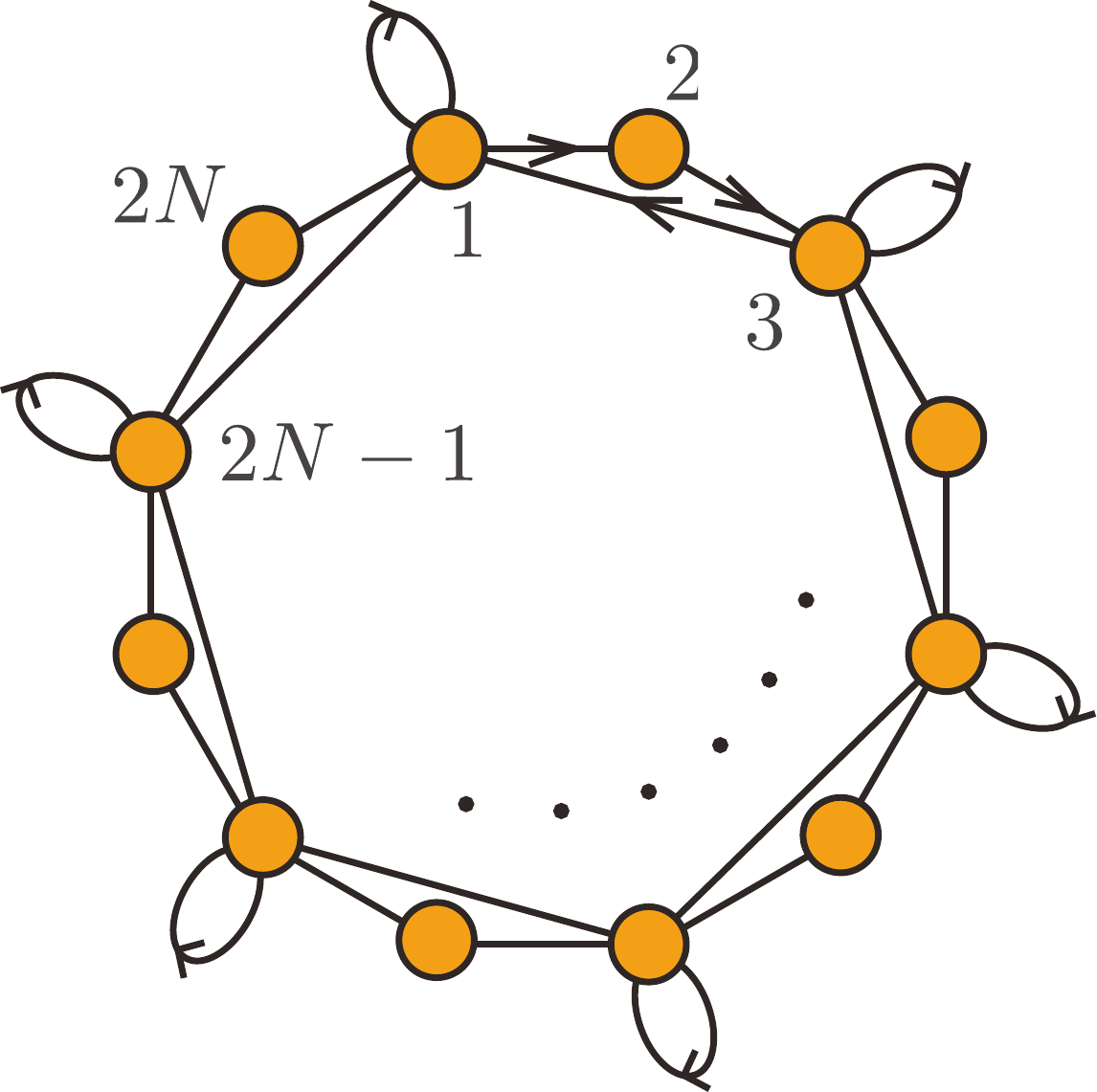}\\
\end{center}
\caption{The quiver Chern-Simons theory of M2-branes probing 
$\mathbb{C}^3/( \mathbb{Z}_N \times \mathbb{Z}_N) \times \mathbb{C}$.
The Chern-Simons level vector is ${}^t k=(1,-1,1,-1,\cdots)$.}
\label{c3znzncquiver}
\end{figure}

Having obtained $\mathbb{C}^3/( \mathbb{Z}_3 \times \mathbb{Z}_3) \times \mathbb{C}$ theory 
by un-Higgsing $\mathbb{C}^2/ \mathbb{Z}_3\times \mathbb{C}$,
we now generalize the analysis for the grandparent $\mathbb{C}^2/\mathbb{Z}_N\times \mathbb{C}$.
The resulting theory describes the theory for the M2-brane probing 
$\mathbb{C}^3/( \mathbb{Z}_N \times \mathbb{Z}_N) \times \mathbb{C}$.
This result gives a proof of the conjecture of \cite{Franco:2008um} that 
the quiver theory Fig.\ref{c3znzncquiver} gives $\mathbb{C}^3/( \mathbb{Z}_N \times \mathbb{Z}_N) \times \mathbb{C}$ theory.
As we shall see, the Kasteleyn matrix is an effective tool to prove it.

\begin{figure}[htbp]
\begin{center}
\includegraphics[width=12cm,bb=0 0 486 90]{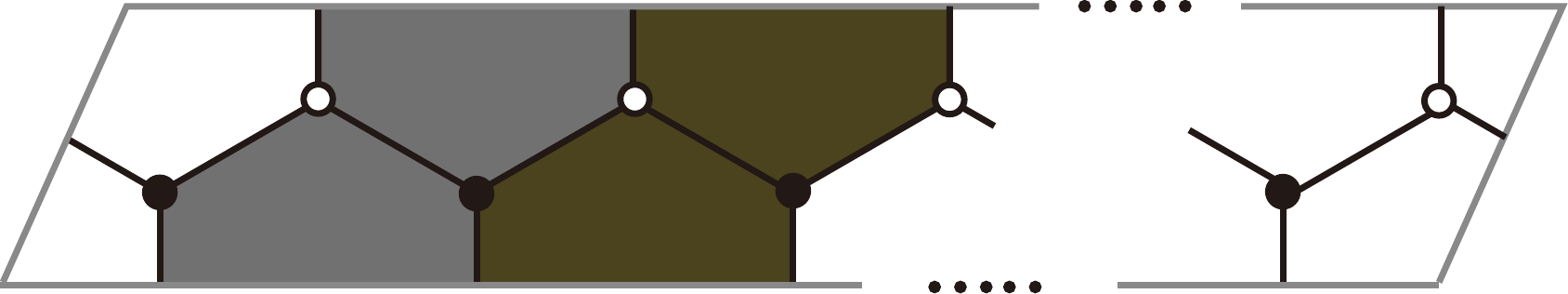}\\
\end{center}
\caption{The dimer model for the toric Calabi-Yau 3-fold  $\mathbb{C}^2/( \mathbb{Z}_N \times \mathbb{Z}_N) \times \mathbb{C}$.}
\label{c3zndimer}
\end{figure}

\begin{figure}[htbp]
\begin{center}
\includegraphics[width=14cm,bb=0 0 486 105]{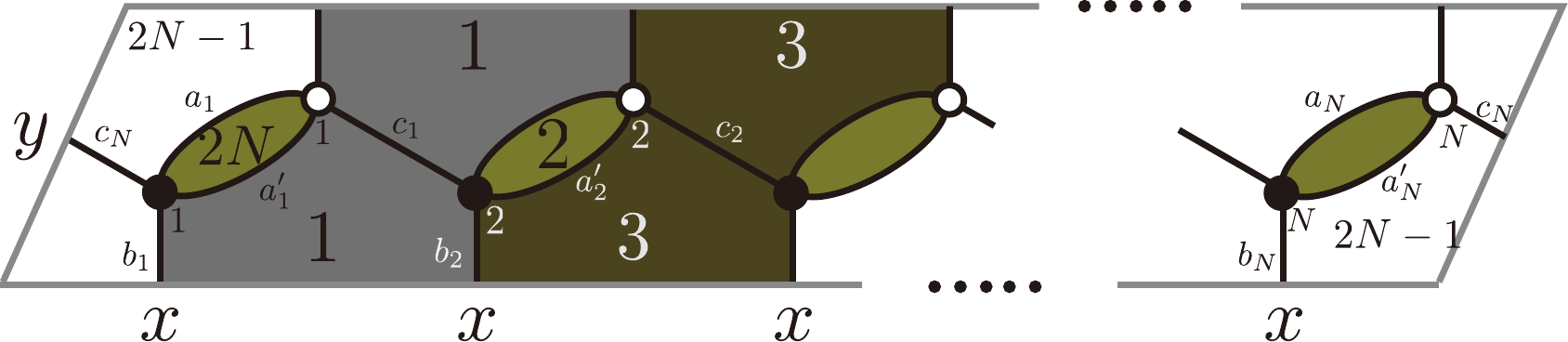}\\
\end{center}
\caption{The un-Higgsed dimer model.}
\label{c3znunhig}
\end{figure}

The dimer model of the quiver gauge theory for D3 branes on the orbifold $\mathbb{C}^2/ \mathbb{Z}_N\times \mathbb{C}$ is shown in Fig.\ref{c3zndimer}.
We repeat the above procedure for $\mathbb{C}^2/ \mathbb{Z}_N\times \mathbb{C}$.
In general there exist many ways to un-Higgs the dimer model of a (3+1)-dimensional grandparent theory.
However we are now interested in the special type of un-Higgsing, 
which is a generalization of the case of $\mathbb{C}^2/ \mathbb{Z}_3\times \mathbb{C}$.
Thus Fig.\ref{c3znunhig} is the un-Higgsed dimer on which we focus in this article.
The superpotential associated with the un-Higgsed dimer model is
\begin{align}
W=\sum_{n=1}^{N}\tr\left( c_n({a^{\prime}}_n a_n b_n-b_{n+1} {a^{\prime}}_{n+1} a_{n+1}) \right).
\end{align}
Here the index $N+1$ means $N+1\equiv 1$.
Let us focus on the following choice of the Chern-Simons levels: 
\begin{align}
\label{nc3zn}
n_i=1 \textrm{ for } i={a^{\prime}}_m, \quad
n_i=1 \textrm{ otherwise}.
\end{align}
The Kasteleyn matrix encodes the toric Calabi-Yau singularity which the mesonic moduli space describes.
From the dimer model Fig.\ref{c3zndimer} we easily determine the Kasteleyn matrix for the grandparent theory:
\begin{align}
\label{kastc3zn}
K(a,b,c;x,y)=\left(\begin{array}{ccccccc}a_1+b_1x & 0 &  \cdots&  & 0 & 0 & c_Ny \\c_1 & a_2+b_2x &  &  & 0 & 0 & 0 
\\ &  & \cdot &  &  &  &  \\ &  &  & \cdot &  &  &  
\\0 & 0 &  &  & a_{N-2}+b_{N-2} & 0 & 0 \\0 & 0 &  &  & c_{N_2} & a_{N-1}+b_{N-1}x & 0 
\\0 & 0 &\cdots  &  & 0 & c_{N_1} & a_N+b_Nx\end{array}\right).
\end{align}
Replacing $a_m$ to $a_m+a^{\prime}_m z$, we obtain the matrix for the un Higgsed theory Fig.\ref{c3znunhig}.
The cofactor expansion implies the permanent of (\ref{kastc3zn}):
\begin{align}
\label{permc3zn}
\textrm{pern}K(a,b,c;x,y)=\prod_{n=1}^{N}(a_n+xb_n)+y\prod_{n=1}^{N}c_n.
\end{align}
\begin{figure}[htbp]
\begin{center}
\includegraphics[width=9cm,bb=0 0 414 255]{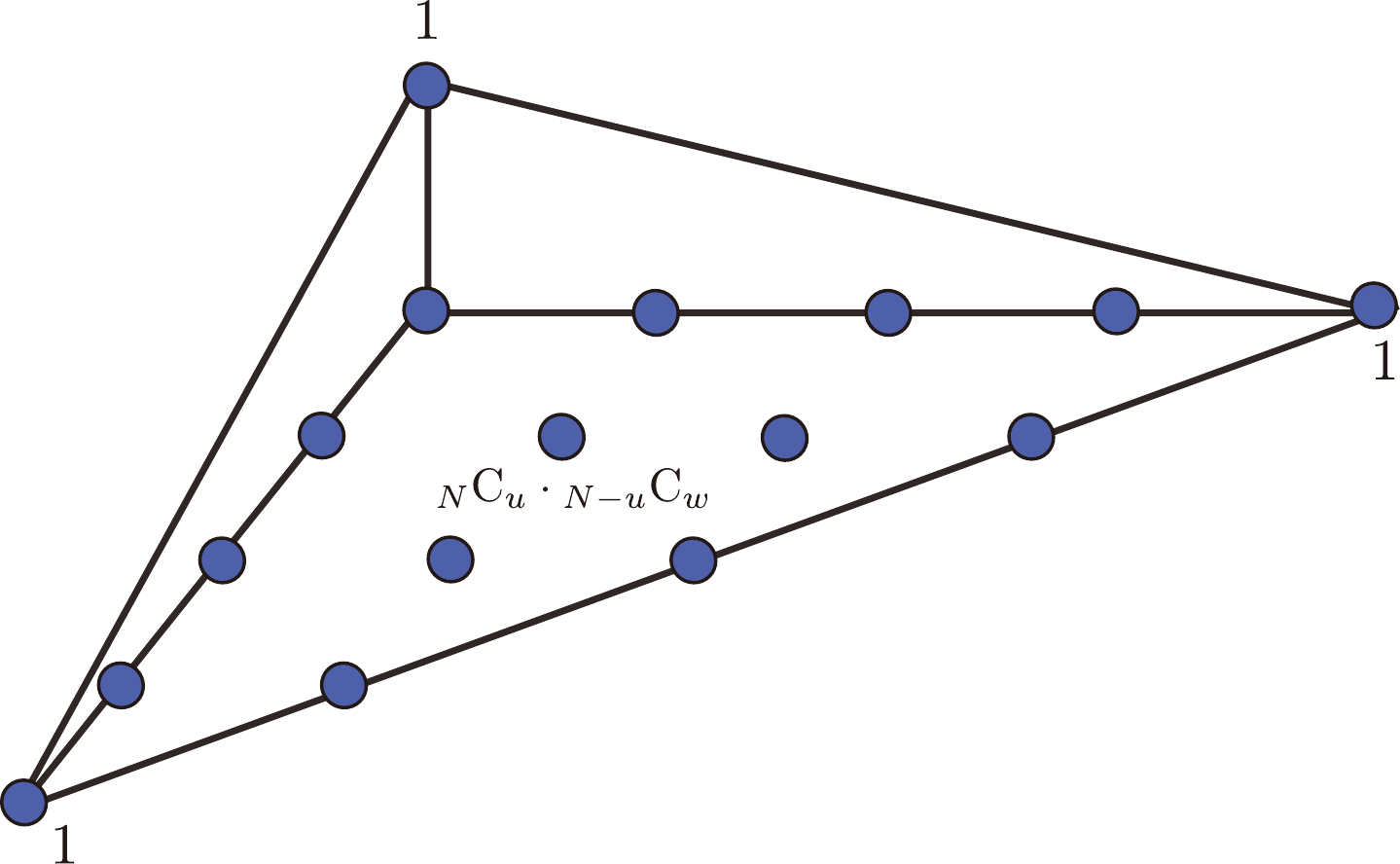}\\
\end{center}
\caption{The toric diagram of the moduli space of the un-Higgsed theory.
It is precisely the abelian orbifold $\mathbb{C}^3/( \mathbb{Z}_N \times \mathbb{Z}_N) \times \mathbb{C}$.}
\label{c3znznctoric}
\end{figure}
We prove it in Appendix.\ref{sec:perm}.
Thus the permanent for the un-Higgsed theory is 
\begin{align}
\label{permkceznunhig}
\textrm{pern}K(a,a^{\prime},b,c;x,y,z)=\prod_{n=1}^{N}(a_n+z{a^{\prime}}_n+xb_n)+y\prod_{n=1}^{N}c_n.
\end{align}
Here we use $n_{{a^{\prime}}_n}=1$.
Each monomial of this polynomial is associated with a  perfect matching, and
a monomial  weighted with $x^uy^vz^w$ describes a point $(u,v,w)$ in the toric diagram of the moduli space.
Thus the moduli space of the quiver Chern-Simons theory is represented by 
the toric diagram Fig.\ref{c3znznctoric}.
This is the orbifold $\mathbb{C}^3/( \mathbb{Z}_N \times \mathbb{Z}_N) \times \mathbb{C}$.
It is an easy combinatorics to show that the expansion of the (\ref{permkceznunhig}) implies the multiplicity
${}_N\textrm{C}_u\cdot {}_{N-u}\textrm{C}_w$ for the point $(u,0,w)$ of the toric diagram.
Hence the total multiplicity of the theory is
\begin{align}
1+\sum_{u=0}^{N}\sum_{w=0}^{N-u}{}_N\textrm{C}_u\cdot {}_{N-u}\textrm{C}_w=1+3^N.
\end{align}

Let us compute the Chern-Simons levels for this theory.
The incidence matrix $d$ is encoded in the dimer model.
For instance the nonzero matrix elements for the GLSM field ${a^{\prime}}_{n+1}$ are
\begin{align}
-d_{{2n}{{a^{\prime}}_{n+1}}}=d_{{2n+1}{{a^{\prime}}_{n+1}}}=1.
\end{align}
Thus the assignment (\ref{nc3zn}) is identical with the following choice of the Chern-Simons levels:
\begin{align}
{}^tk={}^t(d\cdot n)=(1,-1,1,-1,\cdots,1,-1).
\end{align}
Therefore we obtain the world volume theory of a M2 brane probing 
$\mathbb{C}^3/( \mathbb{Z}_N \times \mathbb{Z}_N) \times \mathbb{C}$ as
Fig.\ref{c3znzncquiver}.
This is exactly the theory proposed in  \cite{Franco:2008um}.
They computed the moduli space only for $N=3$.
Here we prove the proposal for general $N$ by utilizing the power of the Kasteleyn matrix method.

\subsection{ $\mathbb{C}^3/ (\mathbb{Z}_N\times \mathbb{Z}_M)$ grandparent}
\begin{figure}[htbp]
\begin{center}
\includegraphics[width=10cm,bb=0 0 504 237]{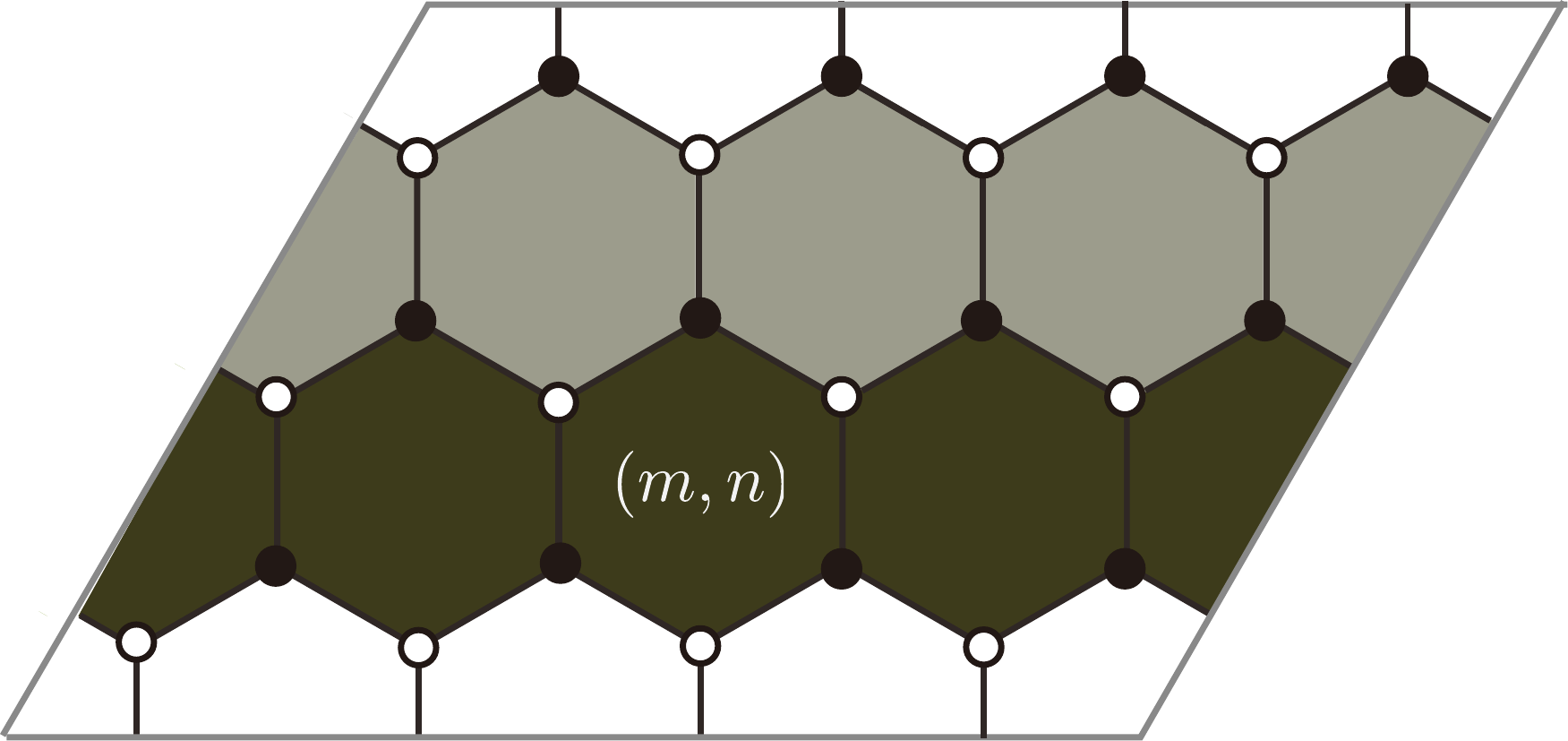}\\
\end{center}
\caption{The dimer model of  $\mathbb{C}^3/ (\mathbb{Z}_N\times \mathbb{Z}_M)$ grandparent theory.
 }
\label{c3znzm}
\end{figure}

\begin{figure}[htbp]
\begin{center}
\includegraphics[width=10cm,bb=0 0 411 238]{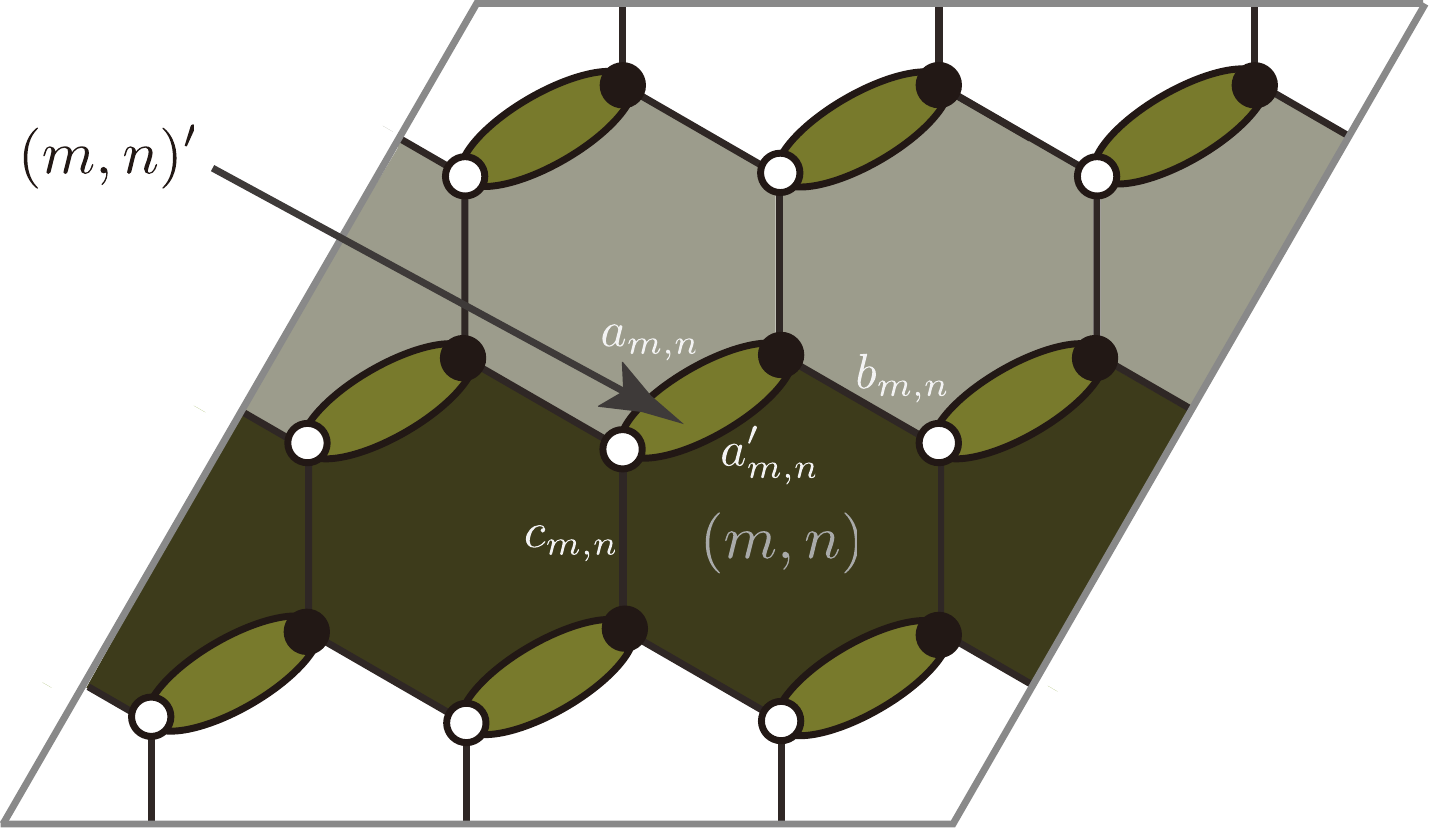}\\
\end{center}
\caption{The un-Higgsed dimer model of  $\mathbb{C}^3/ (\mathbb{Z}_N\times \mathbb{Z}_M)$.}
\label{c3znznunhig}
\end{figure}

Next we study the $\mathbb{C}^3/ (\mathbb{Z}_N\times \mathbb{Z}_M)$ grandparent theory.
The dimer model of the grandparent is shown in  Fig.\ref{c3znzm}, and
the gauge factors are indexed by $(m,n)$.
We consider the un-Higgsed theory denoted in Fig.\ref{c3znznunhig}.
This un-Higgsing is a simple extension of that in the previous subsection.

We study the quiver Chern-Simons theory with the Chern-Simons levels given by $n_{a_{mn}}=1$.
By using the incidence matrix of the un-Higgsed dimer model, this Chern-Simons levels are given by
\begin{align}
k_{(m,n)}=-1,\quad k_{(m,n)^{\prime}}=1.
\end{align}
The Kasteleyn matrix of the theory is given by
\begin{equation}
K=
\begin{pmatrix}
A_1 & B_1 & 0 & \cdots & 0 & 0 
\\0 & A_2 & B_2 &  &  0 &0   
\\  &&\hdotsfor{3}&
\\0 & 0 & 0 &   & A_{M-1} & B_{M-1} 
\\xB_M & 0 & 0 & \cdots & 0 &  A_M
\end{pmatrix}
\end{equation}
The blocks of the matrix are given by
\begin{equation}
A_m=
\begin{pmatrix}
a_{m1}+za^{\prime}_{m1} &0 & 0 & \cdots & 0 & yc_{mN} 
\\ c_{m1} & a_{m2}+za^{\prime}_{m2}& 0 &  &  0 &0   
\\  &&\hdotsfor{3}&
\\0 & 0 & 0 &   \cdots& c_{mN-1} & a_{mN}+za^{\prime}_{mN}
\end{pmatrix}
\end{equation}
and $B_m=\textrm{diag}(b_{m1},b_{m2},\cdots,b_{mN})$.

\begin{figure}[htbp]
\begin{center}
\includegraphics[width=8cm,bb=0 0 413 175]{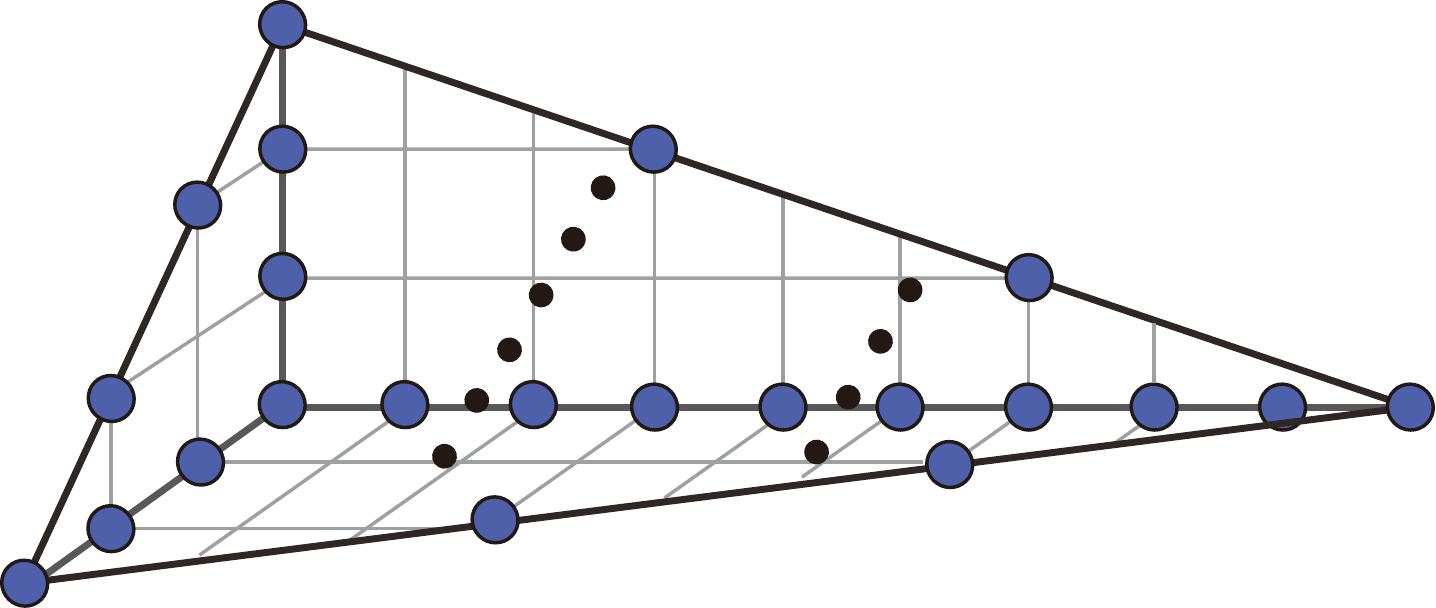}\\
\end{center}
\caption{The toric diagram of the moduli space of the un-Higgsed theory.
We omit internal points from the figure for simplicity.}
\label{c3znzmunhigtoric}
\end{figure}
Let us analyze the specific example with $M=N=3$.
The following formula, which was given in \cite{Hanany:2005ve}, is useful to compute the toric diagram of the moduli space:
\begin{align}
\textrm{det}K=\textrm{det}(A_1A_2A_3)\textrm{det}(1-x{A_1}^{-1}B_1{A_2}^{-1}B_2{A_3}^{-1}B_3).
\end{align}
For generic matrix element $a$, $b$ and $c$,  we can recast it into $\textrm{perm}K$ by forgetting signs.
Thus we obtain the toric diagram Fig.\ref{c3znzmunhigtoric}.

\subsection{ $(\mathbb{C}^2/ \mathbb{Z}_2)\times \mathbb{C}$ grandparent}
We shall discuss the grandparent theory corresponding to the orbifold $(\mathbb{C}^2/ \mathbb{Z}_2)\times \mathbb{C}$.
The dimer model of it is shown in the left of Fig.\ref{c2z2cdimer}.
This dimer is called the $\mathscr{H}_2$ model.
\subsubsection*{ New phase of  $C(dP_3)  \times \mathbb{C}$ theory: The $\mathscr{D}_2 \mathscr{H}_2$ model}
\begin{figure}[htbp]
\begin{center}
\includegraphics[width=8cm,bb=0 0 348 236]{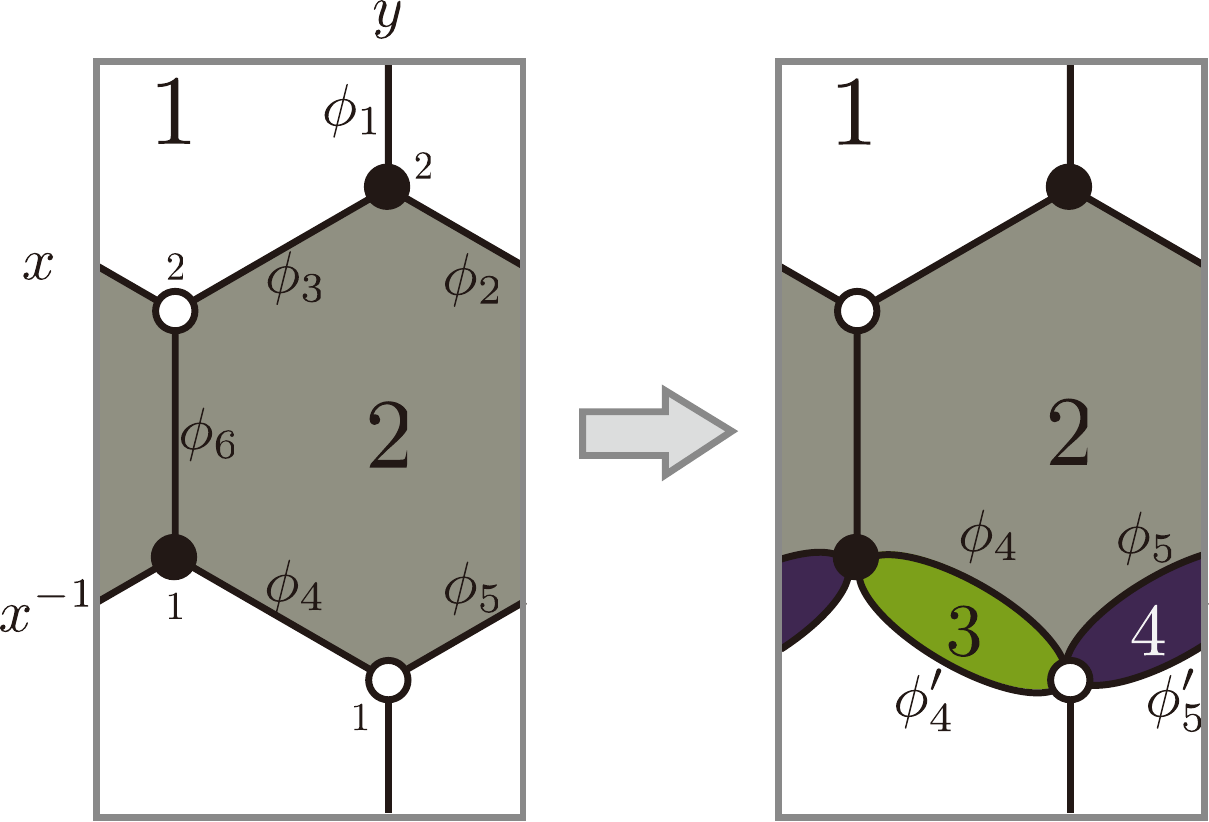}\\
\end{center}
\caption{The dimer model for $\mathbb{C}^2/ \mathbb{Z}_2  \times \mathbb{C}$ (left)
and its un-Higgsing (right).}
\label{c2z2cdimer}
\end{figure}
\begin{figure}[htbp]
\begin{center}
\includegraphics[width=4cm,bb=0 0 160 120]{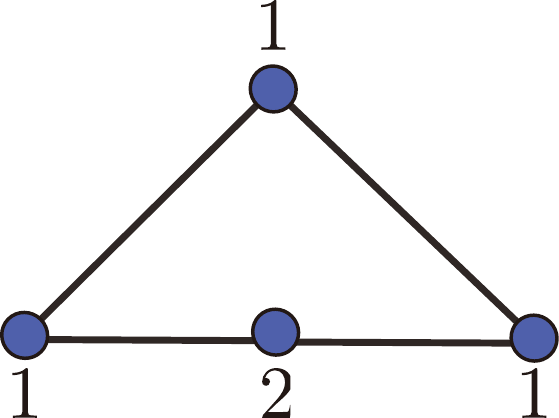}\\
\end{center}
\caption{The toric diagram of $\mathbb{C}^2/ \mathbb{Z}_2  \times \mathbb{C}$.}
\label{c2z2ctoric}
\end{figure}

We shall discuss the grandparent theory corresponding to the orbifold $(\mathbb{C}^2/ \mathbb{Z}_2)\times \mathbb{C}$.
The dimer model of it is shown in the left of Fig.\ref{c2z2cdimer}.
This dimer is called the $\mathscr{H}_2$ model.
The Kasteleyn matrix of the dimer model is the following $2\times 2$ matrix:
\begin{equation}
K=\left(\begin{array}{cc}\phi_5x^{-1}+\phi_4 & \phi_6 \\\phi_1y & \phi_2x+\phi_3\end{array}\right).
\end{equation}
Here the rows and the columns are indexed by the black and white nodes.
The permanent of the matrix consists
of five terms:
\begin{align}
\textrm{perm}K=\phi_2\phi_5+\phi_3\phi_4+x\phi_2\phi_4+x^{-1}\phi_3\phi_5
+y\phi_1\phi_6.
\end{align}
Thus we obtain the perfect matchings matrix:
\begin{equation}
P=\left.\begin{array}{c|ccccc}  
{}& p_1 & p_2 & p_3 & p_4 & p_5 
\\ \hline \phi_1& 0 & 0 & 0 & 0 & 1 
\\ \phi_2 & 1 & 0 & 1 & 0 & 0 
\\ \phi_3 & 0 & 1 & 0 & 1 & 0 
\\ \phi_4 & 0 & 1 & 1 & 0 & 0
 \\ \phi_5 & 1 & 0 & 0 & 1 & 0 
 \\ \phi_6& 0 & 0 & 0 & 0 & 1
 \end{array}\right.
\end{equation}
These five perfect matchings form the toric diagram of the moduli space.
The diagram is show in Fig.\ref{c2z2ctoric}.
This geometry is precisely the orbifold $\mathbb{C}^2/ \mathbb{Z}_2  \times \mathbb{C}$ as expected.

\begin{figure}[htbp]
\begin{center}
\includegraphics[width=12cm,bb=0 0 516 165]{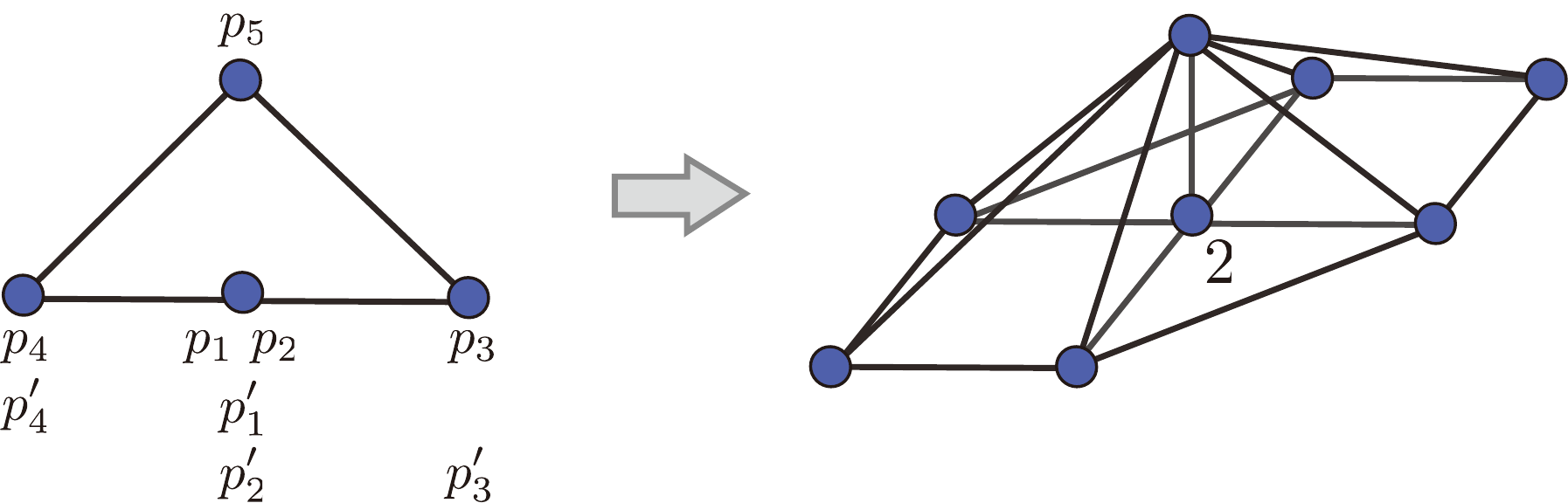}\\
\end{center}
\caption{The toric diagram of the moduli space which describes $dP_3\times \mathbb{C}$ (right) and its projection (left).}
\label{c2z2cunhig}
\end{figure}

\begin{figure}[htbp]
\begin{center}
\includegraphics[width=5cm,bb=0 0 204 238]{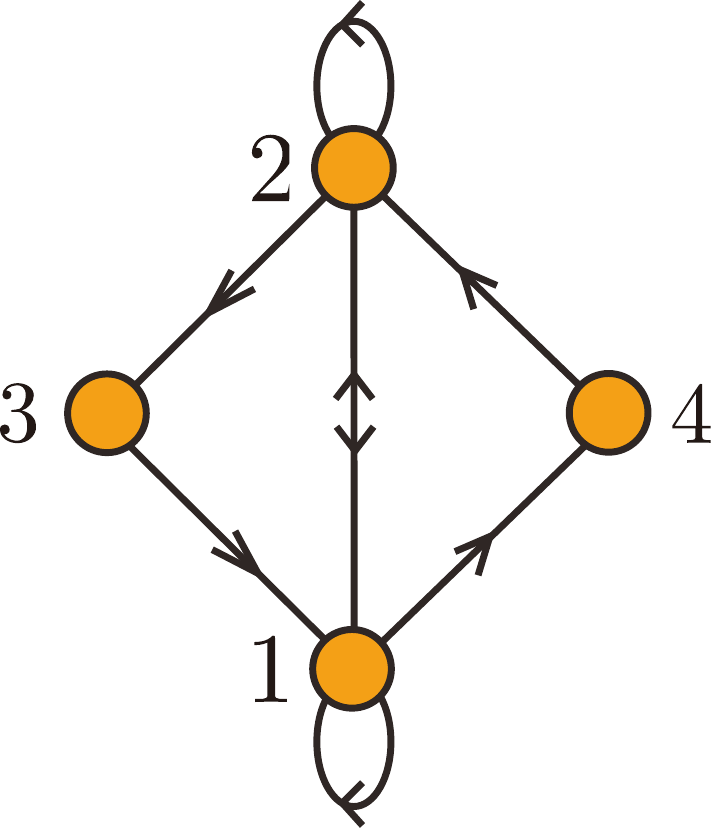}\\
\end{center}
\caption{The quiver diagram of a world volume theory of M2 branes on $dP_3\times \mathbb{C}$.
The Chern-Simons levels are ${}^tk=(2,0,-1,-1)$.}
\label{dp3quiver}
\end{figure}

Let us leave the grandparent theory and turn to investigate un-Higgsings of the grandparent theory.
For the present, we shall concentrate on an un-Higgsing with which adds 2 double bonds.
The un-Higgsed dimer model we study now is drawn in the right of Fig.\ref{c2z2cdimer}.
We shall call this theory the two double-bonded two-hexagon model $\mathscr{D}_2 \mathscr{H}_2$.
Fig.\ref{dp3quiver} indicates the quiver diagram of the  $\mathscr{D}_2 \mathscr{H}_2$ model.
The superpotential is given by
\begin{align}
W=\tr\left( \phi_1(X_{13}X_{32}X_{24}X_{41}-X_{12}X_{21})
-\phi_6(X_{24}X_{41}X_{13}X_{32}-X_{21}X_{12}) \right).
\end{align}
The doubling procedure on the edges of the grandparent increases the multiplicities of the toric diagram.
The additional perfect matchings are
\begin{align}
&p^{\prime}_1=\phi_2\phi^{\prime}_5,\quad p^{\prime}_4=\phi_3\phi^{\prime}_5,\nonumber\\
&p^{\prime}_2=\phi_3\phi^{\prime}_4,\quad p^{\prime}_3=\phi_2\phi^{\prime}_4.
\end{align}
We can indicate this by plotting these perfect matchings on the $xy$-plane as Fig.\ref{c2z2cunhig}.
We collect them into the perfect matching matrix:
\begin{equation}
P=\left.\begin{array}{c|ccccccccc}  
{}& p_1 & p_2 & p_3 & p_4 & p_5 &p^{\prime}_1 & p^{\prime}_2  & p^{\prime}_3 & p^{\prime}_4
\\ \hline \phi_1& 0 & 0 & 0 & 0 & 1         &0&0&0&0
\\ \phi_2 & 1 & 0 & 1 & 0 & 0                   &1&0&1&0
\\ \phi_3 & 0 & 1 & 0 & 1 & 0                   &0&1&0&1
\\ \phi_4 & 0 & 1 & 1 & 0 & 0                    &0&0&0&0
\\ \phi^{\prime}_4 & 0 & 1 & 1 & 0 & 0   &0&1&1&0
 \\ \phi_5 & 1 & 0 & 0 & 1 & 0                    &0&0&0&0
  \\ \phi^{\prime}_5 & 1 & 0 & 0 & 1 & 0    &1&0&0&1
 \\ \phi_6& 0 & 0 & 0 & 0 & 1                    &0&0&0&0
 \end{array}\right.
\end{equation}
Let us focus on a specific choice of Chern-Simons levels.
We chooce
\begin{align}
n_{4^{\prime}}=-n_{5^{\prime}}=1,\quad \textrm{otherwise } n_i=0.
\end{align}
This means that the Chern-Simons levels are given by
\begin{align}
{}^tk=(2,0,-1,-1).
\end{align}
Then the perfect matchings $p^{\prime}_1$ , $p^{\prime}_2$, $p^{\prime}_3$ amd $p^{\prime}_4$ are lifted up.
Their third coordinates, which are given by the formula $q={}^tP\cdot n$, are 
\begin{align}
q_{1^{\prime}}=q_{4^{\prime}}=-1,\quad q_{2^{\prime}}=q_{3^{\prime}}=1.
\end{align}
Thus the uplifted toric diagram forms $dP_3\times \mathbb{C}$ as Fig.\ref{c2z2cunhig}. 
\subsubsection*{ New phase of  $\mathbb{C}^2/\mathbb{Z}_2 \times \mathbb{C}^2$ theory: The $\mathscr{D}_1 \mathscr{H}_2$ model}
\begin{figure}[htbp]
\begin{center}
\includegraphics[width=4cm,bb=0 0 155 237]{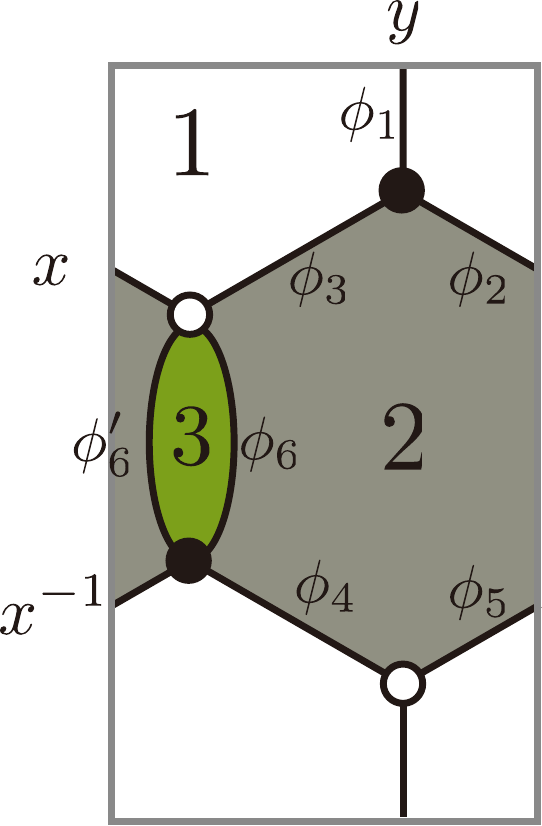}\\
\end{center}
\caption{An un-Higgsing of $\mathbb{C}^2/\mathbb{Z}_2 \times \mathbb{C}^2$ theory.}
\label{c2z2cdimer2}
\end{figure}

Next let us consider the un-Higgsed $\mathbb{C}^2/\mathbb{Z}_2 \times \mathbb{C}^2$ theory Fig.\ref{c2z2cdimer2} of different type.
We refer to it as the $\mathscr{D}_1 \mathscr{H}_2$ model.
The superpotential is given by
\begin{align}
W=\tr\left( \epsilon_{st}\phi_1X^{s}_{12}X^{t}_{21}+\epsilon_{st}X_{23}X_{32}X^{s}_{21}X^{t}_{12}\right).
\end{align}
Here $X^{s}_{12}$ and  $X^{t}_{21}$ transform as the fundamental representation under a global $SU(2)$ symmetry.

\begin{figure}[htbp]
\begin{center}
\includegraphics[width=10cm,bb=0 0 430 135]{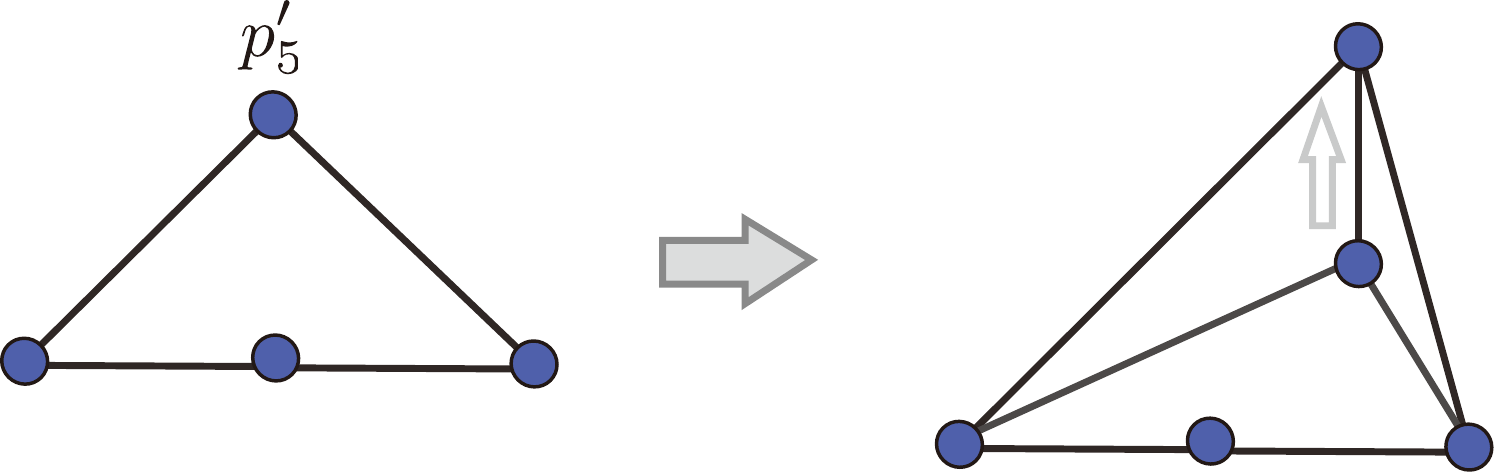}\\
\end{center}
\caption{The lift of the toric diagram.}
\label{c2z2c2toric}
\end{figure}

The Kasteleyn matrix of the dimer is
\begin{equation}
K=\left(\begin{array}{cc}\phi_5x^{-1}+\phi_4 & \phi_6+{\phi_6}^{\prime} \\ \phi_1y & \phi_2x+\phi_3\end{array}\right).
\end{equation}
Therefore its permanent consists of 6 perfect matchings:
\begin{equation}
\textrm{perm}K=\phi_3\phi_4+\phi_3\phi_5+x^{-1}\phi_3\phi_5+x\phi_2\phi_4+y\phi_1(\phi_6+{\phi_6}^{\prime} ).
\end{equation}
The toric diagram of this model therefore has the same shape of $\mathbb{C}^2/\mathbb{Z}_2 \times \mathbb{C}^2$ theory,
however there is an additional perfect matching $p^{\prime}_5=\phi_1{\phi_6}^{\prime}$ as Fig.\ref{c2z2c2toric}.
Let us lift up the point and construct a tetrahedron toric diagram.

\begin{figure}[htbp]
\begin{center}
\includegraphics[width=8cm,bb=0 0 341 53]{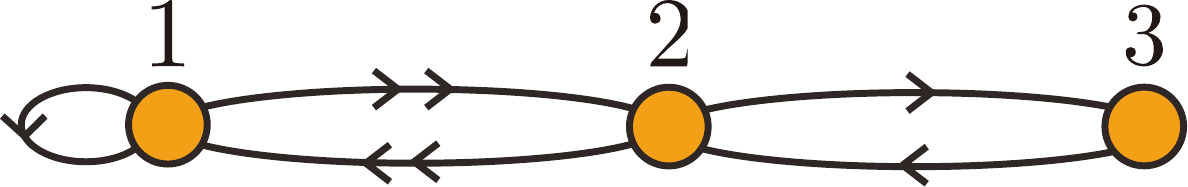}\\
\end{center}
\caption{The quiver diagram of $\mathbb{C}^2/ \mathbb{Z}_2  \times \mathbb{C}^2$ theory.
The Chern-Simons levels are given by ${}^tk=(0,-1,1)$.}
\label{c2z2c2quiver}
\end{figure}

Let us turn on $n_{6^{\prime}}=1$.
This corresponds to the following Chern-Simons levels of the quiver theory Fig.\ref{c2z2c2quiver}:
\begin{align}
{}^tk=(0,-1,1)
\end{align}
This choice of the Chern-Simons levels lifts the perfect matching 
$p^{\prime}_5$ as $q_{p_{5^{\prime}}}=P_{6^{\prime} p_{5^{\prime}}}n_{6^{\prime}}=1$.
Then the toric diagram of the modu space is given by Fig.\ref{c2z2c2toric}.
This is the toric diagram of the abelian orbifold $\mathbb{C}^2/\mathbb{Z}_2 \times \mathbb{C}^2$.
Hence we obtain the worldvolume theory of M2 branes which probe $\mathbb{C}^2/\mathbb{Z}_2 \times \mathbb{C}^2$ Fig.\ref{c2z2c2quiver}.

\subsection{ $\mathbb{C}^3$ grandparent}
\begin{figure}[htbp]
\begin{center}
\includegraphics[width=7cm,bb=0 0 250 125]{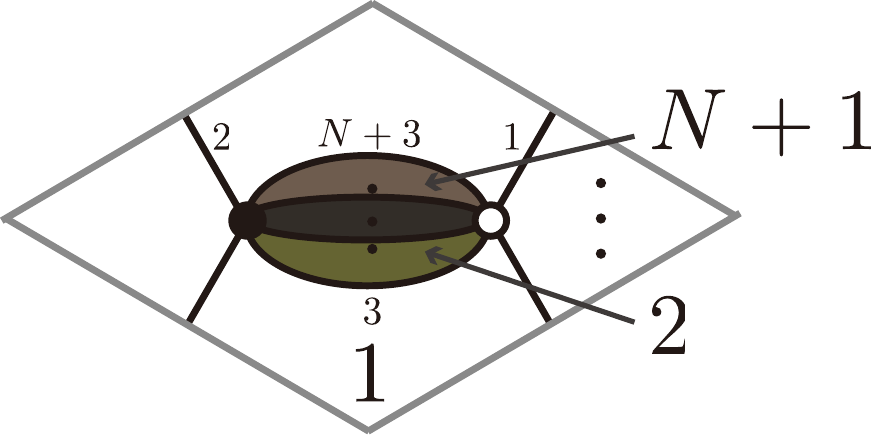}\\
\end{center}
\caption{The dimer model for an un-Higgsed $\mathbb{C}^3$ theory.}
\label{C3multidimer}
\end{figure}
\begin{figure}[htbp]
\begin{center}
\includegraphics[width=5cm,bb=0 0 204 142]{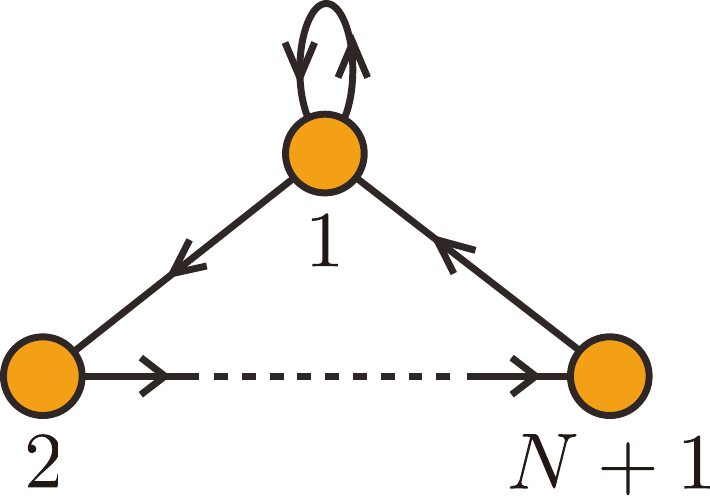}\\
\end{center}
\caption{The quiver diagram for an un-Higgsed $\mathbb{C}^3$ theory.
We study the quiver Chern-Simons theory with ${}^tk=(-N,1,1,\cdots,1)$.}
\label{C3multiquiver}
\end{figure}
\begin{figure}[htbp]
\begin{center}
\includegraphics[width=3cm,bb=0 0 113 152]{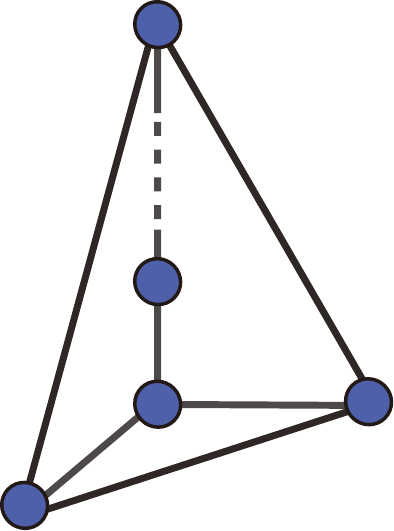}\\
\end{center}
\caption{The toric diagram of the moduli space.
All the multiplicities of the points are one.}
\label{C3multitoric}
\end{figure}

Finally let us examine a very simple grandparent theory corresponding to $\mathbb{C}^3$, 
where the corresponding dimer is a tiling of hexagons $\mathscr{H}_1$.
An un-Higgsing operation we will consider introduces a multi-bond edge $\mathscr{M}$ in a dimer model.
This is a simple extension of the "doubling" procedure, on which we have concentrated in the previous subsections. 
Let us study the un-Higgsed dimer model which  is shown in Fig.\ref{C3multidimer}.
We shall call the dimer model the one multi-bonded one hexagon model $\mathscr{M}_1 \mathscr{H}_1$.
The quiver diagram is drawn in Fig.\ref{C3multiquiver}.
This quiver theory has two adjoint chiral fields and $N+1$ bifundamentals.
The superpotential, which has two terms, is given by
\begin{align}
W=\tr (\left[\Phi_1, \Phi_2\right]X_{12}X_{23}\cdots X_{N+1 1}).
\end{align}
From Fig.\ref{C3multidimer}, we can choose perfect matchings as $p_i=\phi_i$.
Thus the perfect matching matrix is the unit matrix.
Let us consider ${}^tn=(0,0,0,1,2,\cdots,N)$, in other words we discuss the Chern-Simons levels ${}^tk=(-N,1,1,\cdots,1)$.
Using $n_{m+3}=m$, the Kasteleyn matrix of the quiver Chern-Simons theory is given by
\begin{align}
K=xp_1+yp_2+\sum_{n=1}^{N}z^{n-1}p_{2+n}.
\end{align}
Therefore the moduli space of the theory is described by the toric diagram Fig.\ref{C3multitoric}.
This toric diagram describes the orbifold $\mathbb{C}^2/\mathbb{Z}_N\times \mathbb{C}^2$.
Thus we obtain the quiver Chern-Simons theory whose moduli space is the orbifold.
This theory gives a generalization of the dual ABJM theory: 
it recover the Phase II by putting $N=1$.

\section{Phases of $C(Q^{111})$ Theory}

In this section, we study the phases of the world volume theories of a M2 brane probing the singularity $C(Q^{111})$.
This Calabi-Yau singularity is a 4-fold analogue of the conifold $\mathcal{C}=C(T^{11})$.
This 4-fold is a homogenous coset space like the conifold:
\begin{align}
SU(2)\times SU(2)\times SU(2)/U(1)\times U(1).
\end{align}
An important point is that 
the M2 brane theories we discuss here are expected to be AdS/CFT dual to the Freund-Rubin $AdS_4\times Q^{111}$ solution of M-theory vacua.
In this paper we focus on the field theory side and
we construct three quiver Chern-Simons theories whose moduli space is precisely $C(Q^{111})$.
One of these theories is new, and the others have already given in the previous works \cite{Nilsson:1984bj}.
The approach using grandparent theories gives an unified perspective for the construction of these theories.

\subsection{The $\mathscr{D}_2 \mathscr{C}$ model: an un-Higgsing of the conifold grandparent }

\begin{figure}[htbp]
\begin{center}
\includegraphics[width=8cm,bb=0 0 331 187]{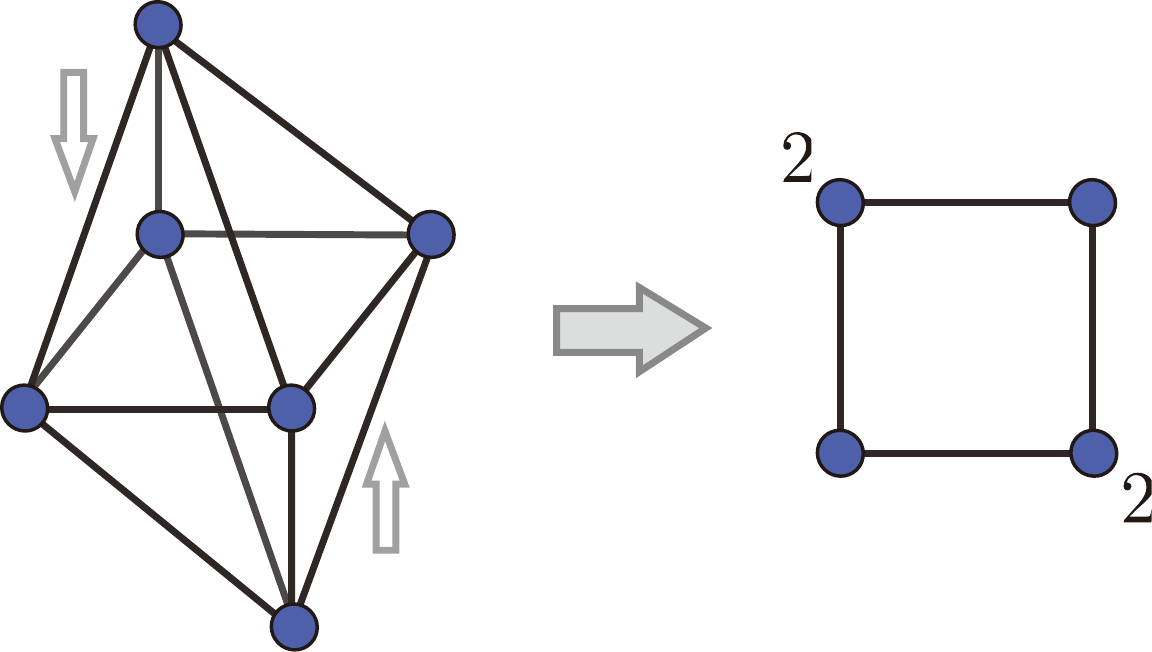}\\
\end{center}
\caption{A projection of the toric diagram which gives the conifold as the grandparent theory for $C(Q^{111})$.}
\label{q111toric1}
\end{figure}

Let us consider what is a grandparent whose toric diagram is contained in a projection of that of $C(Q^{111})$ as a subdiagram.
First we investigate a specific projection shown in Fig.\ref{q111toric1}.
We study other types of projection in the latter part of this section.
\begin{figure}[htbp]
\begin{center}
\includegraphics[width=12cm,bb=0 0 547 350]{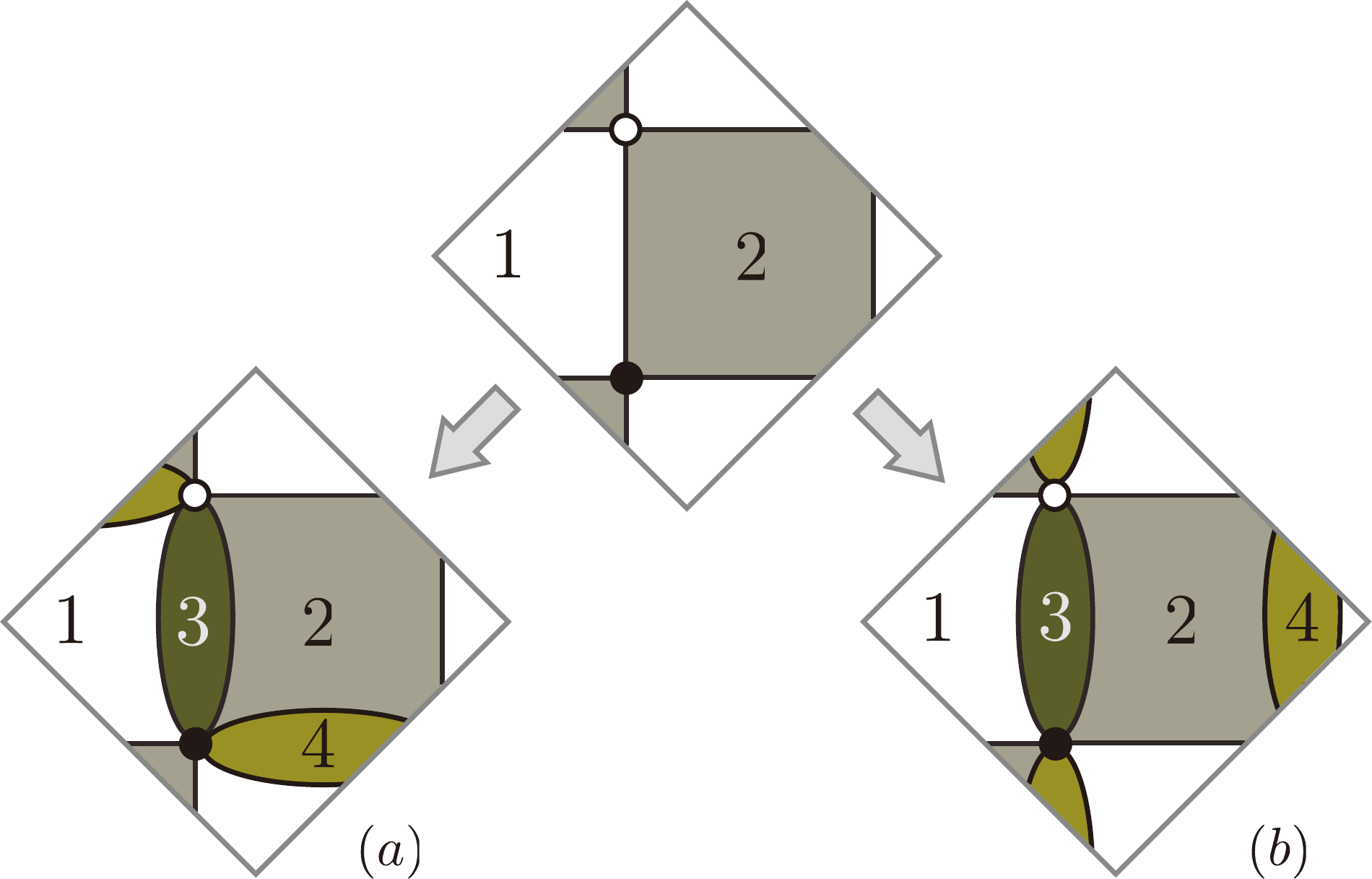}\\
\end{center}
\caption{$\mathscr{D}_2 \mathscr{C}$ models: Un-Higgsings of the $\mathscr{C}$  dimer model.}
\label{conifoldunhig}
\end{figure}
The projected toric diagram in Fig.\ref{q111toric1} is identical with one of the conifold up to multiplicities.
Thus the Calabi-Yau 4-fold $C(Q^{111})$ with this projection involves the conifold theory $\mathscr{C}$ as a grandparent theory which generates a phase of the 4-fold.

\begin{figure}[htbp]
\begin{center}
\includegraphics[width=3cm,bb=0 0 120 117]{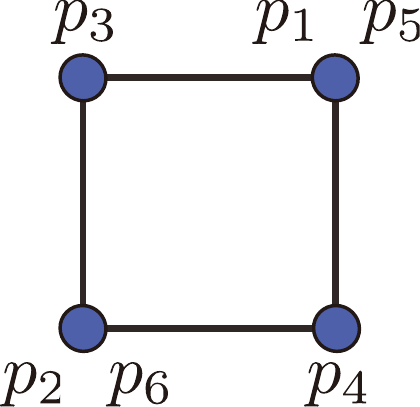}\\
\end{center}
\caption{The projected toric diagram of the un-Higgsed theory (b).}
\label{q111conifold}
\end{figure}

\begin{figure}[htbp]
\begin{center}
\includegraphics[width=5cm,bb=0 0 207 222]{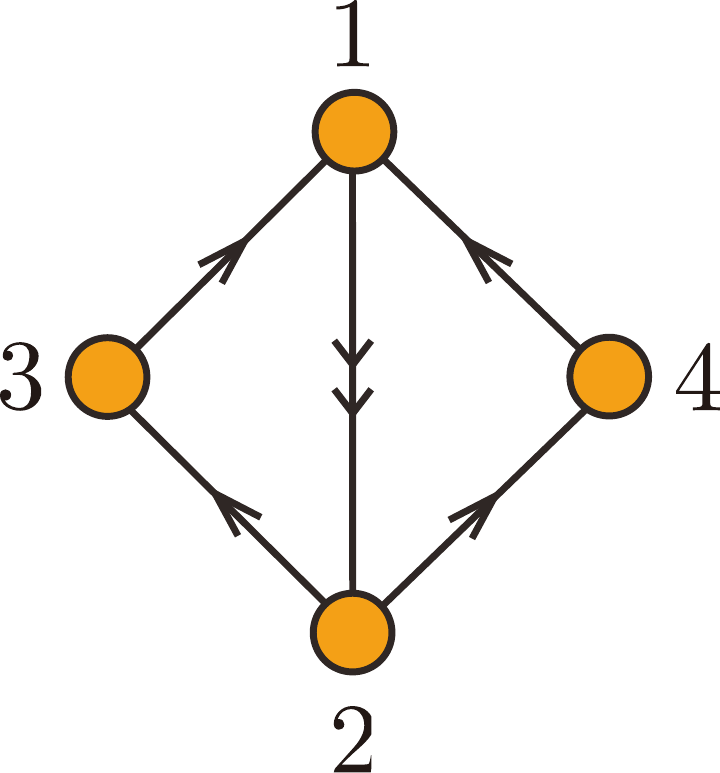}\\
\end{center}
\caption{The quiver diagram of the $\mathscr{D}_2 \mathscr{C}$ model for  $C(Q^{111})$ theory.
The Chern-Simons levels are ${}^tk=\pm(1, 1,- 1, -1 )$ or ${}^tk=\pm(1, -1,0,0 )$.}
\label{q111quiver1}
\end{figure}

Next we have to find an un-Higgsing of the grandparent theory which recovers the 4-fold $C(Q^{111})$ as the moduli space 
after turning on specific Chern-Simons levels.
There exist simple examples of un-Higgsing as Fig.\ref{conifoldunhig}.
The dimer (a) at the feft side of the figure implies a phase of $D_3$ theory, which was shown in \cite{Davey:2009sr}.
Now we investigate the another dimer (b).
The quiver diagram associated with the dimer (b) is indicated in Fig.\ref{q111quiver1}.
There are 6 matter chiral fields in this theory since the number of edges of the dimer is 6.
The superpotential of the theory is
\begin{align}
W=\epsilon_{st}\tr\left( X^{s}_{12}X_{23}X_{31}X^{t}_{12}X_{24}X_{41}\right).
\end{align}
$X^{s}_{12}$'s form a doublet of a global $SU(2)$ symmetry.
The incidence matrix of the dimer (b) is given by
\begin{equation}
\label{dq111d2c}
d=\left.\begin{array}{c|cccc}  
{}& 1 & 2 & 3 & 4  
\\ \hline X_{13}^{1} & 1 & 0 & -1 & 0 
\\ X_{13}^{2} & 0 & 1 & -1 & 0 
\\ X_{34}^{1} & 0 & 0 & 1 & -1 
\\ X_{34}^{2} & 0 & 0 & 1 & -1 
\\ X_{41}& -1 & 0 & 0 & 1
\\ X_{42}& 0 & -1 & 0 & 1 
\end{array}\right.
\end{equation}
The Kasteleyn matrix of the dimer (b) is $1\times 1$ matrix:
\begin{align}
K&=X_{13}^{1}+X_{13}^{2}x^{-1}y^{-1}+X_{34}^{1}x^{-1}+X_{34}^{2}y^{-1}+X_{41}+X_{42}x^{-1}y^{-1}\nonumber\\
&=p_{1}+p_{2}x^{-1}y^{-1}+p_{3}x^{-1}+p_{4}y^{-1}+p_{5}+p_{6}x^{-1}y^{-1}.
\end{align}
This leads to 6 perfect matchings, and we therefore obtain the suitable multiplicities as Fig.\ref{q111conifold}. 
In the following, these perfect matchings will be uplifted and lowered in order to construct an octahedron
as the toric diagram of the moduli space. 
The structure of the perfect matchings $p_{\alpha}$ are encoded in the diagonal perfect matching matrix:
\begin{equation}
P=\left.\begin{array}{c|cccccc}  
{}& p_1 & p_2 & p_3 & p_4 & p_5 & p_6  
\\ \hline X_{13}^{1} & 1 & 0 & 0 & 0 & 0 & 0
\\ X_{13}^{2} & 0 & 1 & 0 & 0 & 0 & 0
\\ X_{34}^{1} & 0 & 0 & 1 & 0 & 0& 0
\\ X_{34}^{2} & 0 & 0 & 0 & 1 & 0 & 0
\\ X_{41}& 0 & 0 & 0 & 0 & 1& 0
\\ X_{42}& 0 & 0 & 0 & 0 & 0 & 1
\end{array}\right.
\end{equation}

There exist 4 choices of uplifting.
Firsr we study the condition that the perfect matchings $p_2$ and $p_5$ (or $p_6$ and $p_1$) are uplifted and lowewed 
and the points form the toric diagram of $C(Q^{111})$.
In other words, the perfect matchings $p_2$ and $p_5$ obtain nonzero $z$-coordinates by turning on Chern-Simons levels:
\begin{align}
q_{p_2}=\pm 1,\quad q_{p_5}=\mp 1, \quad q_{\alpha}=0 \textrm{ otherwise}.
\end{align}
Using the relation $q={}^tP\cdot n$, we see that two components of the vector $n$ are nonzero:
\begin{align}
n_{X_{13}^{2} }=\pm 1,\quad n_{X_{41}}=\mp 1.
\end{align}
This choice of $n$ and the incidence matrix (\ref{dq111d2c}) imply the following choice of Chern-Simons levels:
\begin{align}
\label{cslq1111}
{}^tk=\pm( 1, 1,- 1,- 1 ).
\end{align}
This theory is one of $C(Q^{111})$ theories which was obtained in \cite{Franco:2008um}.

The another theory which was found in \cite{Franco:2008um} corresponds to an another choice of perfect matchings which will be uplifted. 
Next let us  lift up the perfect matchings $p_2$ and $p_1$ (or $p_6$ and $p_5$)
in order that the resulting toric diagram describes the 4-fold of our interest $C(Q^{111})$.
In this case, the nonzero components of $n$ are
\begin{align}
n_{X_{13}^{2} }=\pm 1,\quad n_{X_{13}^{1} }=\mp 1
\end{align}
and therefore
\begin{align}
\label{cslq1112}
{}^tk=\pm(1,- 1,0,0 ).
\end{align}
This is precisely the another choice of the Chern-Simons levels in \cite{Franco:2008um}.

Thus we rederive the quiver Chern-Simons theories Fig.\ref{q111quiver1} 
with the Chern-Simons levels (\ref{cslq1111}) and (\ref{cslq1112}), which were obtained in \cite{Franco:2008um}.
 Our derivation is relied on a projection of the toric diagram and a choice of a grandparent.
 Thus we can derive other theories for $C(Q^{111})$, since an other choice of a grandparent and an un-Higgsing
 involves a new theory.
 In the following, we find other phases of $C(Q^{111})$ theory.
 
 \subsection{The $\mathscr{S}_4$ model: Phase I of $C(\mathbb{F}_0)$ as a grandparent }
 
\begin{figure}[htbp]
\begin{center}
\includegraphics[width=12.5cm,bb=0 0 582 181]{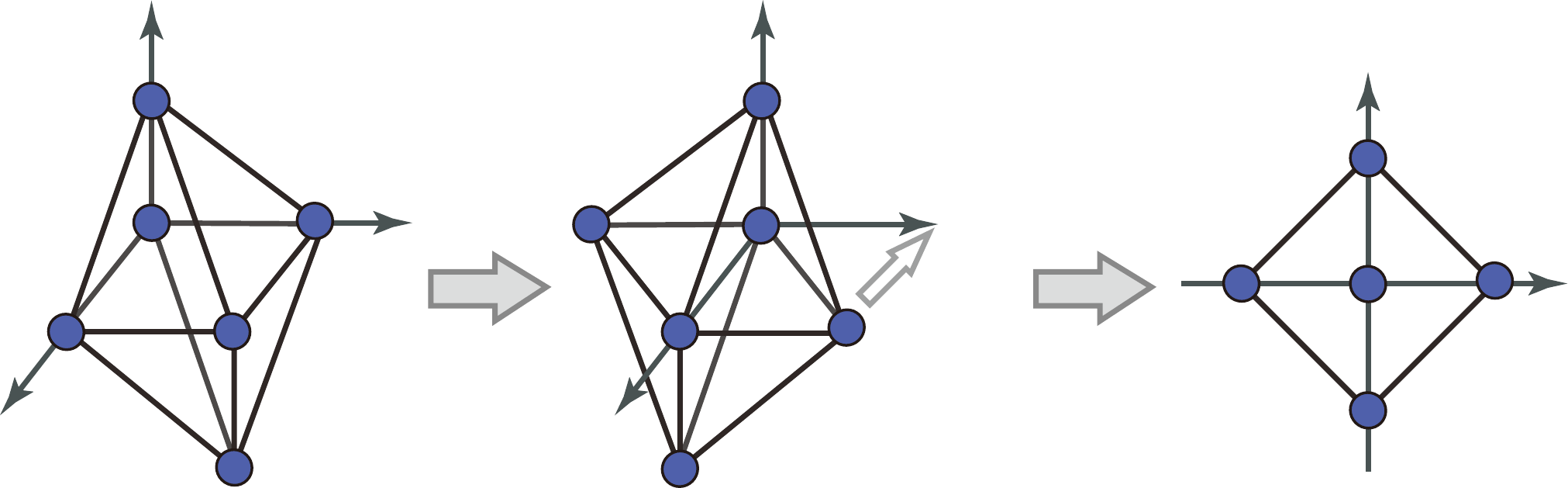}\\
\end{center}
\caption{A $SL(3, \mathbb{Z})$ transformation and projection of the toric diagram which give $C(\mathbb{F}_0)$.}
\label{q111toric2}
\end{figure}

 Let us consider the element of $SL(3, \mathbb{Z})$ transformation
 \begin{align}
 M=\left(\begin{array}{ccc}1 & -1 & 0 \\ 1 & 0 & 0 \\0 & 0 & 1\end{array}\right).
\end{align}
It transform the points of the toric diagram of $C(Q^{111})$ as follows:
\begin{align}
\left(\begin{array}{cccccc}
0 & 1 & 0 & 0 & 1 & 1 
\\0 & 0 & 1 & 0 & 1 & 1 
\\0 & 0 & 0 & 1 & 0 & -1\end{array}\right)
\to
\left(\begin{array}{cccccc}
0 & 1 & -1 & 0 & 0 & 0 
\\0 & 1 &  0& 0 & 1 & 1 
\\0 & 0 & 0 & 1 & 0 & -1\end{array}\right).
\end{align}
Here we collect the points of the toric diagram in the columns of this matrix.
The transformation is indicated with the right arrow of Fig.\ref{q111toric2}.
In this way, by rotating and projecting the toric diagram of $C(Q^{111})$, we obtain the diagram of $C(\mathbb{F}_0)$ as Fig.\ref{q111toric2}.
The $\mathbb{F}_0$ theory may be therefore able to fill the role of the grandparent theory of $C(Q^{111})$.
As we will see in this section, the $\mathbb{F}_0$ theory actually leads to quiver Chern-Simons theories whose moduli spaces are $C(Q^{111})$.

\begin{figure}[htbp]
\begin{center}
\includegraphics[width=5.5cm,bb=0 0 218 218]{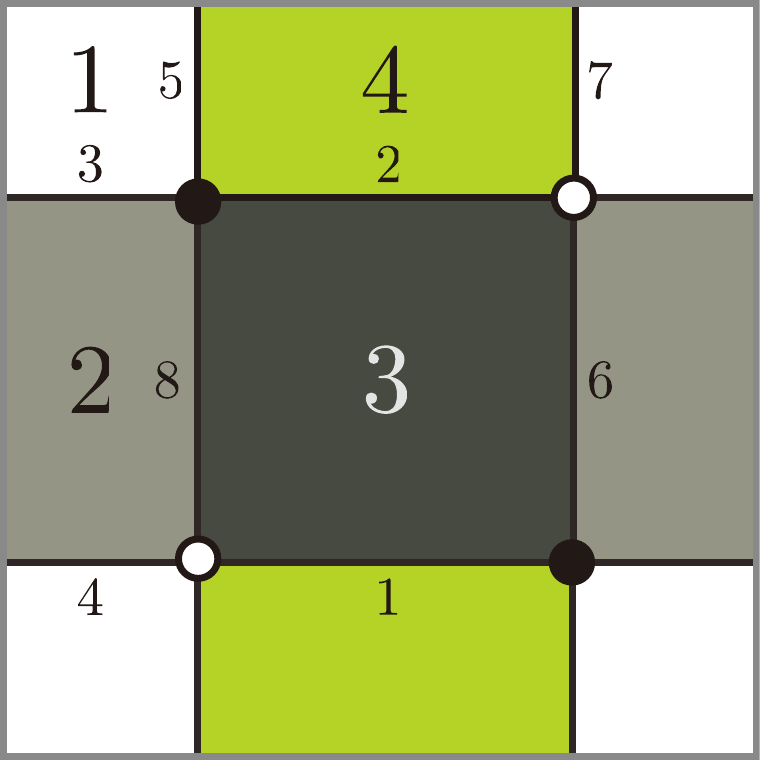}\\
\end{center}
\caption{The dimer model for .}
\label{f0phase1dimer}
\end{figure}

\begin{figure}[htbp]
\begin{center}
\includegraphics[width=4.2cm,bb=0 0 196 182]{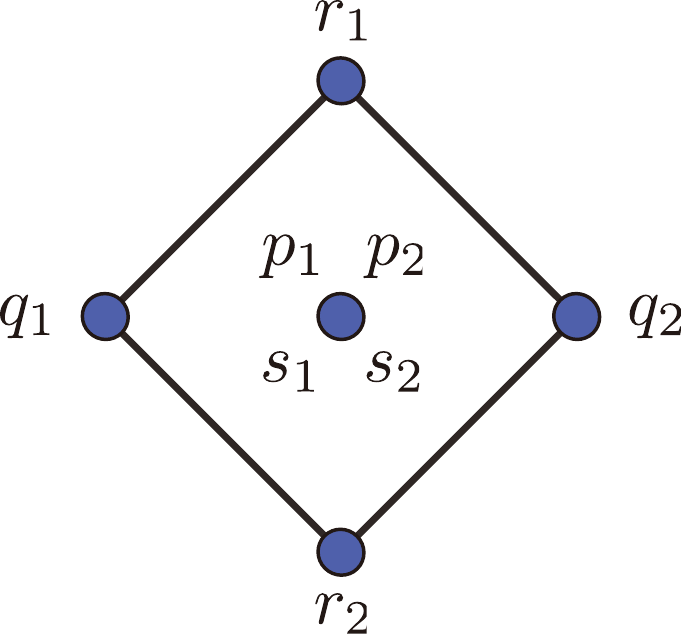}\\
\end{center}
\caption{The toric diagram of the $\mathbb{F}_0^{\textrm{I}}$ grandparent.}
\label{f0phase1}
\end{figure}

\begin{figure}[htbp]
\begin{center}
\includegraphics[width=12cm,bb=0 0 408 144]{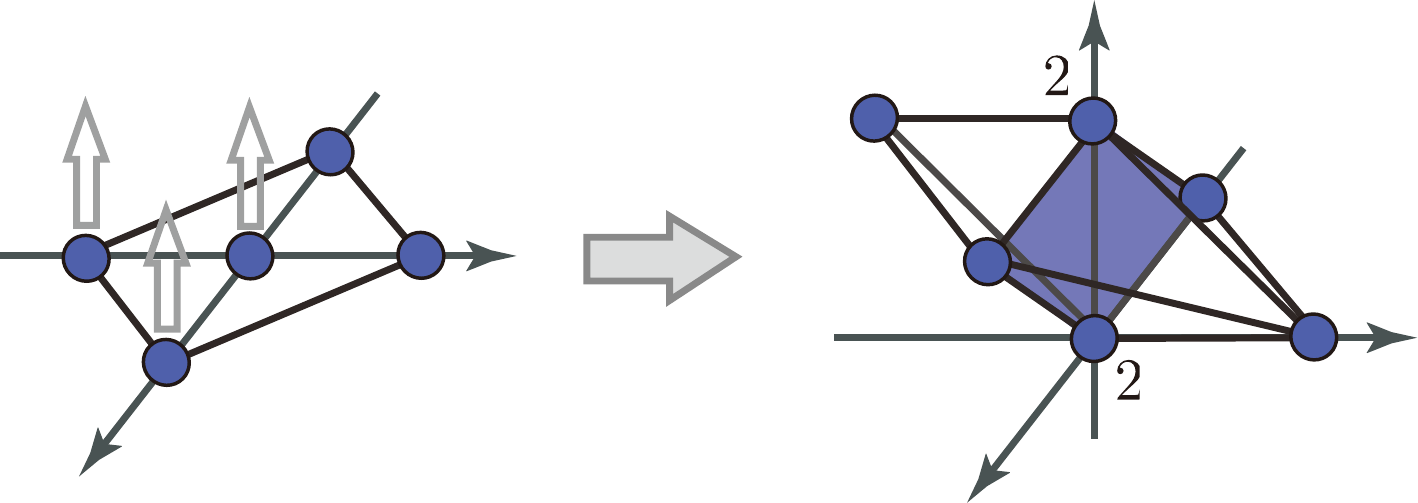}\\
\end{center}
\caption{The uplift of the toric diagram.
Two points have the multiplicitie 2.}
\label{q111toric22}
\end{figure}

We begin by studying the phase I of $\mathbb{F}_0$ \cite{Feng:2000mi}\cite{Feng:2001xr}.
Fig.\ref{f0phase1dimer} is the dimer model of this $\mathbb{F}_0^{\textrm{II}}$ theory.
There are 8 matter fields, which correspond to 8 edges of the dimer.
The indices $s,t$ and $u,v$ label the fundamentals of two global $SU(2)$'s.
The superpotential is given by
\begin{align}
W=\epsilon_{st}\epsilon_{uv}\tr\left(X^{s}_{12}X^{u}_{23}X^{t}_{34}X^{v}_{41}\right).
\end{align}
We can construct the $2\times2$ Kasteleyn matrix of the grandparent:
\begin{align}
K&=\left(\begin{array}{cc} X^{1}_{34}+X^{2}_{12}x & X^{1}_{23}+X^{2}_{41}y^{-1}
\\X^{2}_{23}+X^{1}_{41}y & X^{2}_{34}+X^{1}_{12}x^{-1}\end{array}\right)\nonumber\\
&=\left(\begin{array}{cc} \phi_{1}+\phi_{4}x & \phi_{6}+\phi_{7}y^{-1}
\\\phi_{8}+\phi_{5}y & \phi_{2}+\phi_{3}x^{-1}\end{array}\right).
\end{align}
The permanent of the matrix is given by
\begin{align}
\textrm{K}&=\phi_{1}\phi_{2}+\phi_{3}\phi_{4}+x^{-1}\phi_{1}\phi_{3}+x\phi_{2}\phi_{4}\nonumber\\
&\qquad+\phi_{5}\phi_{7}+\phi_{6}\phi_{8}    +y\phi_{5}\phi_{6}+y^{-1}\phi_{7}\phi_{8}.
\end{align}
We thus obtain 8 perfect matchings which form a square toric diagram with an internal point as Fig.\ref{f0phase1}.
The monomials of the polynomial give the perfect matching matrix:
\begin{equation}
\label{pmmf0i}
P=\left.\begin{array}{c|cccccccc}  
{}& p_1 & p_2 & q_1 & q_2 & r_1 & r_2  & s_1& s_2
\\ \hline \phi_{1} & 1& & 1& & & & &
\\ \phi_{2} & 1& & &1 & & & &
\\ \phi_{3} & & 1& 1& & & & &
\\ \phi_{4} & & 1& & 1& & & &
\\ \phi_{5} & & & & & 1& & 1&
\\ \phi_{6} & & & & & 1& & &1
\\ \phi_{7} & & & & & & 1&1 &
\\ \phi_{8} & & & & & & 1& &1
\end{array}\right.
\end{equation}
These perfect matchings form the toric diagram of $C(\mathbb{F}_0)$ as Fig.\ref{f0phase1}.
There exist many possibilities for a lift of the toric diagram according to Fig.\ref{q111toric22}.
The structure of the perfect matchings (\ref{pmmf0i}) involves the following two cases.

\subsubsection*{(i) lift of $p_1$, $s_1$(or $p_2$, $s_2$), $q_2$ and $r_2$}
First we consider the case where the points $p_1$, $s_1$, $q_2$ and $r_2$ are uplifted.
In this case, we have to choose $n$ in order that the third coordinates of the lifted toric diagram are
\begin{align}
{}^tq=(1,0,0,1,0,1,1,0).
\end{align}
Using the relation $q={}^tP\cdot n$, we find that the vector $n$ must satisfy the following equations:
\begin{align}
&n_1+n_2=1,\quad n_1+n_3=0,\quad n_3+n_4=0,\quad n_2+n_4=1\nonumber\\
&n_5+n_6=0,\quad n_5+n_7=1,\quad n_7+n_8=1,\quad n_6+n_8=0.\nonumber
\end{align}
These constraints have the following integral solutions for $l,m\in \mathbb{Z}$:
\begin{align}
{}^tn=(l,1-l,-l,l,m,-m,1-m,m).
\end{align}
Recall the incidence matrix for the dimer Fig.\ref{f0phase1dimer}:
\begin{equation}
d=\left.\begin{array}{c|cccccccc}  
{}& p_1 & p_2 & q_1 & q_2 & r_1 & r_2  & s_1& s_2
\\ \hline {1} & & & 1& 1& -1& & -1&
\\ {2} & & &-1 &-1 & & 1& &1
\\  {3} &1 & 1& & & & -1& &-1
\\  {4} & -1& -1& & & 1& & 1&
\end{array}\right.
\end{equation}
Then these choices of the vector $n$ give the unique Chern-Simons level vector:
\begin{align}
{}^tk=(-1,0,1,0).
\end{align}
The point here is that the Chern-Simons levels are independent of the choice of the integers $l$ and $m$.

\subsubsection*{(ii) lift of $p_1$, $s_2$(or $p_2$, $s_1$),  $q_2$ and $r_2$}
Next let us study the case where the third coordinate of the lifted toric diagram is given by
\begin{align}
{}^tq=(1,0,0,1,0,1,0,1).
\end{align}
We realize this uplift by imposing the following constraints for $n$.
\begin{align}
&n_1+n_2=1,\quad n_1+n_3=0,\quad n_3+n_4=0,\quad n_2+n_4=1\nonumber\\
&n_5+n_6=0,\quad n_5+n_7=0,\quad n_7+n_8=1,\quad n_6+n_8=1.\nonumber
\end{align}
They have the following integral solutions for $l,m\in \mathbb{Z}$:
\begin{align}
{}^tn=(l,1-l,-l,l,m,-m,-m,1+m).
\end{align}
These choices of the vector $n$ lead to the unique Chern-Simons level vector:
\begin{align}
{}^tk=(0,1,0,-1).
\end{align}
The Chern-Simons levels are also independent of the choice of the integers $l$ and $m$.

\begin{figure}[htbp]
\begin{center}
\includegraphics[width=4cm,bb=0 0 152 160]{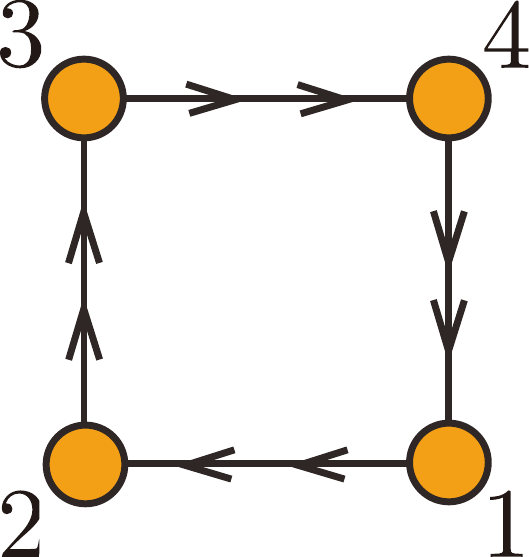}\\
\end{center}
\caption{The quiver diagram of the $C(Q^{111})$ theory.
The Chern-Simons levels are ${}^tk=\pm (1,0,-1,0)$ or $\pm (0,1,0,-1)$.}
\label{q111quiver2}
\end{figure}

We thus obtain the quiver Chern-Simons theories for $C(Q^{111})$ as is shown in Fig.\ref{q111quiver2}.
These theories have been constructed in \cite{Aganagic:2009zk}.
Our prescription gives the new derivation of the Aganagic theory by the argument of toric geometry of the moduli space.
Notice that the moduli space become the orbifold $C(Q^{111})/\mathbb{Z}_2$, as was shown in \cite{Franco:2009sp}, 
if we choose the Chern-Simons levels as follows: 
\begin{align}
{}^tk=(1,1,-1,-1).
\end{align}

 \subsection{The $\mathscr{S}_2 \mathscr{O}_2$ model: Phase II of $C(\mathbb{F}_0)$ as a grandparent }
\begin{figure}[htbp]
\begin{center}
\includegraphics[width=6cm,bb=0 0 205 201]{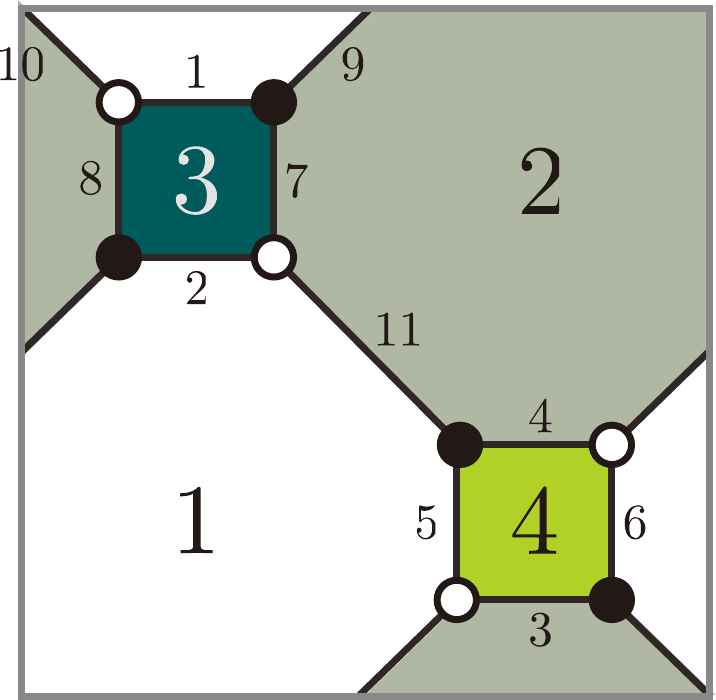}\\
\end{center}
\caption{The dimer model for the $\mathbb{F}_0^{\textrm{II}}$ theory.}
\label{f0phase2dimer}
\end{figure}

\begin{figure}[htbp]
\begin{center}
\includegraphics[width=5cm,bb=0 0 208 220]{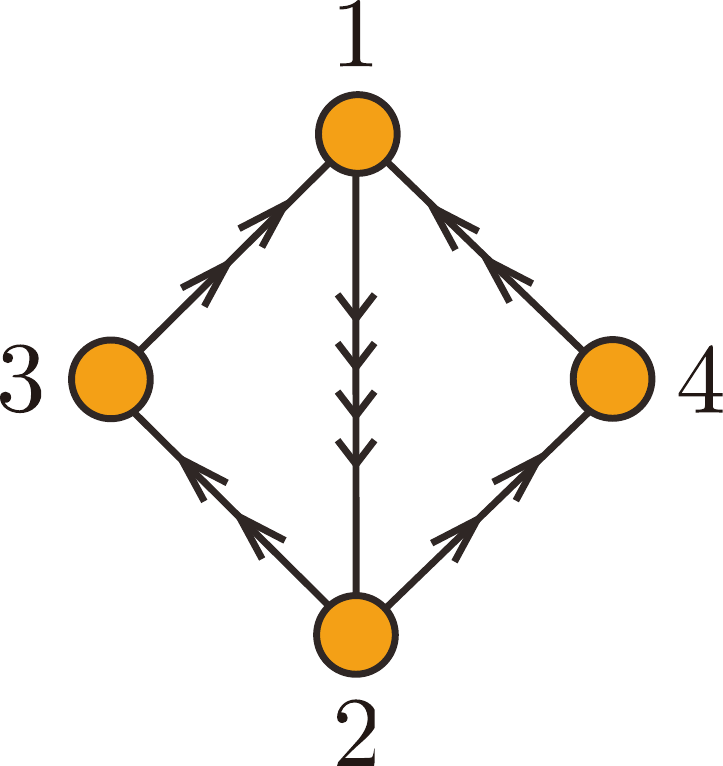}\\
\end{center}
\caption{The quiver diagram of the  $Q^{111}$ theory with the $\mathscr{S}_2 \mathscr{O}_2$ dimer model.
The Chern-Simons levels are ${}^tk=(1,-1,0,0)$ or $(-1,1,0,0)$.}
\label{q111quiver3}
\end{figure}

At the end of this section, we extend the above arguments for Phase II of the $\mathbb{F}_0$ theory \cite{Feng:2000mi}\cite{Feng:2001xr}.
The dimer model of the Phase II $\mathbb{F}_0^{\textrm{II}}$ theory is shown in Fig.\ref{f0phase2dimer}.
This theory is called the $\mathscr{S}_2 \mathscr{O}_2$ model 
since this fundamental domain consists of two squares and two octagons.
The superpotential is given by
\begin{align}
W=\epsilon_{st}\epsilon_{uv}\tr\left(X^{su}_{12}X^{v}_{23}X^{t}_{31}\right)-\epsilon_{st}\epsilon_{uv}\tr\left(X^{us}_{12}X^{v}_{24}X^{t}_{41}\right).
\end{align}
The quiver encoded in the dimer is drawn in Fig.\ref{q111quiver3}.

Let us study the $\mathbb{F}_0^{\textrm{II}}$ theory with the Kasteleyn matrix analysis.
The Kasteleyn matrix of Fig.\ref{f0phase2dimer} is given by
\begin{equation}
K=\left(\begin{array}{cccc} 
\phi_8&  \phi_1 & 0&  x^{-1}y^{-1}\phi_{10}
\\ \phi_2&  \phi_7 &  \phi_{11} & 0 
\\  0&   x\phi_{12}& \phi_5  & \phi_3  
\\ y \phi_9& 0  & \phi_4  &\phi_6
\end{array}\right).
\end{equation}
The permanent of the matrix is the following $12\times 9$ matrix:
\begin{align}
\textrm{perm}K&=\phi_1\phi_2\phi_5\phi_6+\phi_3\phi_4\phi_7\phi_8
+\phi_1\phi_2\phi_3\phi_4+\phi_5\phi_6\phi_7\phi_8+\phi_9\phi_{10}\phi_{11}\phi_{12}\nonumber\\
&\qquad +x\phi_6\phi_8\phi_{11}\phi_{12}+x^{-1}\phi_5\phi_7\phi_9\phi_{10}
+y\phi_1\phi_3\phi_9\phi_{11}+y^{-1}\phi_2\phi_4\phi_{10}\phi_{12}.
\end{align}
We find therefore 9 perfect matchings for the $\mathbb{F}_0^{\textrm{II}}$ theory.
Recall that the number of perfect matchings for the $\mathbb{F}_0^{\textrm{I}}$ theory is 8.
The additional perfect matching corresponds to the internal point at the origin of the 2 dimensional toric diagram.
Thus the multiplicity of the internal point is 5 for the $\mathbb{F}_0^{\textrm{I}}$ theory.

The structure of the perfect matchings is also different from these of  the $\mathbb{F}_0^{\textrm{I}}$ theory
and therefore we can construct $C(Q^{111})$ theory with different way.
We can see the difference by computing the  perfect matching matrix.
The perfect matching matrix is given by
\begin{equation}
P=\left.\begin{array}{c|ccccccccc}  
{}& p_1 & p_2 & q_1 & q_2 & r_1 & r_2  & s_1& s_2 &s_3
\\ \hline \phi_{1} &1 &  & & & 1& & 1& &
\\ \phi_{2}   & 1&  &  &  & & 1& 1& &
\\  \phi_{3}  &  & 1&  &  & 1& & 1& &
\\  \phi_{4}  &  & 1&  &  & & 1&1& &
\\ \phi_{5}   & 1&  & 1&  & & & &1 &
\\ \phi_{6}   & 1&  &  & 1& & & &1 &
\\ \phi_{7}   &  & 1& 1&  & & & &1 &
\\ \phi_{8}   &  &1 &  &1  & & & &1 &
\\ \phi_{9}   &  &  & 1 &  &1 & & & &1
\\ \phi_{10} &  &  & 1 &  & & 1& & &1
\\ \phi_{11} &  &  &   & 1 & 1& & & &1
\\ \phi_{12} &  &  &   & 1 & & 1& & &1
\end{array}\right.
\end{equation}

\begin{figure}[htbp]
\begin{center}
\includegraphics[width=12cm,bb=0 0 383 166]{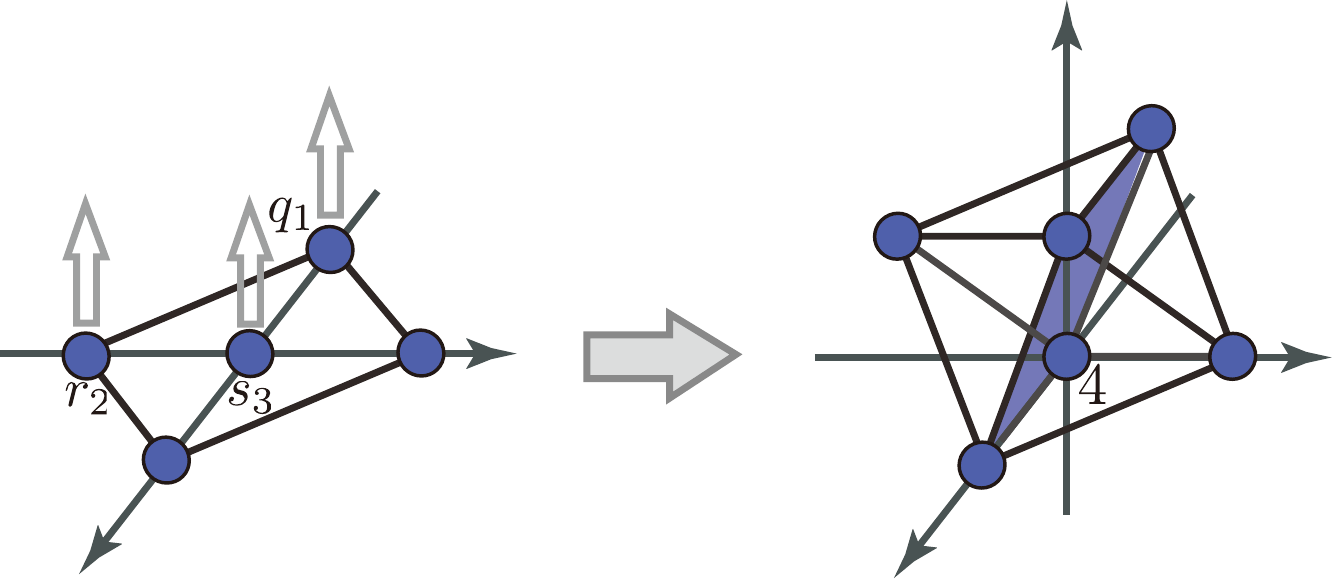}\\
\end{center}
\caption{The uplift of the toric diagram.
One of the multiplicities of the resulting toric diagram is 4.}
\label{q111toric3}
\end{figure}

In order to obtain the octahedron toric diagram of $C(Q^{111})$ as the moduli space of the resulting Chern-Simons theory, 
we have to uplift $q_1$, $r_2$, and some internal points for instance. 
First let us turn on $n_{10}$ for this purpose,
and 3 points thereby get nonzero third coordinates:
\begin{align}
q_{\alpha}=1 \textrm{ for } \alpha=q_1, r_2 ,s_3.
\end{align}
This uplift is drawn in Fig.\ref{q111toric3}.
In this way we obtain an octahedron which describes the toric data of $C(Q^{111})$.
This choice of $n$ corresponds to the following Chern-Simons levels:
\begin{align}
{}^tk=(1,-1,0,0).
\end{align}

We can also obtain the same moduli space by turning on $n_2$ and $n_7$.
We choose $n$ as
\begin{align}
n_2=n_7=1,\quad n_i=0 \textrm{ otherwise},
\end{align}
thereby lifting six points along the direction of the z-axis.
The nonzero z-coordinates are given by
\begin{align}
q_{\alpha}=1 \textrm{ for } \alpha=q_1,r_2,p_1,p_2,s_1,s_2.
\end{align}
In this way  we obtain an octahedron diagram which is $SL(3,\mathbb{Z})$ equivalent to the previous one.
The Chern-Siomons levels associated with the choice of $n$ are
\begin{align}
{}^tk=(-1,1,0,0).
\end{align}

Thus we get the $\mathcal{N}=2$ quiver Chern-Simons theory whose moduli space is the Calabi-Yau cone of the
Sasaki-Enstein manifold $Q^{111}$.
The quiver diagram and the Chern-Simons levels are shown in Fig.\ref{q111quiver3}.
We propose that this theory describes a new phase of the $Q^{111}$ theory which is a toic dual of the previous two phases.

\section{Conclusion}
In this article we study $\mathcal{N}=2$ quiver Chern-Simons theories whose moduli spaces are toric Calabi-Yau 4-fold.
We discuss the relation between Aganagic's stringy derivation of M2 brane theories and the forward algorithm of quiver Chern-Simons theories,
and we observe that the forward algorithm implies the set-up of Aganagic if the  Chern-Simons theory has a consistent 
parent theory in $3+1$ dimensions.
Meanwhile we see that Chern-Simons theories without
superconformal parent theory give fractional GLSM $U(1)$ charge, which might be a sign of inconsistency.
It would be of interest to give stringy understanding of this property.

We also construct quiver Chern-Simons theories which do not have a superconformal parent but have a Calabi-Yau moduli spaces.
In order to construct a quiver Chern-Simons theory associated with a specific toric 4-fold,
2-dimensional diagram obtained by projection the toric diagram onto plane is important.
These projected toric diagram, in general,  can not be realized as the moduli space of a  $3+1$ dimensional quiver theory,
and  thus we can not find parent theory for generic situation.
We find that the grandparent theory that emerges from the projected toric diagram is a useful starting point.
The moduli space of this grandparent has the toric diagram which is a sub-diagram of the projected one. 
By un-Higgsing the grandparent theory, i.e. by adding points to the toric diagram, 
we can construct the quiver theory whose toric toric diagram is the same as the projected one.
Then we can recover the 3 dimensional toric diagram by turning on an appropriate Chern-Simons levels.
Using this scheme, we give many quiver Chern-Simons theories: M2 brane theory for $\mathbb{C}^2/\mathbb{Z}_2\times \mathbb{C}^2$, 
$C(dP_3)\times \mathbb{C}$ and $C(Q^{111})$ for instance.
The grandparent theory gives a unified perspective for the derivation of toric phases of these theories.

Understanding the stringy meaning of quiver Chern-Simons theories without parent theory is an important issue.
Dualities, such as mirror symmetry\cite{Feng:2005gw}, might play a key role to derive these theory from string theory set-up.
It is also important to study AdS/CFT duality for our M2 brane theories.
Since $C(Q^{111})$, for instance, has a well-studied gravity dual, we might judge the right or wrong of our proposal.
We expect that these approach from string theory would give hint
to understand the observation that there exist many toric phases for M2 brane theories, especially theories without consistent parents.
We leave these for future work.


\section*{Acknowledgements}
M.T. is supported by JSPS Grant-in-Aid for Creative Scientific Research, No.19GS0219.

\appendix

\section*{Appendix}
\section{The cofactor expansion of permanents}
\label{sec:perm}

The permanent of a matrix $K$ is, roughly speaking, a modification  of the determinant, which is a sum over permutations without weighting by sign.
The definition is given by
\begin{align}
\textrm{perm}K=\sum_{\sigma\in S_N}K_{1\sigma(1)}K_{2\sigma(2)}\cdots K_{N\sigma(N)}.
\end{align}
It is easy to prove that we can expand the permanent using the cofactor:
\begin{align}
\label{cofact}
\textrm{perm}K&=\sum_{n=1}^{N}K_{1N}\sum_{\tilde{\sigma}\in S_N| \tilde{\sigma}(1)=n}K_{2\tilde{\sigma}(2)}\cdots K_{N\tilde{\sigma}(N)}\nonumber\\
&=\sum_{n=1}^{N}K_{1N} \textrm{ perm} \tilde{K}_{n}.
\end{align}
Here $\tilde{K}_{n}$ is the cofactor of $K$ with respect to the $1$-th row and $n$-th column.

By using the cofactor expansion (\ref{cofact}), we can compute the permanent of various Kasteleyn matrices.
In this appendix, we show the equation (\ref{permc3zn}).
The application of the cofactor expansion (\ref{cofact}) to the Kasteleyn matrix (\ref{kastc3zn}) implies
the following relation:
\begin{eqnarray}
\nonumber
\textrm{perm}K(a,b,c;x,y)
&\equiv& \textrm{perm}
\begin{pmatrix}
a_1+b_1x & 0 &  \cdots&  & 0 & 0 & c_Ny 
\\c_1 & a_2+b_2x &  &  & 0 & 0 & 0 
\\ &  & \hdotsfor{3} &  &  
\\0 & 0 &  &  & a_{N-2}+b_{N-2} & 0 & 0 
\\0 & 0 &  &  & c_{N_2} & a_{N-1}+b_{N-1}x & 0 
\\0 & 0 &\cdots  &  & 0 & c_{N_1} & a_N+b_Nx
\end{pmatrix}\\
\nonumber
&&=
yc_N\textrm{ perm}
\begin{pmatrix}
c_1 & a_2+b_2x &  &  & 0 & 0 
\\ &  & \hdotsfor{2} &  &   
\\0 & 0 &  &  & a_{N-2}+b_{N-2} & 0  
\\0 & 0 &  &  & c_{N-2} & a_{N-1}+b_{N-1}x  
\\0 & 0 &\cdots  &  & 0 & c_{N-1} 
\end{pmatrix}\\
\nonumber
&&\quad +(a_N+xb_N)
\textrm{ perm}
\begin{pmatrix}
a_1+b_1x & 0 &  \cdots&  & 0 & 0 
\\ c_1 & a_2+b_2x &  &  & 0 & 0  
\\ &  & \hdotsfor{2} &  &  &  
\\0 & 0 &  &  & a_{N-2}+b_{N-2} & 0  
\\0 & 0 & \cdots  &  & c_{N-2} & a_{N-1}+b_{N-1}x  
\end{pmatrix}.
\end{eqnarray}
Meanwhile we can compute the permanent of the following matrix by using the cofactor expansion:
\begin{equation}
\textrm{ perm}
\begin{pmatrix}
X_1 & Y_1 &  \hdotsfor{2} & 0 & 0 
\\ 0 & X_2 &Y_2 & &  0& 0 
\\ &  & \hdotsfor{2} &  &   
\\0 & 0 &  \hdotsfor{2}  & X_{N-1} & Y_{N-1}
\\0 & 0 &\hdotsfor{2}& 0 & X_{N} 
\end{pmatrix}
=\prod_n X_n.
\end{equation}
By applying this formula to $\textrm{perm}K(a,b,c;x,y)$, we find
\begin{align}
\textrm{perm}K(a,b,c;x,y)=y\prod_{n}c_n+\prod_n (a_n+xb_n).
\end{align}


\end{document}